\documentclass[a4paper,10pt]{article}
\usepackage{jheppubMod}
\usepackage[utf8x]{inputenc}
\usepackage{makecell}
\usepackage{subfigure}
\usepackage{ulem}
\usepackage[svgnames,dvipsnames]{xcolor}
\usepackage{pdflscape}
\usepackage{afterpage}
\usepackage{multirow}

\hypersetup{
  unicode=true,
  pdftoolbar=true,
  pdfmenubar=true,
  pdffitwindow=false,
  pdfstartview={FitH},
  pdftitle={Hard processes in multi-TeV ion collisions --
  Fuks, et al, PRD111, 014025 (2025) [arXiv:2405.19399]},
  pdfauthor={Fuks, et al},
  pdfsubject={Subject},pdfcreator={Creator},pdfproducer={Producer},
  pdfkeywords={Hadron Colliders}{Ion Collisions}{Standard Model}{Multiboson Production},
  pdfnewwindow=true,
  colorlinks=true,
  linkcolor=blue,
  citecolor=ForestGreen,
  filecolor=magenta,
  urlcolor=ForestGreen
}

\graphicspath{{Plots/}}

\newcommand{\orcid}[1]{\,\href{https://orcid.org/#1}{\includegraphics[width=9pt]{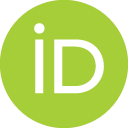}}}
\newcommand{\orcidBF}{0000-0002-0041-0566} %
\newcommand{\orcidRR}{0000-0002-3316-2175} %
\newcommand{\orcidAS}{0009-0008-0008-2815} %

\newcommand{\confirm}[1]{{\color{black}#1}}

\newcommand{\mgamc}{\textsc{mg5amc}}

\newcommand{\GeV}{{\rm ~GeV}}
\newcommand{\TeV}{{\rm ~TeV}}
\newcommand{\ub}{{~\mu\rm b}}

\title{Hard processes in multi-TeV ion collisions}

\author[a]{Benjamin Fuks\orcid{\orcidBF}}
\affiliation[a]{Laboratoire de Physique Th\'{e}orique et Hautes \'{E}nergies (LPTHE), UMR 7589,\\ Sorbonne Universit\'{e} \& CNRS, 4 place Jussieu, 75252 Paris Cedex 05, France}
\emailAdd{fuks@lpthe.jussieu.fr}

\author[b]{\!\!, Fotios Marougkas}
\affiliation[b]{Physics Department, University of Ioannina, University Campus,
45110 Ioannina, Greece}
\emailAdd{f.marugas@gmail.com}
\author[c]{\!\!, Richard Ruiz\orcid{\orcidRR}}
\affiliation[c]{Institute of Nuclear Physics Polish Academy of Sciences, 31342 Krakow, Poland}
\emailAdd{rruiz@ifj.edu.pl}

\author[d]{\!, and Alicja Sztandera\orcid{\orcidAS}}
\affiliation[d]{Faculty of Fundamental Problems of Technology,\\ Wrocław University of Science and Technology, 50370 Wrocław, Poland}

\abstract{Motivated by the ion-collision program at the Large Hadron Collider, plans for its high-luminosity upgrade, and on-going discussions for multi-TeV future hadron colliders, we systematically investigate hard-scattering, Standard Model processes in many-TeV ion-ion collisions. We focus on the symmetric beam configurations $^{208}$Pb-$^{208}$Pb, $^{131}$Xe-$^{131}$Xe, $^{12}$C-$^{12}$C, and $pp$, and we catalog total and fiducial cross sections for dozens of processes, ranging from associated-Higgs and multiboson production to  associated-top pair production, at next-to-leading order in QCD   for nucleon-nucleon collision energies from $\sqrt{s_{NN}}=1$ to $100$ TeV. We report the residual scale uncertainties at this order as well as the uncertainties originating from fits of nuclear parton densities. We also discuss the propagation of nuclear dynamics (as encoded in nuclear parton densities) into parton luminosities, and ultimately into predictions for cross sections. Finally, we report on the emergence of trends and the reliability of extrapolating cross sections across different nuclei. For Pb-Pb collisions at a hypothetical Future Circular Collider with $\sqrt{s_{NN}}=39$ TeV,
$\mathcal{O}(10^{8})$ weak bosons,
$\mathcal{O}(10^5)$ diboson pairs,
$\mathcal{O}(10^4)$ $WH$ and $ZH$ pairs,
$\mathcal{O}(10^3)$ triboson events,
$\mathcal{O}(10^5)$ high-$p_T$ photons events,
and
$\mathcal{O}(10^7)$ $t\overline{t}$ pairs can be
produced with  $\mathcal{L}=33$~nb$^{-1}$ of data.
At $\sqrt{s_{NN}}=5.52$ TeV, one can expect $\mathcal{O}(10-10^6)$ single, multiboson, and top events per $1$ nb$^{-1}$.
Decay rates and experimental selection/acceptance rates will impact final event yields,	and merits further study;
as an illustrative example,
we focus on select diboson and triboson channels
in lead-lead collisions
and discuss their observability at the high-luminosity phase of the Large Hadron Collider and the Future Circulate Collider.
}

\preprint{IFJPAN-IV-2024-7, COMETA-2024-09}
\arxivnumber{2405.19399}

\begin{document}
\maketitle
\setcounter{page}{2}

\section{Introduction}\label{sec:intro}
\begin{figure}[!t]
\includegraphics[width=.95\textwidth]{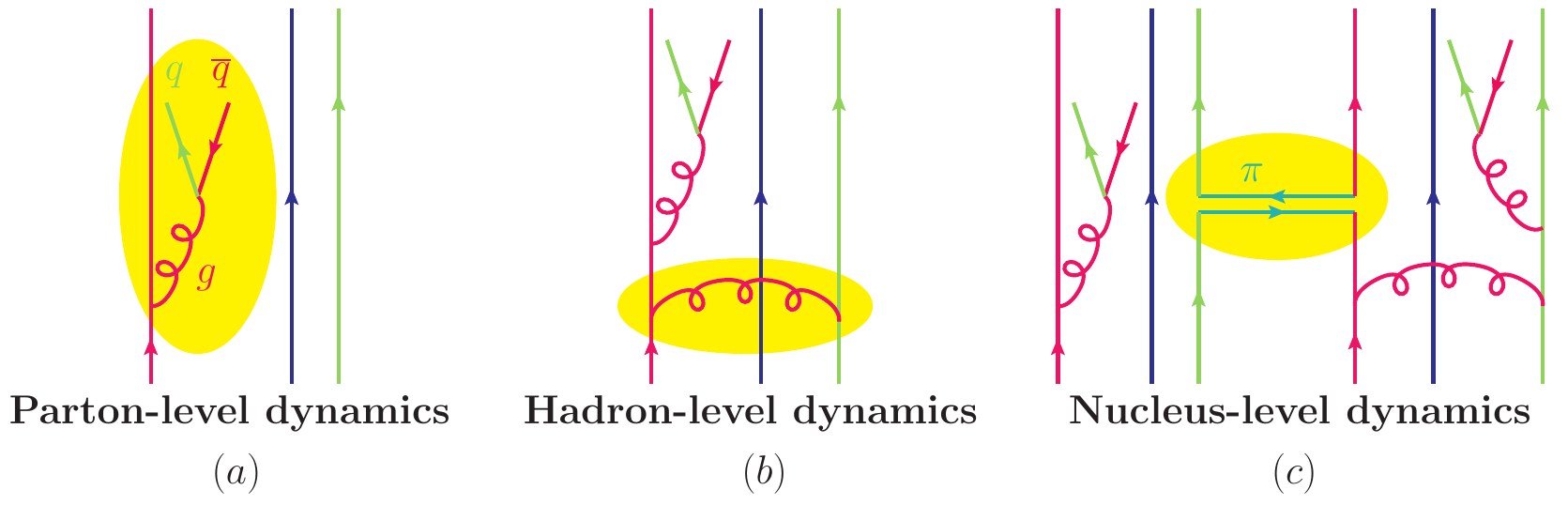}
\label{fig:diagram_QCDexchanges}
\caption{Typical QCD dynamics at (a) the parton level, \textit{e.g.}, QCD emission by one of the constituting partons of a specific nucleon, (b) the nucleon level, \textit{e.g.}, parton exchange between several of the constitute partons of a given nucleon), and (c) the nucleus level, \textit{e.g.}, pion exchange between two nucleons.}
\end{figure}

High-intensity collisions between nuclei at center-of-mass energies of many TeVs is an immensely powerful probe of the partonic structure of nuclei and the interplay of different dynamics~\cite{PHENIX:2003nhg,ALICE:2008ngc,Albacete:2016veq,Dainese:2016gch,FCC-ionsstudygroup:2017glf,Citron:2018lsq,Brewer:2021kiv}. The latter includes nucleus-level dynamics, non-perturbative dynamics at the level of individual nucleons/hadrons, as well as perturbative contributions at the partonic level. Such typical QCD interactions at the three levels are illustrated in Fig.~\ref{fig:diagram_QCDexchanges}. Modern fits of nuclear parton density functions (nPDFs) to data from lepton-nucleus deep-inelastic scattering (DIS) experiments and  nucleus-nucleus (or nucleus-proton) collision experiments show clearly that the structure of heavy, medium, and light nuclei differ both qualitatively and quantitatively from that of the proton~\cite{Kovarik:2010uv,Kovarik:2015cma,Eskola:2016oht,Walt:2019slu,Khanpour:2020zyu,Helenius:2021tof,Eskola:2021nhw,Eskola:2022rlm,Duwentaster:2022kpv,AbdulKhalek:2022fyi,nCTEQ:2023cpo}. We refer to Refs.~\cite{Muller:2012zq,Klasen:2023uqj,Achenbach:2023pba,Arslandok:2023utm} for recent reviews on nPDFs and the associated phenomenology.

At the level of scattering cross sections, differences between nucleus-nucleus collisions and proton-proton collisions are further accentuated by several additional effects. For example: the number of nucleons $(A)$ in a nucleus; the impact of Dokshitzer-Gribov-Lipatov-Altarelli-Parisi (DGLAP) scale evolution; kinematic/phase space requirements (like strong cuts imposed on the transverse momentum $[p_T]$ and the rapidity $[y]$ of particles produced in the hard-scattering process); and the opening of possibly important partonic sub-channels at next-to-leading order (NLO) in perturbation theory and beyond. The consequence is that we cannot naively extrapolate scattering rates and kinematic distributions of high-energy processes in many-TeV nuclear collisions from proton-proton collisions in a trivial but justified way.

Motivated by the ion collision programs at the Relativistic Heavy Ion Collider (RHIC) and the Large Hadron Collider (LHC), operation plans at the LHC's high-luminosity upgrade (HL-LHC)~\cite{Citron:2018lsq,Brewer:2021kiv}, and long-term community discussions regarding ion activities at a hypothetical future circular collider at even higher energies~\cite{Dainese:2016gch,FCC-ionsstudygroup:2017glf}, we explore the aforementioned interplay by systematically cataloging the scattering rates of dozens\footnote{For a full list of all 42 processes
at representative collision energies,
see Tables~\ref{tab:summary_pb208}, \ref{tab:summary_pb208_bis}, \ref{tab:summary_xe131}, \ref{tab:summary_xe131_bis}, \ref{tab:summary_cx12}, \ref{tab:summary_cx12_bis}, \ref{tab:summary_hx1} and \ref{tab:summary_hx1_bis} in Sec.~\ref{sec:xsec_summary}.} of hard-scattering / high-momentum-transfer processes in many-TeV nucleus-nucleus collisions for various nuclei. As a first step in this program, we focus on the symmetric beam configurations $^{208}$Pb-$^{208}$Pb, $^{131}$Xe-$^{131}$Xe, $^{12}$C-$^{12}$C, and (for reference) $pp$. While Pb-Pb and Xe-Xe collisions have been demonstrated at the LHC, the choice of C-C collisions is motivated by  plans for $^{16}$O-$^{16}$O collisions at the LHC~\cite{Dainese:2016gch,FCC-ionsstudygroup:2017glf}. Presently, the knowledge of nPDFs for $^{12}$C  is  far superior to that of $^{16}$O due to the availability of muon DIS data on solid
carbon~\cite{Klasen:2023uqj,Gomez:1993ri,NewMuon:1995cua,E665:1995xur,NewMuon:1995tgs,NewMuon:1996gam,NewMuon:1996yuf,Kovarik:2015cma}.
Therefore, due to their proximity on the period table and island of stability, $^{12}$C is taken as a proxy for $^{16}$O.

We consider Standard Model processes ranging from single and multiboson boson production to Higgs- and top-associated production channels. Throughout this work, we assume that the Collinear Factorization Theorem for hard, inclusive proton collisions can also describe the inclusive production of heavy states in nuclear collisions. (The observation of $t\overline{t}$ pairs and weak boson in nucleus-nucleus and proton-nucleus collisions supports this assumption~\cite{ATLAS:2012qdj,CMS:2011zfr,CMS:2012fgk,ATLAS:2014sic,CMS:2017hnw,CMS:2020aem,ATLAS:2024qdu}.) However, due to the subtleties of defining jets in a medium, we do not consider processes with high-$p_T$ jets at Born level. Furthermore, we carry out this work up to NLO in quantum chromodynamics (QCD) for nucleon-nucleon collision energies spanning the range $\sqrt{s_{NN}}=1-100$ TeV.
As an illustration of our findings, for the
benchmark scenario~\cite{FCC-ionsstudygroup:2017glf}
of a hypothetical Future Circular Collider
with $\sqrt{s_{NN}}=39$ TeV, we find that
$\mathcal{O}(10^{8})$ weak bosons,
$\mathcal{O}(10^5)$ diboson pairs,
$\mathcal{O}(10^4)$ $WH$ and $ZH$ pairs,
$\mathcal{O}(10^3)$ triboson events,
$\mathcal{O}(10^5)$ high-$p_T$ photons events,
and
$\mathcal{O}(10^7)$ $t\overline{t}$ pairs can be
produced in  Pb-Pb collisions
with  $\mathcal{L}=33$ nb$^{-1}$ of data.
At the  HL-LHC with $\sqrt{s_{NN}}=5.52$ TeV, one can expect $\mathcal{O}(10-10^6)$ single, multiboson, and top events per $1$ nb$^{-1}$.

The remainder of this study continues in the following manner: In Sec.~\ref{sec:setup} we describe the computational setup and tool chain on which our calculations are based. We then review the differences between nPDFs at different scales by comparing and contrasting PDFs of individual parton species in Sec.~\ref{sec:xpdf} and parton luminosities in Sec.~\ref{sec:lumi}. Our main results are reported in Sec.~\ref{sec:xsec}, where we present various cross sections and ratios of cross sections for the processes of interest. We take special care to report theoretical uncertainties wherever available. In Sec.~\ref{sec:extrapolation} we discuss qualitatively and quantitatively the reliability of extrapolating cross sections for one pair of nuclei from predictions for a second pair of nuclei. Finally, we summarize our findings and conclude in Sec.~\ref{sec:summary}.

\section{Computational Setup}\label{sec:setup}
Calculations of  total cross sections for nucleon-nucleon collisions (as driven by nucleus-nucleus collisions) relies on
the \textsc{MadGraph5aMC@NLO} (\mgamc) framework~\cite{Stelzer:1994ta,Alwall:2014hca}, version~\confirm{3.4.2}, and its built-in implementation of the Standard Model.
For each of the processes that we consider, rates are calculated  at leading order (LO) and NLO in QCD.
Hard-scattering matrix elements are convolved with the  \texttt{CT18NLO} (\texttt{lhaid=14400}) parton densities~\cite{Hou:2019qau} for $pp$ collisions,
 while different NLO sets of the \texttt{nCTEQ15HQ} parton densities~\cite{Kovarik:2015cma, Duwentaster:2022kpv}
are used for
Pb-Pb (\texttt{nCTEQ15HQ\_FullNuc\_208\_82}),
Xe-Xe (\texttt{nCTEQ15HQ\_FullNuc\_131\_54}),
and C-C (\texttt{nCTEQ15HQ\_FullNuc\_12\_6})
collisions.
Like other fitting groups, the \texttt{nCTEQ} collaboration does not provide LO nPDFs, meaning that we compute LO cross sections with NLO nPDFs.
Access to the PDF and nPDF libraries is provided in an automated manner through the \textsc{Lhapdf6} package~\cite{Buckley:2014ana}. These libraries also handle the running of the strong coupling constant $\alpha_s(\mu_r)$ and DGLAP evolution.

We choose our central $(\zeta=1)$ collinear factorization $(\mu_f)$ and renormalization $(\mu_r)$ scales to be
half the sum of the transverse energy of final-state particles
$(k)$ (\texttt{dynamical\_scale\_choice=-1}):
\begin{align}
    \mu_f,\ \mu_r = \zeta \times \mu_0,\quad\text{where}\quad
     \mu_0 \equiv \frac12 \sum_k \sqrt{m_k^2 + p_{Tk}^2}\ .
\end{align}
We include theoretical uncertainties originating from two distinct sources. Scale uncertainties are estimated through
the so-called nine-point method, which consists of varying
$\zeta$ discretely and independently over the range $\zeta\in\{0.5,1.0,2.0\}$.
PDF uncertainties are evaluated with the Hessian method, as required when using the \texttt{CT18} and \texttt{nCTEQ15HQ}
parton densities~\cite{Pumplin:2001ct,Kovarik:2015cma}. In these cases, the PDF libraries are shipped with an orthogonal set of $n$ pairs of PDF eigenvectors, which allows for a determination of the PDF error $\Delta\sigma_\mathrm{PDF}$ through the formula
{\small{\begin{equation}
    \Delta\sigma_\mathrm{PDF} = \sqrt{\sum_{i=1}^n\Big[ \max\big(|\sigma_{+i}-\sigma_0|, |\sigma_{-i}-\sigma_0|\big) \Big]^2}\,,
\end{equation}}}%
where $\sigma_0$ stands for the cross section value obtained when using the best PDF fit. For concreteness, we present PDF uncertainties at the 68\% confidence level (CL).

\paragraph{Accessibility:} In accordance with Findable, Accessible, Interoperable, Reusable (FAIR) principles, our simulation scripts, cross sections, and plotting routines
are available from the public repository:
\begin{center}
\href{https://gitlab.cern.ch/riruiz/public-projects/-/tree/master/IonsNLO}{https://gitlab.cern.ch/riruiz/public-projects/-/tree/master/IonsNLO}\ .
\end{center}

\section{Parton Densities}\label{sec:xpdf}

Before presenting our survey of cross sections in Sec.~\ref{sec:xsec}, we start with a review of parton densities for our representative nuclei. We focus on the four nuclei:
\begin{align}
 ^{208}\text{Pb},\
 ^{131}\text{Xe},\
 ^{12}\text{C},\ \ \text{and}\ \
 ^{1}\text{H\ (proton)},\
\end{align}
which span over two orders of magnitude in atomic mass number $(A)$ and nearly two in atomic number / proton charge $(Z)$. We do so for the following reasons: First, the LHC and its detector experiments were designed for high-energy $^{208}$Pb collisions~\cite{Pettersson:1995yyq,Bruning:2004ej}. The experimental collaborations will continue to collect lead collision data well into the high-luminosity era. It is therefore pragmatic and beneficial to focus on predictions for such collisions. Second, during Run II, a $^{131}$Xe program was initiated at the LHC~\cite{Schaumann:2018qat}. Like lead, the xenon program is slated to continue, so providing predictions is in order. Next, there are plans to initiate an $^{16}$O program at the LHC~\cite{Citron:2018lsq,Brewer:2021kiv}. However, at the moment, the oxygen data available for nPDF fits are limited. This is a qualitatively different situation for oxygen's nearby neighbor, $^{12}$C, for which there is an abundance of
muon deep-inelastic scattering on solid
carbon~\cite{Gomez:1993ri,NewMuon:1995cua,E665:1995xur,NewMuon:1995tgs,NewMuon:1996gam,NewMuon:1996yuf,Kovarik:2015cma}.
Consequently, we use carbon as a proxy for oxygen. Finally, the proton allows us to define a normalization in ratios that isolate nuclear dynamics from hadronic dynamics; predictions for (unbounded) protons will be used as a reference throughout our study.

The precise nPDFs we use are listed in Sec.~\ref{sec:setup} and consist of \texttt{nCTEQ15HQ} densities, which we analyze in detail in Sec.~\ref{sec:xpdf_densities}. Importantly, the \texttt{nCTEQ15HQ} distributions are fit to heavy quark and quarkonium data that place strong constraints on the gluon nPDF down to $x\sim10^{-5}$~\cite{Duwentaster:2022kpv}. In order to further quantify uncertainties associated with nPDF modeling, we compare our nPDFs choices to those from  contemporary families in Sec.~\ref{sec:xpdf_others}.

\subsection{Nuclear parton densities}\label{sec:xpdf_densities}
\begin{figure}
\subfigure[]{\includegraphics[width=0.48\textwidth]{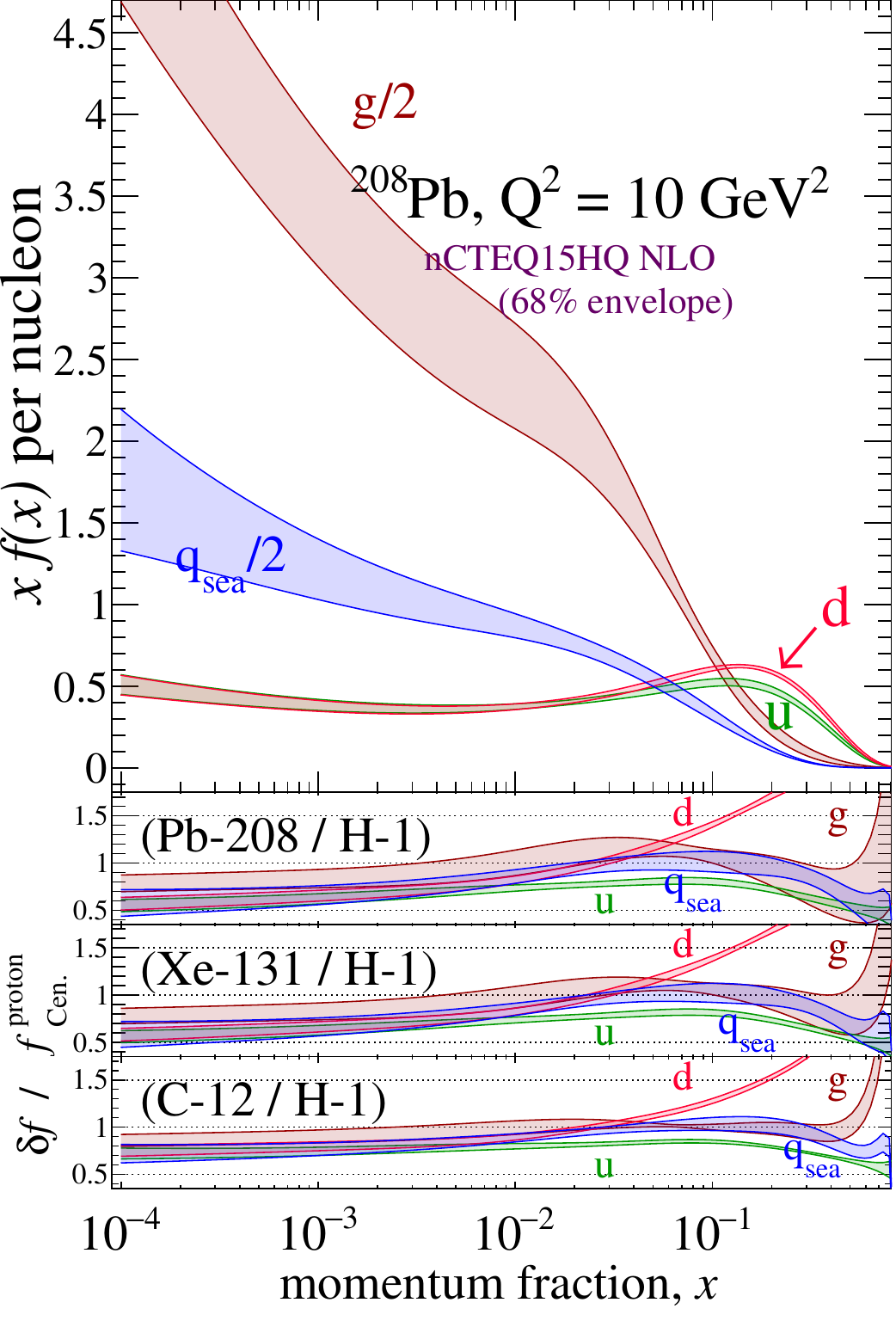}
\label{fig:ionPDFx_MultiIon_Q_xx3GeV}}
\subfigure[]{\includegraphics[width=0.48\textwidth]{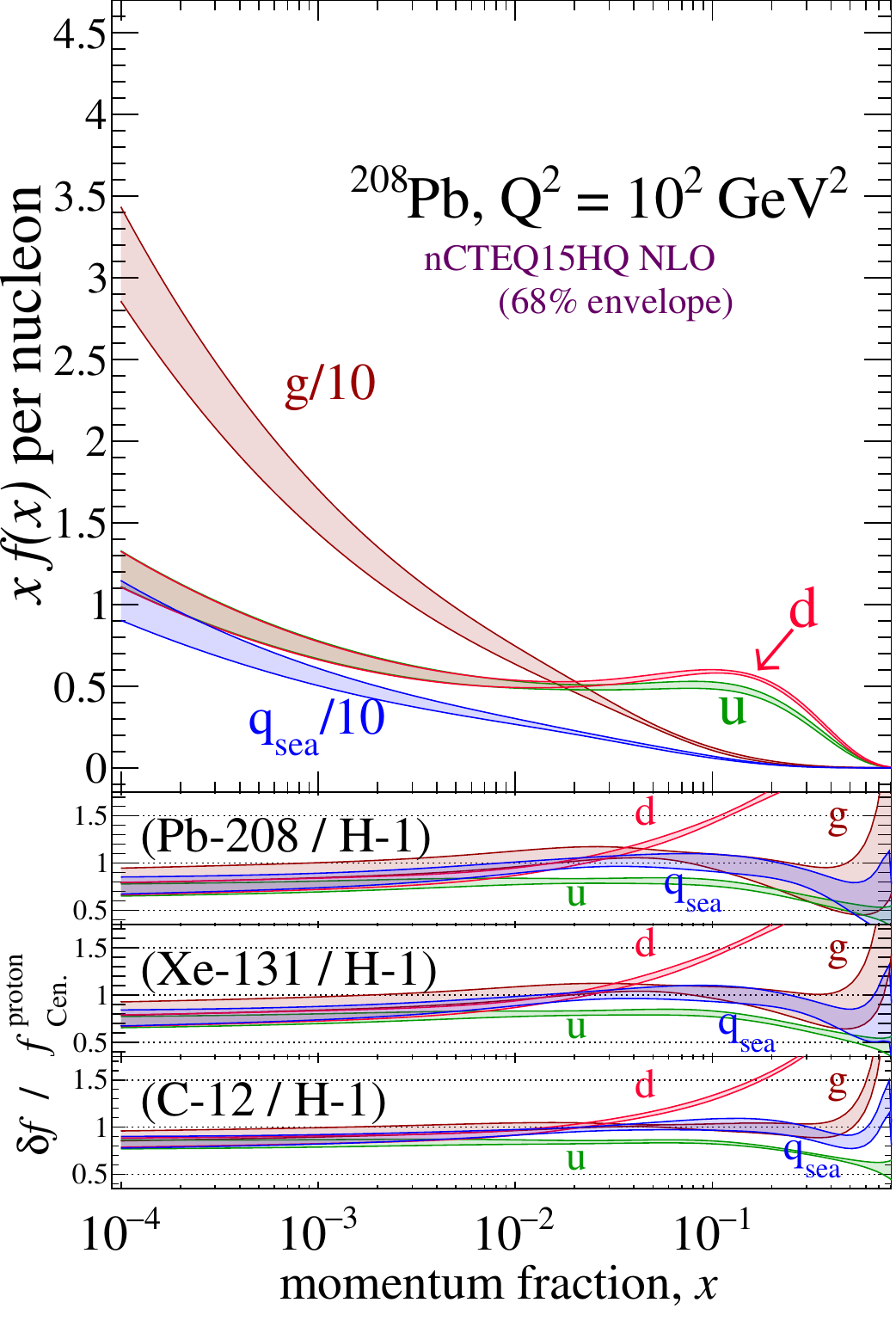}
\label{fig:ionPDFx_MultiIon_Q_x10GeV}}
  \caption{
  Top panel: For a collinear factorization scale of $Q=\sqrt{10}\GeV\approx3.2\GeV$ [\ref{fig:ionPDFx_MultiIon_Q_xx3GeV}] and $10\GeV$ [\ref{fig:ionPDFx_MultiIon_Q_x10GeV}], the quantity $x f(x)$ per nucleon as function of (averaged) nucleon momentum fraction $x$ for the gluon $(g)$, up quark $(u)$, down quark $(d)$, and sea quark $(q_{\rm sea})$ contributions to $^{208}$Pb. Band thickness corresponds to the nPDF uncertainty at 68\% CL. Lower three panels: the ratio of $x f(x)$ per nucleon relative to the corresponding quantity in the proton for $^{208}$Pb (second panel), $^{131}$Xe (third panel) and $^{12}$C (fourth panel).}
  \label{fig:ionPDFx_MultiIon_loQ}
\end{figure}

In this section, we delve into the characteristics of the collinear parton densities $(f)$ that go into our study. In the two figures shown in Fig.~\ref{fig:ionPDFx_MultiIon_loQ}, the top panel serves as our starting point, depicting the quantity $x f(x)$ per nucleon as a function of the averaged momentum fraction $x$ of a parton within a nucleon.\footnote{\label{foot:momentum_fraction}It is important to note that the averaged momentum fraction of a parton within a nucleon covers the domain $x\in[0,A]$, with a strong suppression for $x>1$. This quantity is related to the momentum fraction of a parton in the whole nucleus $x_A\in[0,1]$ by $x=A \cdot x_A$. For further discussion on momentum rescaling in nuclei, see Ref.~\cite{Ruiz:2023ozv}.}
The utility of quantifying nPDFs in terms of the weighted densities ``$xf(x)$'' stems from its robustness / invariance against rescaling:
\begin{align}
    \label{eq:pdfrescaling}
    x_N\ f^N(x_N)\ =\ (x_A A)\ \left(\frac{1}{A} f^A(x_A)\right)\ =\ x_A\ f^A(x_A)\ ,
\end{align}
where we have affixed $A$ and $N$ labels to denote nucleus- and nucleon-level quantities. This means that the weighted density per nucleon is the same weighted density at the nucleus level. Throughout the remainder of this study, we focus on nucleon-level quantities, which we denote as $x$ and $f(x)$ for brevity,  unless needed for clarity.

The contributions of the gluon $(g)$, up quark $(u)$, down quark $(d)$, and sea quark $(q_{\rm sea})$ densities for $^{208}$Pb are displayed at a collinear factorization scale of $Q=\sqrt{10}\GeV\approx3.2\GeV$ [Fig.~\ref{fig:ionPDFx_MultiIon_Q_xx3GeV}] and $10\GeV$ [Fig.~\ref{fig:ionPDFx_MultiIon_Q_x10GeV}]. Here, the sea quark distribution is defined as the sum of all quark and antiquark species in the $n_f=5$ flavor number scheme, minus the $u$ and $d$ valence distributions $u_v$ and $d_v$. Explicitly, these distributions are given by
\begin{subequations}
\begin{align}
    &q_{\rm sea} = (u-u_v) + \overline{u} +  (d-d_v) + \overline{d}
    \label{eq:def_qsea} + s + \overline{s} + c + \overline{c} + b + \overline{b}\,,\\
    &\qquad\text{where} \quad
    u_v  = u - \overline{u},\quad
    d_v  = d - \overline{d}\ .
    \label{eq:def_qval}
\end{align}
\end{subequations}
We neglect contributions from the top quark and the photon as partons of nuclei due to their absence in nPDF fits. Finally, the band thickness corresponds to the nPDF uncertainty at 68\% CL. In the second panel from the top, we show the ratio of $x f(x)$ per nucleon for $^{208}$Pb and $^{1}$H, \textit{i.e.}, $x f(x)$ of the proton. Similarly, the third and bottom panels display the same ratio with respect to the proton's densities but for $^{131}$Xe and $^{12}$C respectively.

We focus first on the $Q=\sqrt{10}\GeV\approx3.2\GeV$ case [\ref{fig:ionPDFx_MultiIon_Q_xx3GeV}], which corresponds roughly to the low-energy boundary at which PDFs and perturbative QCD are meaningfully defined. For lead, we report that the weighted density $x f_g(x)$ for the gluon reaches \confirm{$xf(x)\sim 5~(2)$ for $x\sim10^{-2}~(10^{-1})$} and with uncertainties at the \confirm{$10\%-15\%$ level for $x\lesssim10^{-2}$}. The sea quark distribution is about a third as large but carries comparable uncertainty for \confirm{$x\lesssim10^{-3}$}. The $u$ and $d$ distributions both maintain a value of \confirm{$x f(x)\sim 0.5$ for $x\sim10^{-4}-10^{-1}$}, with corresponding uncertainties at the \confirm{$20\%-5\%$ level}, and become strongly suppressed for \confirm{$x\gtrsim0.2$}. For \confirm{$x\lesssim10^{-2}$}, the $u$ and $d$ distributions are nearly identical in shape, normalization, and uncertainty; this suggests that they originate from $g\to q\overline{q}$ splittings, which are flavor blind for massless quarks. For \confirm{$x\gtrsim10^{-2}$}, the $d$ density is larger than the $u$ density, with local maxima at around \confirm{$x\sim0.15$}, and reflects that lead is a neutron-rich nucleus. For $^{208}$Pb, the average nucleon is about $(126/208)\approx60\%$ neutron-like and $(82/208)\approx40\%$ proton-like, implying  more valence-like $d$ quarks than valence-like $u$ quarks. The uncertainties for $g$ and $q_{\rm sea}$ parton densities are much larger at this value of $Q$ than for $u$ and $d$ densities, which largely reflects the available types of measurements used to constrain nPDFs.

In comparison to the proton [second panel of Fig.~\ref{fig:ionPDFx_MultiIon_Q_x10GeV}], we observe a variety of behaviors. For the \textit{average nucleon}, all four parton species displayed have densities that are smaller than those of the proton for \confirm{$x\lesssim 10^{-3}$}. This observation is subject to the sizable $g$ and $q_{\rm sea}$ uncertainties in this regime.
Furthermore,
due to the weak dependence on $A$ and the
relative ``largeness'' of $x$ at $\mathcal{O}(10^{-3})$,
it is unclear whether the suppression is related to saturation~\cite{Kutak:2003bd,Jalilian-Marian:2005ccm,Albacete:2014fwa,STAR:2021fgw}.
For larger momentum fractions of  \confirm{$x\sim 10^{-3}-10^{-1}$}, the $g$ and $d$ distributions exceed those of the proton, while the $q_{\rm sea}$ one is comparable to that of the proton and the $u$ density remains well below unity. For \confirm{$x\gtrsim 0.1$}, this trend continues for $d$, $u$, and $q_{\rm sea}$ and again reflects the neutron-rich nature of $^{208}$Pb. Moving from $^{208}$Pb to $^{131}$Xe (third panel) and $^{12}$C (bottom panel), we observe that the distributions of individual parton species each begin to converge and mirror that of the proton, particularly for the gluon and sea distributions and \confirm{$x\lesssim 0.1$}. The $d$ and $u$ quark distributions maintain smaller normalizations but similar shapes to the proton's distributions for \confirm{$x\lesssim 10^{-2}$} as atomic number decreases. Likewise, they maintain differences in normalization and shape for larger $x$ as atomic number decreases. In the $^{12}$C panel, predictions exhibit PDF uncertainties that are sufficiently small to observe that the $d$ and $g$ ratios continue to grow in the $x\to1$ limit. While such a trend is evident for $d$ in other nuclei, the gluon uncertainties are too large to identify this clearly. The relative increases in $d$ and $g$ for $^{12}$C with respect to the proton are larger than the relative decreases in $u$ and $q_{\rm sea}$.

The behavior in the ratios as $x\to1$ illustrates the well-known Fermi motion model for large-$x$ partons in heavy nuclei. In essence, intra-nuclear exchanges of, \textit{e.g.}, pions as illustrated in Fig.~\ref{fig:diagram_QCDexchanges}(c), between nucleons in a large nucleus will cause some nucleons to carry an energy that is larger than the average nucleon energy (and others to carry a below-average energy). A larger $A$ number corresponds to a larger variation of momenta carried by individual nucleons. In more energetic nucleons, the most energetic partons will subsequently carry an above-average energy, and hence populate the large-$x$ region. However, in single-nucleon systems like the proton, momentum conservation strongly suppresses the large-$x$ region. This leads to a ratio of PDFs (per nucleon) that is larger than unity. It then follows from momentum sum rules that the distributions should be smaller at small momentum fractions. That is to say, momentum and probability conservation dictate that \textit{increases} in parton densities at large momentum fractions should be compensated by \textit{decreases} at small momentum fractions.

Turning to Fig.~\ref{fig:ionPDFx_MultiIon_Q_x10GeV}, we show the same parton densities [in terms of the quantity $x f(x)$ per nucleon] and the same ratios to densities in the proton, but for an evolution scale of $Q=10\GeV$. At this higher scale, several qualitative and quantitative changes can be observed. Foremost is a reduction in PDF uncertainties, particularly in the range of \confirm{$x\lesssim10^{-3}-10^{-2}$} for the $q_{\rm sea}$ and $g$ distributions. Over the extended range of momentum fractions \confirm{$x\sim10^{-4}-10^{-1}$}, the $q_{\rm sea}$ uncertainty and $g$ uncertainties reduce to \confirm{$10\%\ (5\%)$ at low (high) values of $x$}. For the $u$ and $d$ distributions, the reduction in uncertainties is more modest, reaching similar values of \confirm{$10\%\ (5\%)$ at low (high) $x$} for the same range of momentum fractions. In increasing $Q\approx3\GeV$ to $Q=10\GeV$, the weighted densities $x f(x)$ themselves also increase. For example: at \confirm{$x\sim10^{-4}$} the $u$ and $d$ densities \confirm{increase by a factor of $2$} while they \confirm{remain largely the same at $x\sim0.1$}. Overall, the reduction in uncertainty is due to the growing importance of perturbative $g\to q\overline{q}$ and $q\to qg$ splitting, which is under good theoretical control, and the shrinking importance of the primordial distribution of partons in a nucleus.

\begin{figure}
\subfigure[]{\includegraphics[width=0.48\textwidth]{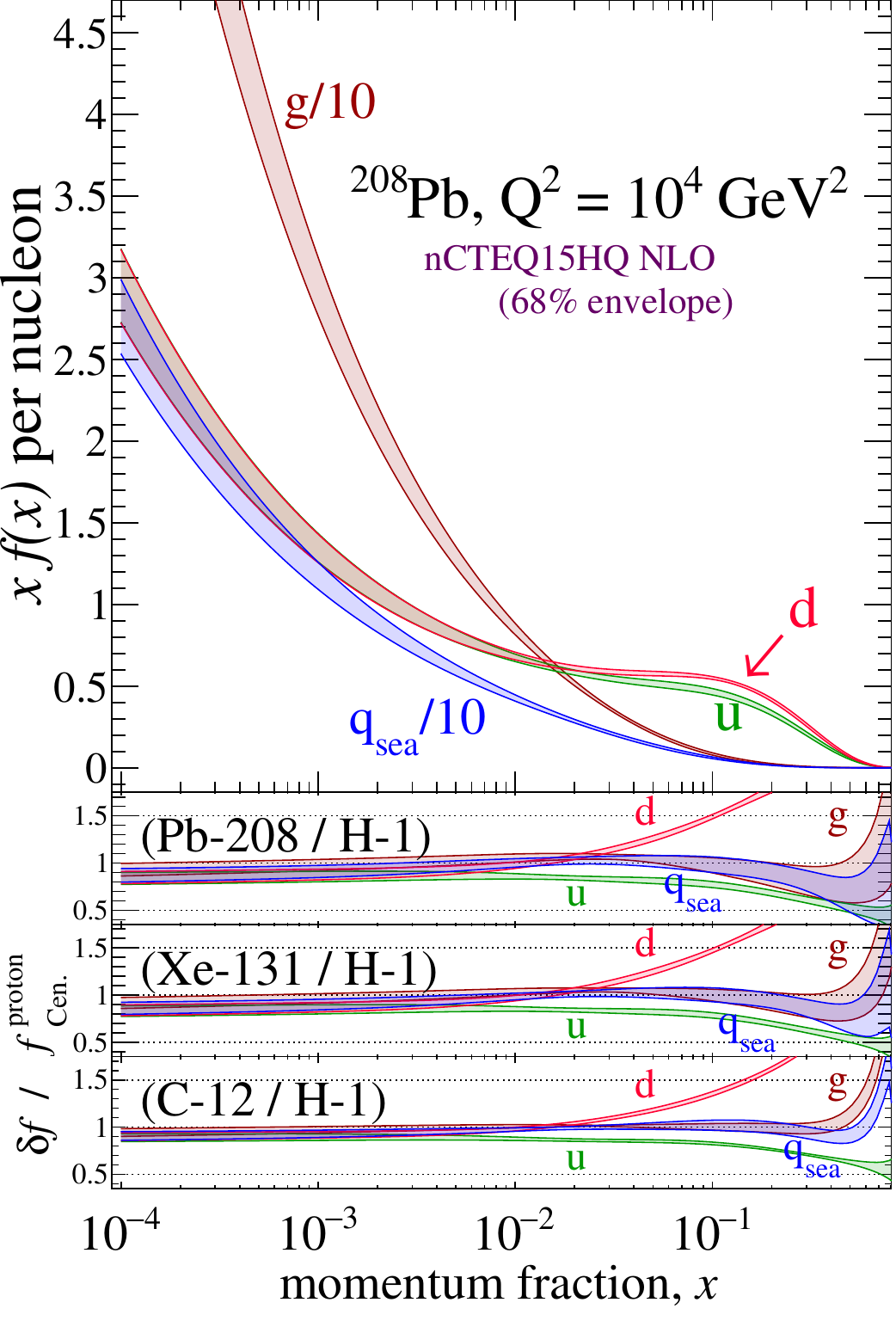}
\label{fig:ionPDFx_MultiIon_Q_100GeV}}
\subfigure[]{\includegraphics[width=0.48\textwidth]{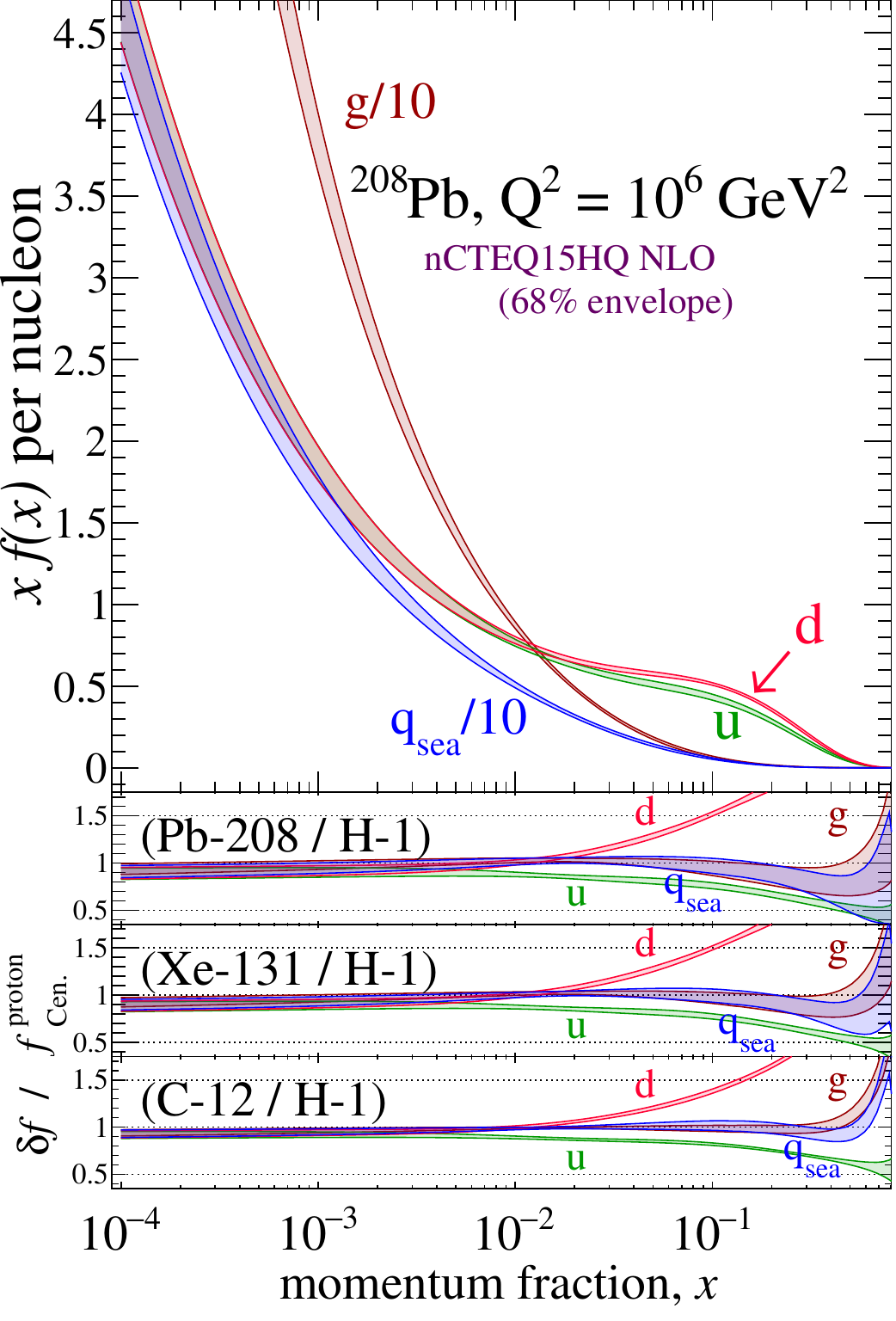}
\label{fig:ionPDFx_MultiIon_Q_xx1TeV}}
  \caption{Same as Fig.~\ref{fig:ionPDFx_MultiIon_loQ} but for $Q=100\GeV$ [\ref{fig:ionPDFx_MultiIon_Q_100GeV}] and $Q=1\TeV$ [\ref{fig:ionPDFx_MultiIon_Q_xx1TeV}].}
  \label{fig:ionPDFx_MultiIon_hiQ}
\end{figure}

Focusing on the lower panels of Fig.~\ref{fig:ionPDFx_MultiIon_loQ}, we observe the same qualitative behavior at $Q=10\GeV$ as at $Q\approx3\GeV$. Notably, however, the smaller PDF uncertainties reveal that for \confirm{$x\lesssim10^{-2}$} the shapes of ratios are largely flat and convergent, indicating that average parton densities in light and heavy nuclei differ from a free proton in only the normalization. A reduced dependence on $A$ and $Z$ factors is consistent with parton densities at low $x$ being driven more by perturbative QCD than nuclear dynamics. For \confirm{$x\gtrsim10^{-2}$}, the shapes and normalizations of densities remain qualitatively and quantitatively different from the proton. Finally, in the $x\to1$ limit, the emergence of an enhanced  $q_{\rm sea}$ distribution that mirrors the $g$ distribution at $x\to1$ can be observed. This is clearest in the  $^{12}$C ratio due to its smaller uncertainties but is also present for $^{131}$Xe and $^{208}$Pb. We again attribute the similarity of $g$ and $q_{\rm sea}$ to perturbative $g\to q\overline{q}$ splitting in  QCD.

In Fig.~\ref{fig:ionPDFx_MultiIon_hiQ} we show the same information as in Fig.~\ref{fig:ionPDFx_MultiIon_loQ} but for $Q=100\GeV$ [\ref{fig:ionPDFx_MultiIon_Q_100GeV}] and $Q=1\TeV$ [\ref{fig:ionPDFx_MultiIon_Q_xx1TeV}]. Quantitatively, the PDFs show considerable growth and a reduction of uncertainties  at these larger scales. Qualitatively, the behavior of densities and ratios of densities are the same as observed at lower $Q$, and  therefore do not need further discussion. One distinction is that the $q_{\rm sea}$ density is, to a good approximation, \confirm{$10$ times larger} than the $u$ and $d$ distributions for \confirm{$x\lesssim10^{-3}$}; in this range, the $u$ and $d$ distributions and their uncertainties are also approximately equal. This can be interpreted as $u$ and $d$ being  sea-like for \confirm{$x\lesssim10^{-3}$} and generated entirely from perturbative $g\to q\overline{q}$ splittings. Under this assumption, all five quarks and all five antiquarks have the same densities for such  $x$ and $Q$, which by the definition of $q_{\rm sea}$ in Eq.~\eqref{eq:def_qsea} predicts the observed tenfold difference.

In summary, for momentum transfers typical of LHC collisions, the densities of partons $g$, $u$, $d$, and $q_{\rm sea}$ for \confirm{$x\lesssim10^{-3}$} in an average nucleon from heavy, medium, and light nuclei are highly comparable to those found in the proton. In this region, the contribution from perturbative QCD dominates over nuclear dynamics. For the range \confirm{$x\sim10^{-3}-10^{-1}$}, there is an enhancement of $d$ quarks and a suppression of $u$ quarks, while the $g$ and $q_{\rm sea}$ distributions are very proton-like. In this region, the relative abundance of protons and neutrons in a nucleus, and hence valence-like $u$ and $d$ quarks drive the parton densities. As $x\to1$, the $d$, $g$, and $q_{\rm sea}$ distributions are enhanced relative to the proton while the $u$ distribution is suppressed. This follows from the interplay of relative proton-neutron abundance, intra-nucleon exchanges, and gluon splittings perturbative QCD.

\subsection{Alternative nPDF sets}\label{sec:xpdf_others}

\begin{figure}
  \centering
\subfigure[]{\includegraphics[width=0.32\textwidth]{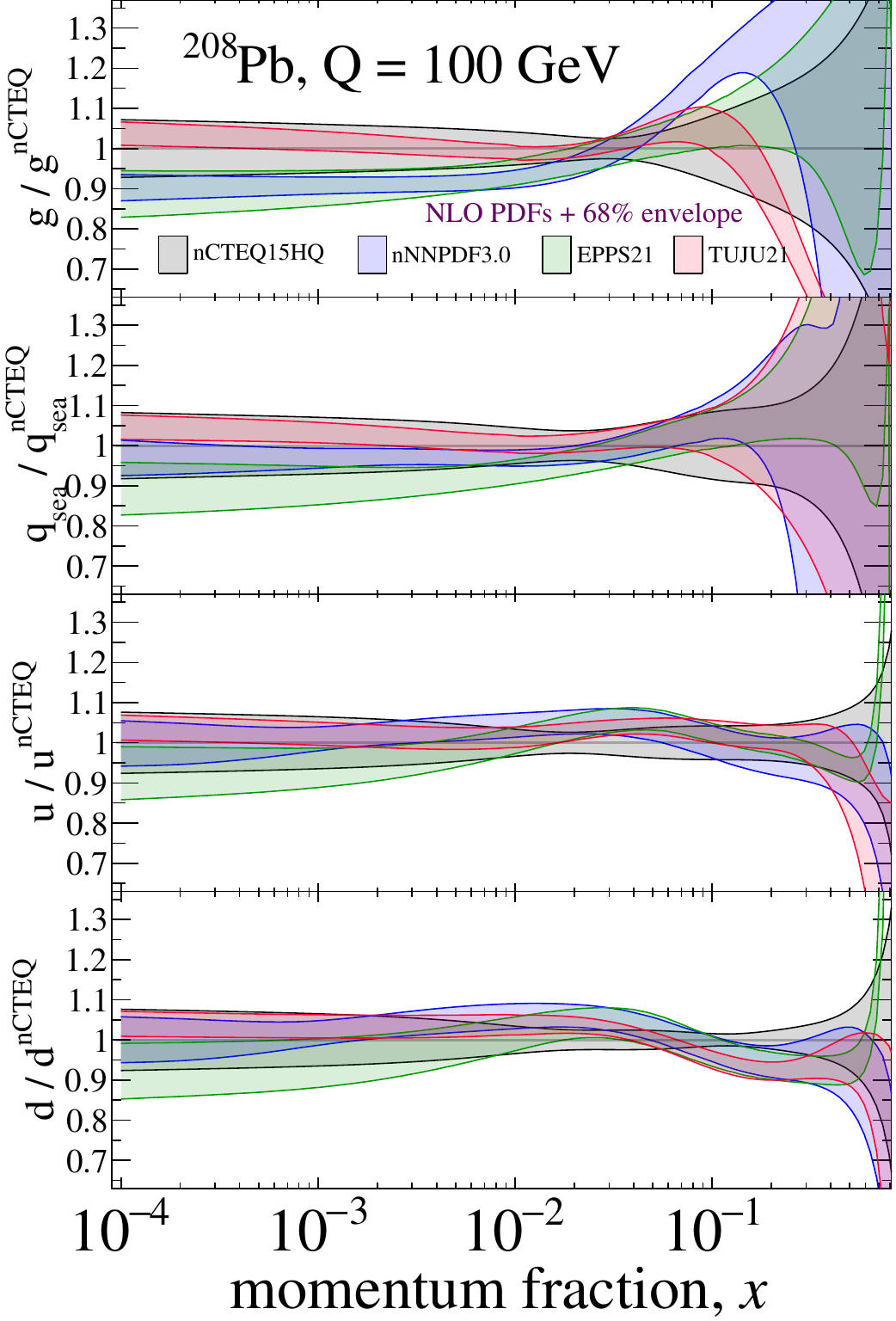}
\label{fig:ionPDFxRatio_MultiSet_Pb208_Q_100GeV}}
\subfigure[]{\includegraphics[width=0.32\textwidth]{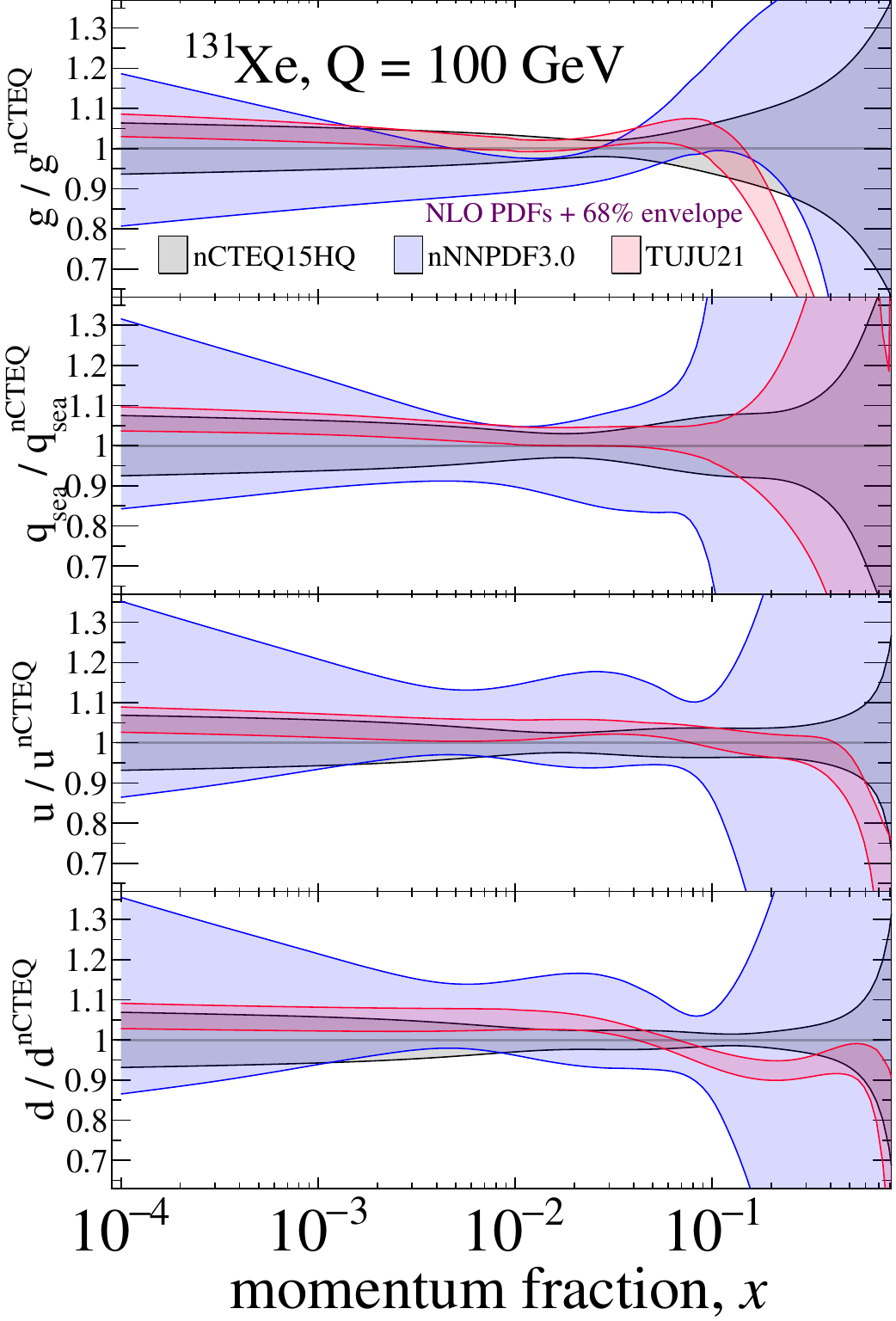}
\label{fig:ionPDFxRatio_MultiSet_Xe131_Q_100GeV}}
\subfigure[]{\includegraphics[width=0.32\textwidth]{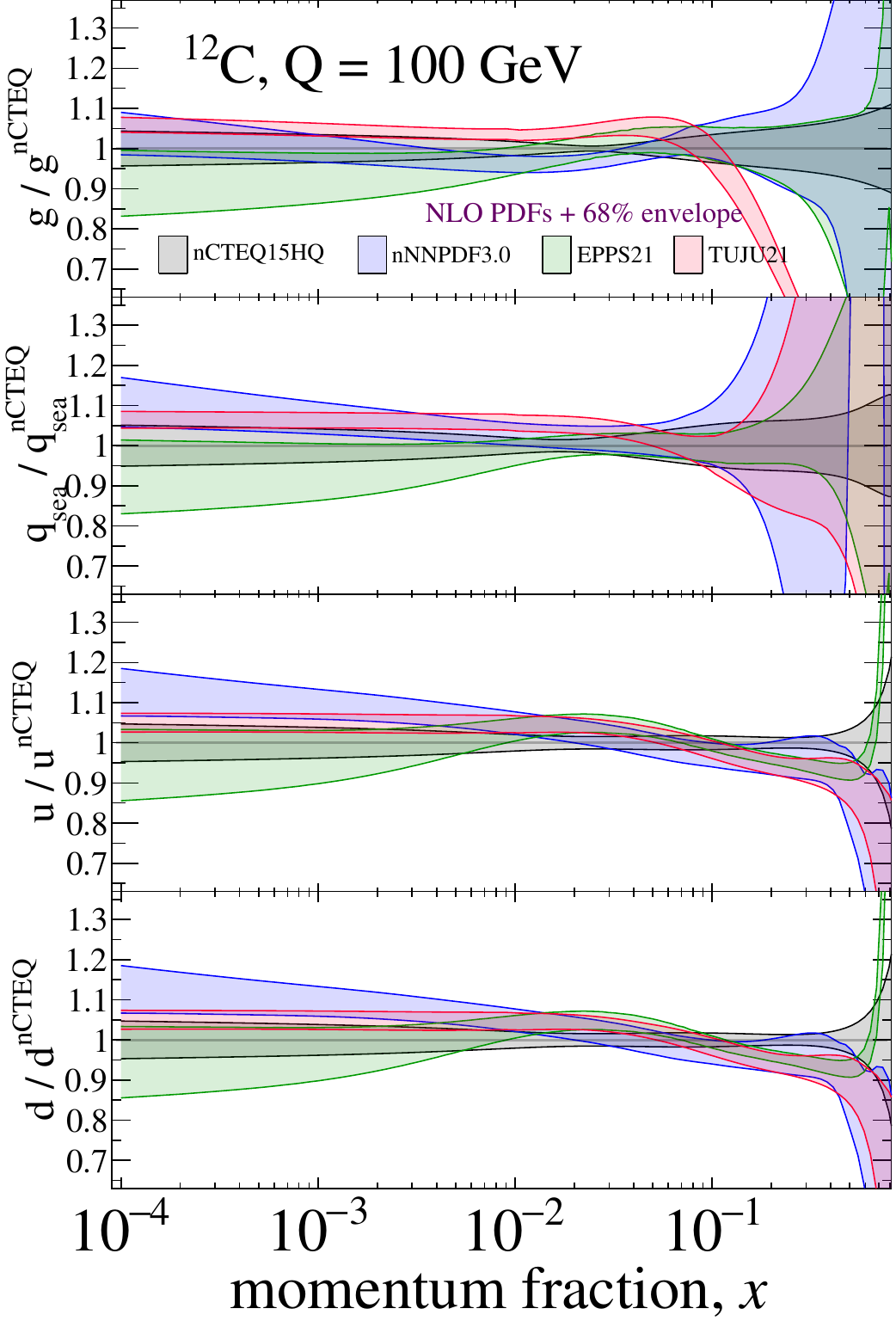}
\label{fig:ionPDFxRatio_MultiSet_xxC12_Q_100GeV}}
  \caption{As a function of (average) momentum fraction $x$, the ratio of nPDFs (per nucleon) of several families relative to  the central set of the \texttt{nCTEQ15HQ} densities, with their 68\% CL uncertainty envelope at $Q=100\GeV$, of the $g$ (top panel), $q_{\rm sea}$ (second panel), $u$ (third panel), and $d$ (bottom panel) parton species, and for  $^{208}$Pb~[\ref{fig:ionPDFxRatio_MultiSet_Pb208_Q_100GeV}], $^{131}$Xe [\ref{fig:ionPDFxRatio_MultiSet_Xe131_Q_100GeV}], and $^{12}$C [\ref{fig:ionPDFxRatio_MultiSet_xxC12_Q_100GeV}].}
  \label{fig:ionPDFxRatio_MultiSet_Q_100GeV}
\end{figure}

Presently, several collaborations have published up-to-date nPDF fits with complementary methodologies and decisions on which data sets to incorporate. These include the families \texttt{KSASG20}~\cite{Khanpour:2020zyu}, \texttt{TuJu21}~\cite{Walt:2019slu,Helenius:2021tof}, \texttt{EPPS21}~\cite{Eskola:2009uj,Eskola:2021nhw}, \texttt{nCTEQ15HQ}~\cite{Kovarik:2015cma, Duwentaster:2022kpv}, and \texttt{nNNPDF3.0}~\cite{Ethier:2020way,AbdulKhalek:2022fyi}. Differences across the sets represent a type of systematic uncertainty associated with modeling and fitting nPDFs.

In order to quantify the impact of this uncertainty on the parton luminosities that we report in Sec.~\ref{sec:lumi} and the cross sections that we report in Sec.~\ref{sec:xsec}, we show in Fig.~\ref{fig:ionPDFxRatio_MultiSet_Q_100GeV} the ratio of nPDFs (per  nucleon) relative to our baseline, which we take to be the central set of the \texttt{nCTEQ15HQ} nPDF family,
\begin{align}
    \frac{x\ f_i(x)\ \text{per nucleon}}{x\ f^{\rm nCTEQ}_i(x)\ \text{per nucleon}}\ =\  \frac{f_i(x)\ \text{per nucleon}}{f^{\rm nCTEQ}_i(x)\ \text{per nucleon}}.
\end{align}
In the figures, we present results, together with the associated 68\% CL uncertainty envelope and at $Q=100\GeV$, as a function of (average) momentum fraction $x$ for the $g$ (top panel), $q_{\rm sea}$ (second panel), $u$ (third panel), and $d$ (bottom panel) parton species. As before, we consider $^{208}$Pb [Fig.~\ref{fig:ionPDFxRatio_MultiSet_Pb208_Q_100GeV}], $^{131}$Xe [Fig.~\ref{fig:ionPDFxRatio_MultiSet_Xe131_Q_100GeV}], and $^{12}$C [Fig.~\ref{fig:ionPDFxRatio_MultiSet_xxC12_Q_100GeV}]. For concreteness, we compare the \texttt{nNNPDF3.0}, \texttt{EPPS21}, and \texttt{TuJu21} nPDF sets, which are accessible with the \texttt{LHAPDF} framework. We do not give a full treatise on the origin of any potential discrepancies (or likeness) between various nPDFs due to the nuanced nature of PDF fitting. Such discussions can be found  in Refs.~\cite{Khanpour:2020zyu, Helenius:2021tof, Eskola:2021nhw, Kovarik:2015cma, Duwentaster:2022kpv, AbdulKhalek:2022fyi}.

\begin{figure}
  \centering
\subfigure[]{\includegraphics[width=0.32\textwidth]{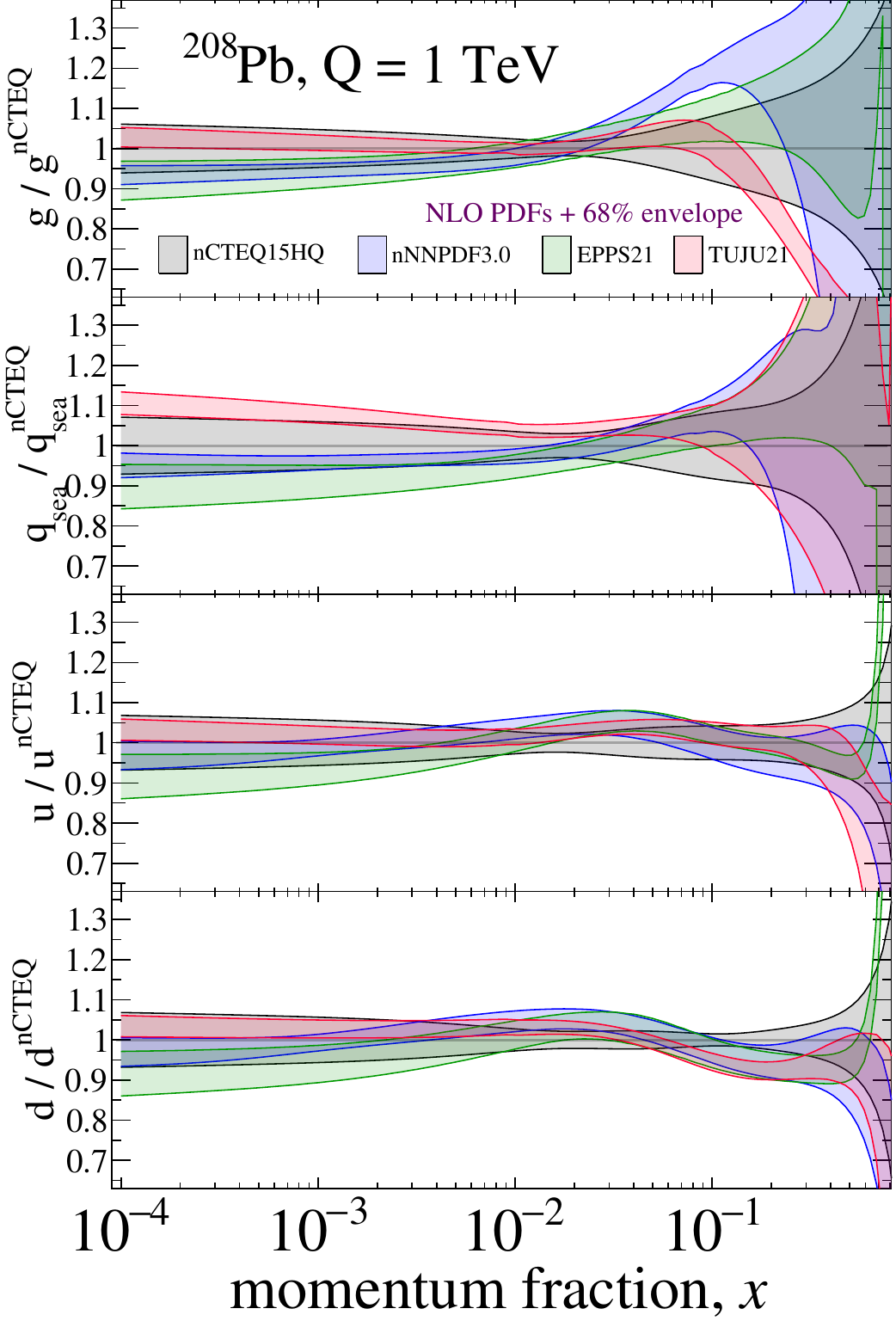}
\label{fig:ionPDFxRatio_MultiSet_Pb208_Q_xx1TeV}}
\subfigure[]{\includegraphics[width=0.32\textwidth]{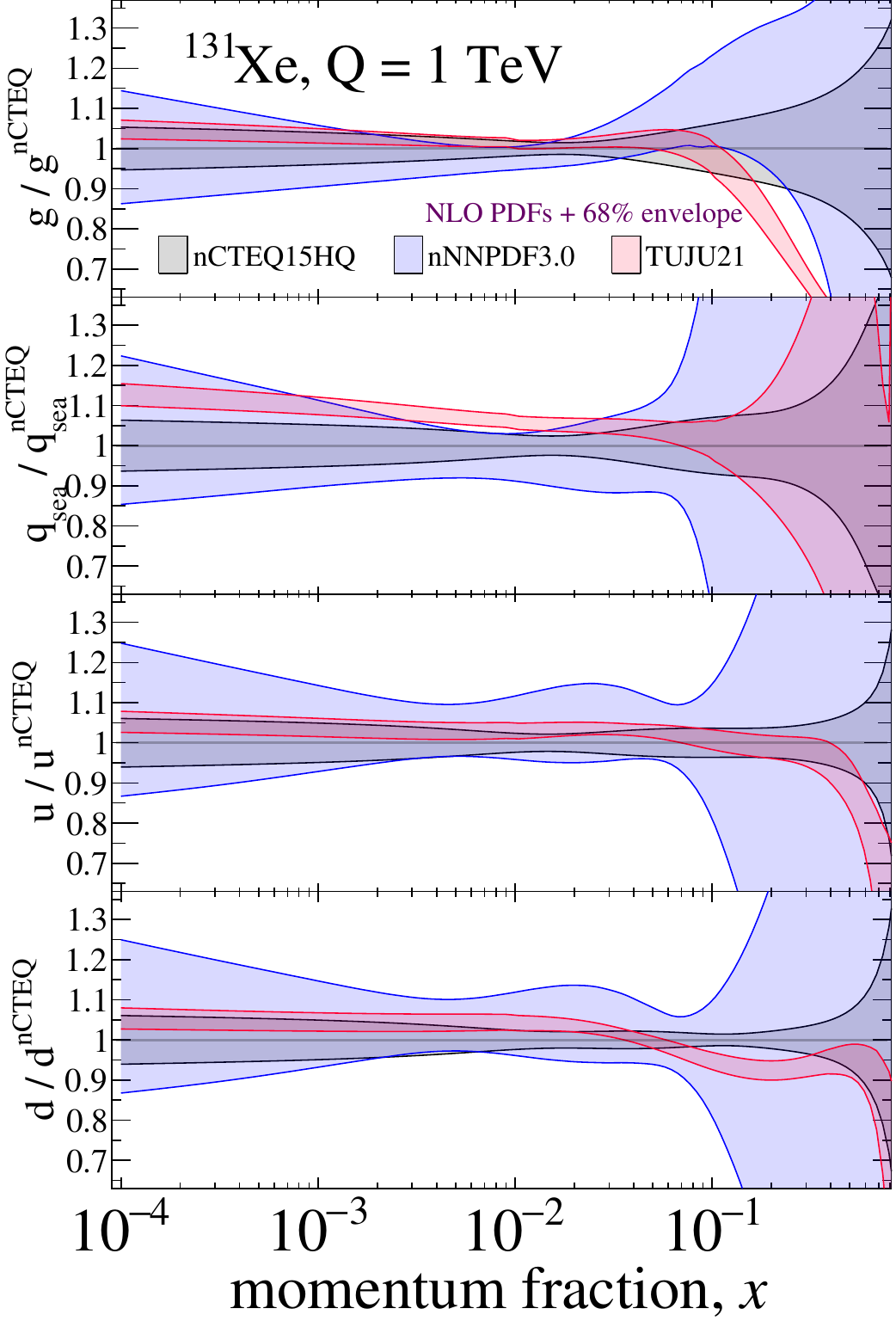}
\label{fig:ionPDFxRatio_MultiSet_Xe131_Q_xx1TeV}}
\subfigure[]{\includegraphics[width=0.32\textwidth]{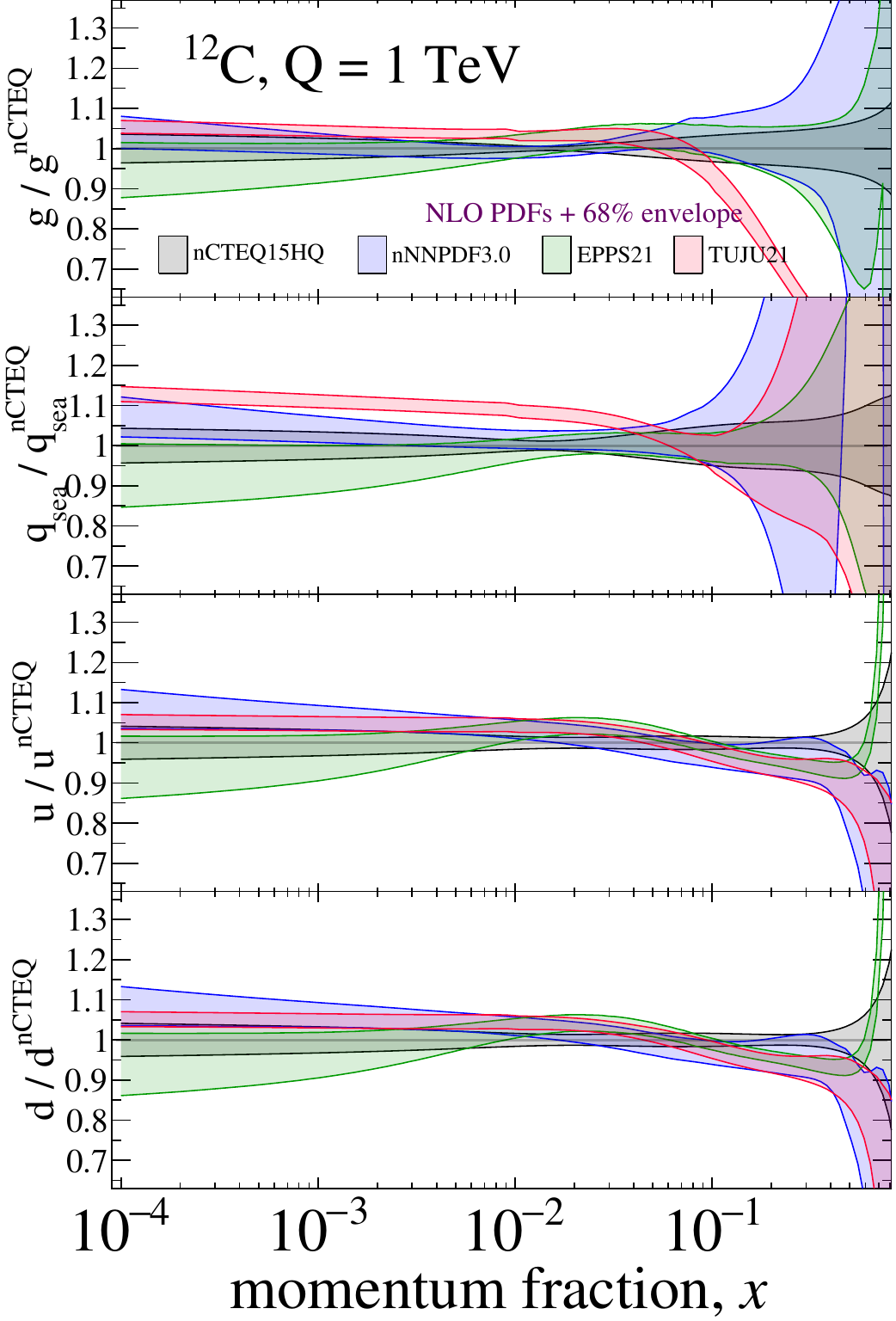}
\label{fig:ionPDFxRatio_MultiSet_xxC12_Q_xx1TeV}}
  \caption{Same as Fig.~\ref{fig:ionPDFxRatio_MultiSet_Q_100GeV} but for $Q=1\TeV$.}
  \label{fig:ionPDFxRatio_MultiSet_Q_xx1TeV}
\end{figure}

Starting with the top panel of Fig.~\ref{fig:ionPDFxRatio_MultiSet_Pb208_Q_100GeV}, we report that the \texttt{nCTEQ15HQ} reference nPDF for the gluon at $Q=100\GeV$ in $^{208}$Pb reaches a baseline uncertainty of \confirm{$\delta f_g/f_g^{\rm nCTEQ} \sim 10\%\ (5\%)$ for $x\sim10^{-4}\ (3\times10^{-2})$.} At larger $x$, the uncertainty quickly exceeds \confirm{$\delta f_g/f_g^{\rm nCTEQ} \sim 10\%\ (20\%)$ when $x\gtrsim0.1\ (0.4)$.} For \confirm{$x\lesssim10^{-2}$}, the three alternative sets exhibit comparable central values and uncertainties for the gluon. Specifically for this range of $x$, \texttt{TuJu21} sits just above unity but within the baseline's uncertainty band, while \texttt{nNNPDF3.0} and \texttt{EPPS21} have central values that reach as low as  \confirm{$f_g/f_g^{\rm nCTEQ} \sim0.9$}, which is just outside the lower edge of the \texttt{nCTEQ15HQ} uncertainty band. For $x\gtrsim0.1$, the four nPDF families exhibit large qualitative differences, tending to values above and below unity, with uncertainties that \confirm{exceed $30\%$ for $x\gtrsim0.3$.}

In the second panel of the same figure, we study the ratio of $q_{\rm sea}$ distributions as predicted by different families of nPDF sets with respect to the reference set, still for $^{208}$Pb and $Q = 100\GeV$. We find that for \confirm{$x\lesssim3\times10^{-2}$} the ratio of $q_{\rm sea}$ distributions and the associated uncertainty bands mirror those of the gluon case for the four nPDF families under consideration. This behavior is expected as low-$x$ sea partons at this $Q$ are largely generated perturbatively via $g\to q\overline{q}$ splitting, and should therefore reflect the behavior of the initial gluons (top panel). On the other hand, for \confirm{$x\gtrsim0.1$}, the shapes and normalizations of $q_{\rm sea}$ partons are qualitatively more comparable to the baseline than in the gluon case but still exhibit uncertainties that \confirm{exceed  $30\%$ for $x\gtrsim0.3$.} In the third and bottom panels of the figure, we report ratios of the $u$ and $d$ distributions as well as their uncertainties. They are all  comparable over the range \confirm{$x\sim 10^{-4}-0.3$}. While some qualitative features emerge at around \confirm{$x\sim 3-4\times10^{-2}$}, the ratios remain within \confirm{$\mathcal{O}(10\%)$ of unity over this larger range.} On the other hand, for larger $x$ values, the ratios differ qualitatively and quantitatively as do their uncertainties. \texttt{EPPS21}, for example, exhibits distributions that are significantly larger than the baseline while \texttt{TuJu21}'s distributions are significantly smaller.

Turning to Fig.~\ref{fig:ionPDFxRatio_MultiSet_Xe131_Q_100GeV}, we present the same information as in the left panel but for the case of $^{131}$Xe, and without \texttt{EPPS21} as this family of nPDF sets does not contain an nPDF for $^{131}$Xe. Overall, we find qualitatively similar ratios for $^{131}$Xe as  found for $^{208}$Pb. More specifically, our baseline ratios exhibit uncertainties that stay within \confirm{$\delta f/f^{\rm nCTEQ} \sim 10\%$ for $x\sim10^{-4}-0.1$.} For this same $x$ range, both \texttt{nNNPDF3.0} and \texttt{TuJu21} exhibit largely flat ratios but different normalizations. The \texttt{TuJu21} ratios systematically sit  \confirm{at around $f_i/f_i^{\rm nCTEQ} \sim 1.05$, or about $5\%$ higher} for all parton species. The \texttt{nNNPDF3.0} ratios sit both slightly above ($u$ and $d$) and slightly below ($g$ and $q_{\rm sea}$) unity but have uncertainties that are \confirm{$2-3$ times larger than the baseline uncertainties.} For larger $x$ values, the trends for \texttt{nNNPDF3.0} and the baseline continue, albeit with larger uncertainties, while \texttt{TuJu21}'s distributions for $g$, $u$, and $d$ are significantly smaller than the baseline. In Fig.~\ref{fig:ionPDFxRatio_MultiSet_xxC12_Q_100GeV}, we find that the uncertainties and spread of distribution ratios for partons in $^{12}$C are qualitatively similar to those already reported for $^{208}$Pb and $^{131}$Xe, and therefore do not need to be discussed further. Quantitatively, the uncertainties are slightly smaller than for $^{208}$Pb. And even with this reduced uncertainty, the various nPDF families \confirm{remain in good agreement for $x\lesssim0.1$.} With the exception of \texttt{TuJu21}'s gluon distribution, this agreement also holds for  \confirm{$x\gtrsim0.1$.}

For completeness, we show in Fig.~\ref{fig:ionPDFxRatio_MultiSet_Q_xx1TeV} the same information as presented in Fig.~\ref{fig:ionPDFxRatio_MultiSet_Q_100GeV} but for a hard scale of $Q=1\TeV$. Qualitatively and quantitatively, we find that the uncertainties and spread of distribution ratios at this higher scale largely remain the same as for $Q=100\GeV$. A small reduction in uncertainties can be observed in most panels and is due to the increased importance of perturbative quark and gluon splittings in populating parton densities, which are under good theoretical control. Overall, parton uncertainties span approximately \confirm{$\delta f_i/f_i^{\rm nCTEQ} \sim 5\%-20\%$ over the range $x\sim 10^{-4}-5\times10^{-2}$.}

\section{Parton Luminosities}\label{sec:lumi}
Parton luminosities $(\Phi_{ij})$ quantify the number incoming parton pairs $(ij)$ in hadron-hadron collisions that go on to scatter at a partonic center-of-mass energy $\sqrt{\hat{s}}$. Since $\sqrt{\hat{s}}=\sqrt{x_1 x_2 s}$ is sourced from a continuum of momentum fractions $x_i$, luminosities also provide a means of quantifying the net impact of PDFs and their uncertainties, which have strong $x$ dependencies. Furthermore, cross sections scale with parton luminosities. Therefore, when multiplied by coupling constants and other na\"ive scaling factors, luminosities provide order-of-magnitude estimates for cross section of complicated processes. Such estimates are particularly reliable for $2\to1$ and $2\to2$ processes.

Much like the relationship between parton densities at the \textit{nucleus} and \textit{nucleon} levels, we assume there is a connection between parton luminosities at the nuclear level ($\Phi_{ij}$) and their nucleon counterpart ($\Phi_{ij,NN}$) derived through averaging. In the context of symmetric collisions of two nuclei with atomic number $A$, the relationship between these two classes of parton luminosities can be expressed as
\begin{equation}
  \Phi_{ij}(\tau_{AA})\ =\ \Phi_{ij,NN}(\tau)\ \times\ A\ \times\ A\ .\
\end{equation}
Here, the dimensionless threshold variables $\tau$ and $\tau_{AA}$ are defined\footnote{It is worth stressing the distinction between $\tau$, which is defined at the nucleon level, and $\tau_{AA}$, which is defined at the nucleus level and is bound by unity, \textit{i.e.}, $\max(\tau_{AA}) = 1$. See footnote \ref{foot:momentum_fraction} and Eq.~\eqref{eq:pdfrescaling} for further details.}, respectively, as (a) the ratio between the squared partonic center-of-mass energy $\hat{s}$ and the squared nucleon-nucleon center-of-mass energy $s_{NN}$, and (b) the ratio between $\hat{s}$ and the squared nucleus-nucleus center-of-mass energy $s$. Symbolically, these are given by
\begin{align}
   \tau\ &=\ \frac{\hat{s}}{s_{NN}}\
   =\ \frac{x_{A1}x_{A2}s}{(s/AA)}\
   =\ \frac{(x_1A)(x_2A)s}{(s/AA)}\
   =\ x_1x_2\ ,
   \\
   \tau_{AA}\ &= \frac{\hat{s}}{s}\
   =\ x_{A1}x_{A2}\
   =\ \tau\ AA\ ,
\end{align}
where $x_{Ai}\in[0,1]$ is the nucleus-level momentum fraction carried by parton $i$, and $x_i=(A x_{Ai})\in[0,A]$ is the nucleon-level analogue (and is obtained via a rescaling by $A$). Formally, $\tau$ can take on values up to $\max(\tau)=A^2$. However, for average momentum fractions in the range $x_i\in[1,A]$ nuclear parton densities are exponentially suppressed~\cite{BCDMS:1994ala,CLAS:2003eih,Fomin:2010ei,Freese:2015ebu,Segarra:2020gtj,Ruiz:2023ozv}. We therefore follow convention by neglecting the $x_i\in[1,A]$ and $\tau\in[1,A^2]$ regions of nPDFs and phase space. For concreteness, we evaluate parton luminosities at a collinear factorization scale equal to the partonic center-of-mass energy: $\mu_f = \sqrt{\hat{s}} = \sqrt{\tau s_{NN}}$.

In the following, we focus on four sets of parton luminosities at the nucleon level. The first one, $\Phi_{gg,NN}$, is relevant for processes induced by gluon fusion, \textit{e.g.}, $gg\to t\overline{t}$, and is defined by
\begin{equation}
  \Phi_{gg,NN}(\tau)\ =\  \int_{\tau/A}^{A} \frac{dx}{x}\ g(x,\mu_f)\ g\Big(\frac{\tau}{x},\mu_f\Big)\
  \approx\
  \int_\tau^1 \frac{dx}{x}\ g(x,\mu_f)\ g\Big(\frac{\tau}{x},\mu_f\Big)\ .
\label{eq:phigg}
\end{equation}
Next, we consider the $qg$ luminosity,  which is relevant for prompt photon production $(qg\to\gamma q)$ and $\mathcal{O}(\alpha_s)$ corrections to $W/Z$ production $(qg\to Vq)$. Summing over all five active quark flavors ($q_i = u, d, s, c, b$), we have:
\begin{subequations}\label{eq:phiqg}
\begin{align}
  \Phi_{qg,NN}(\tau) =  \sum_{i=u,d,\dots}\ \int_{\tau/A}^{A} \frac{dx}{x}\ &\bigg[  g(x,\mu_f)\  q_i\Big(\frac{\tau}{x},\mu_f\Big)\
  +\ g(x,\mu_f)\ \bar{q}_i\Big(\frac{\tau}{x},\mu_f\Big)\
  \nonumber\\
  &\ +\ q_i(x,\mu_f)\ g\Big(\frac{\tau}{x},\mu_f\Big)
     +\ \bar{q}_i(x,\mu_f)\ g\Big(\frac{\tau}{x},\mu_f\Big)\bigg],\\
\approx
\sum_{i=u,d,\dots}\ \int_\tau^1 \frac{dx}{x}\ &\bigg[  g(x,\mu_f)\  q_i\Big(\frac{\tau}{x},\mu_f\Big)\
+\ g(x,\mu_f)\ \bar{q}_i\Big(\frac{\tau}{x},\mu_f\Big)\
    +\ \left(x\leftrightarrow\frac{\tau}{x}\right)
    \bigg],
\end{align}
\end{subequations}
followed the neutral-current  $q\bar q$ luminosity, again for five active quark flavors,
\begin{subequations}
\label{eq:phiqq}
\begin{align}
  \Phi_{q\bar{q},NN}(\tau) = &\ \sum_{i=u,d,\dots} \int_{\tau/A}^{A} \frac{dx}{x}\ \bigg[ q_i(x,\mu_f)\ \bar{q}_i\Big(\frac{\tau}{x},\mu_f\Big) + \bar{q}_i(x,\mu_f)\ q_i\Big(\frac{\tau}{x},\mu_f\Big)\bigg]
  \\
  \approx &\ \sum_{i=u,d,\dots} \int_\tau^1 \frac{dx}{x}\ \bigg[ q_i(x,\mu_f)\ \bar{q}_i\Big(\frac{\tau}{x},\mu_f\Big) + \bar{q}_i(x,\mu_f)\ q_i\Big(\frac{\tau}{x},\mu_f\Big)\bigg],
\end{align}
\end{subequations}
and the  charged-current $q\bar{q}'$ luminosity
\begin{subequations}
\label{eq:phiqqbar}
\begin{align}
  \Phi_{q\bar{q}',NN}(\tau) = &\ \sum_{i=u,\dots}\ \int_{\tau/A}^{A} \frac{dx}{x}\ \bigg[ q_i(x,\mu_f)\ \bar{q}'_i\Big(\frac{\tau}{x},\mu_f\Big) + \bar{q}_i(x,\mu_f)\ q'_i\Big(\frac{\tau}{x},\mu_f\Big)\bigg]
  \\
  \approx &\ \sum_{i=u,\dots}\ \int_\tau^1 \frac{dx}{x}\ \bigg[ q_i(x,\mu_f)\ \bar{q}'_i\Big(\frac{\tau}{x},\mu_f\Big) + \bar{q}_i(x,\mu_f)\ q_i'\Big(\frac{\tau}{x},\mu_f\Big)\bigg]\ .
\end{align}
\end{subequations}
In the charged-current luminosity,
the sum runs over SU$(2)_L$ doublet pairs,
\textit{i.e.}, $q_i^{\phantom{\prime}}\overline{q}_i' = u\overline{d}$, $c\overline{s}$, $\dots$,
but excludes third-generation quarks due to the absence $t~(\overline{t})$ (anti)quarks in modern nPDF sets.

\begin{figure}
\subfigure[]{\includegraphics[width=0.48\textwidth]{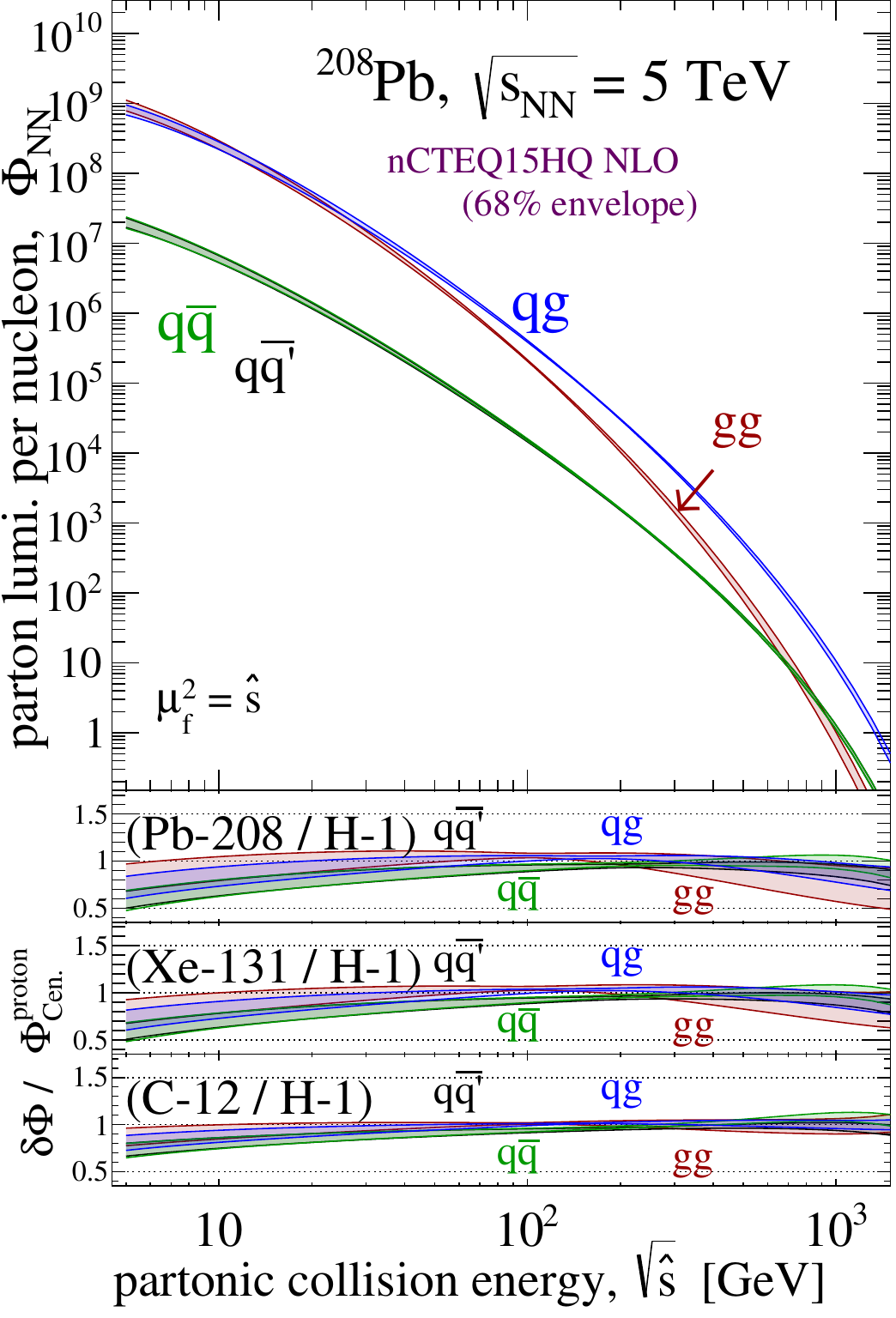}
\label{fig:ionLumi_MultiIon_Q_xx5TeV}}
\hfill
\subfigure[]{\includegraphics[width=0.48\textwidth]{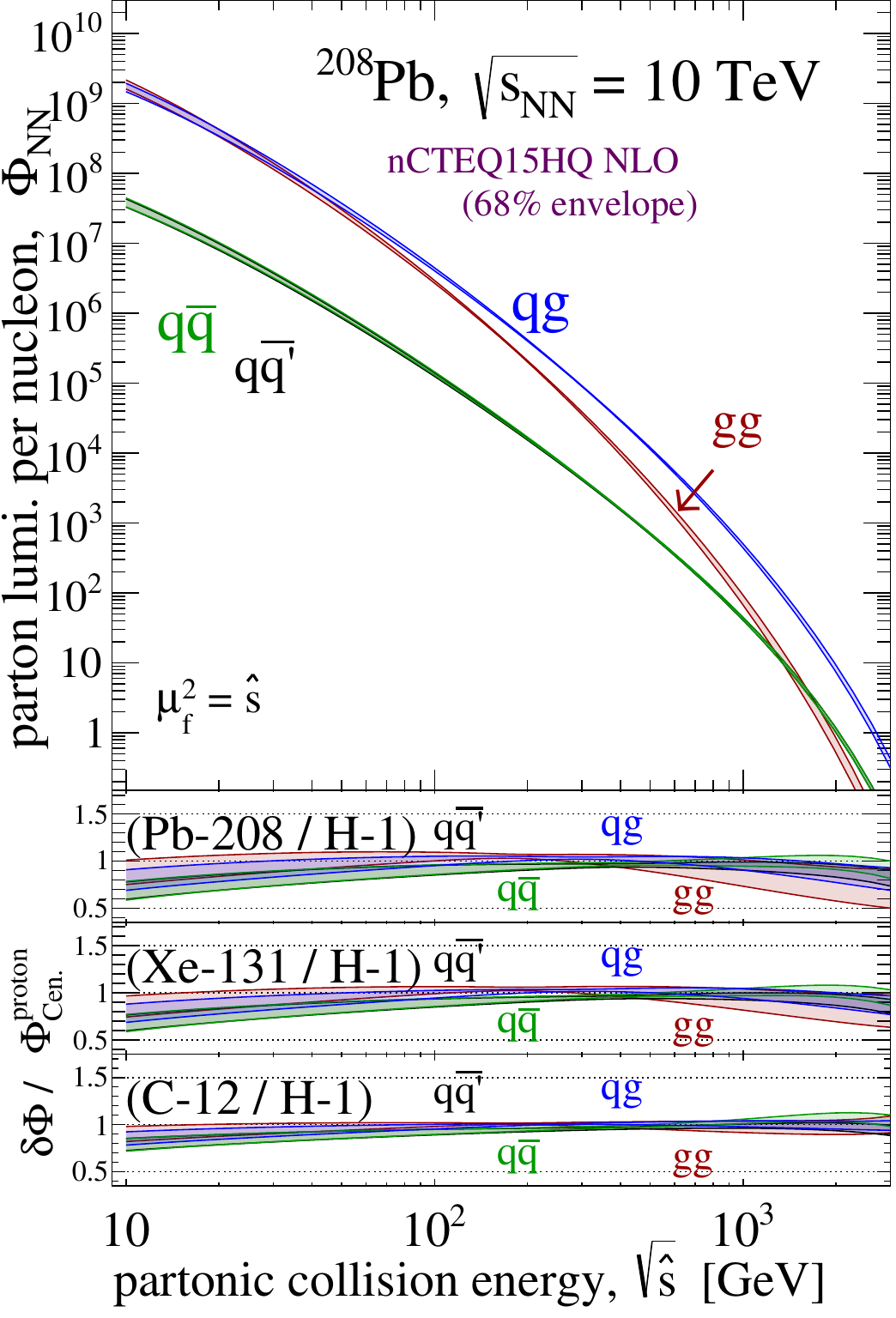}
\label{fig:ionLumi_MultiIon_Q_x10TeV}}
  \caption{Upper panel: For a nucleon-nucleon center-of-mass energy $\sqrt{s_{NN}}=5\TeV$ (a) and $10\TeV$ (b), average parton luminosities (per nucleon) in symmetric $^{208}$Pb collisions for the $gg$ (red), neutral-current $q\bar q$ (green), charged-current $q\bar q'$ (black) and $qg$ (blue) initial states. Band thickness corresponds to the 68\% CL PDF uncertainty. Lower panels: Ratios of luminosities relative to those in proton-proton collisions for symmetric $^{208}$Pb (second panel), $^{131}$Xe (third panel), and $^{12}$C (bottom panel) collisions.}
  \label{fig:ionLumi_MultiIon_loQ}
\end{figure}

In Fig.~\ref{fig:ionLumi_MultiIon_Q_xx5TeV}, we provide predictions for the (average) $gg$, $qg$, $q\bar{q}$ and $q\bar{q}'$ parton luminosities per nucleon as a function of the partonic center-of-mass energy $\sqrt{\hat{s}}$. In the upper panel of the figure, these predictions are presented specifically for symmetric collisions of $^{208}$Pb nuclei at a nucleon-nucleon collision energy of $\sqrt{s_{NN}} = 5\TeV$.

Reflecting the behavior of the $g$ density in lead for different factorization scales in Sec.~\ref{sec:xpdf_densities}, the $gg$ luminosity per nucleon is large at small partonic collision energies and rapidly falls with increasing values of $\sqrt{\hat{s}}$ in a way that is steeper than for the other luminosities considered. Quantitatively, at $\sqrt{s_{NN}} = 5\GeV$, \confirm{$\Phi_{gg,NN}$ reaches up to $10^{9}$ at small $\sqrt{\hat{s}}$} with an uncertainty of about \confirm{20\%}, which follows from the large uncertainty of the gluon density for \confirm{$x\lesssim 10^{-3}$} and scales \confirm{$\mu_f \lesssim 10\GeV$}. The uncertainties, however, decrease with increasing values of $\sqrt{\hat{s}}$ as the bulk of the integral in Eq.~\eqref{eq:phigg} increasingly involves the gluon density in the regime where it is known best (see the upper panels of Figs.~\ref{fig:ionPDFx_MultiIon_loQ} and \ref{fig:ionPDFx_MultiIon_hiQ}). In particular, for partonic center-of-mass energies around \confirm{$\sqrt{\hat{s}}\simeq100\GeV$ $(\tau\simeq 10^{-4})$,} which typically correspond to the production of a single weak or Higgs boson, the uncertainties reach a minimum of \confirm{$2\%-3\%$} for a luminosity per nucleon of \confirm{$\Phi_{NN}\sim2\times 10^5$}. With  further increasing $\sqrt{\hat{s}}$, the luminosity continues to drop due to the suppression of the gluon nPDF at larger momentum fractions of \confirm{$x\gtrsim 0.1$}; ultimately, this is a consequence of momentum conservation and the decreasing likelihood of a single partonic collision occurring at a scale of $\sqrt{\hat{s}}\rightarrow\sqrt{s_{NN}}\ (\tau\to1)$. Likewise, the increasing uncertainty at large $\sqrt{\hat{s}}$ corresponds to the poorer constraints on the gluon nPDF,  particularly  for \confirm{$x\gtrsim0.2$} regardless of the scale.

The neutral-current $q\bar q$ and charged-current $q\bar q'$ luminosities exhibit qualitatively similar behavior as the $gg$ luminosity for small partonic center-of-mass energies, although with a smaller normalization by a factor of about \confirm{$50 - 100$} for $\sqrt{\hat{s}}\lesssim 100\GeV$. Such behavior originates from multiple sources. For example: the definitions of $\Phi_{ij,NN}$ in Eqs.~\eqref{eq:phiqg}-\eqref{eq:phiqqbar} sum over all quark species and therefore are largely insensitive to the difference between protons and neutrons, \textit{i.e.}, it is approximately isospin symmetric. Another example: for the regime under consideration, the dynamics of the quark densities are mostly driven by $g\to q\bar q$ perturbative splittings. At some point, however, contributions from valence quarks impact the up and down quark densities, and therefore the $q\bar q$ and $q\bar q'$ luminosities. This reduces the steepness by which the two luminosities decreases with increasing partonic energy. In other words: the $gg$ luminosity decrease quicker  with increasing $\sqrt{\hat{s}}$ than the $q\bar q$ and $q\bar q'$ luminosities because of the presence of valence quarks at large $x$ (or $\tau$). At collision energies of about \confirm{$\sqrt{\hat{s}}\sim750\GeV\ (\tau\sim0.02)$}, the $gg$, $q\bar q$ and $q\bar q'$ luminosity curves cross, and for larger energies $\Phi_{q\bar{q},NN}, \Phi_{q\bar{q}',NN} > \Phi_{gg,NN}$. We report uncertainties of about \confirm{$\delta\Phi_{NN}/\Phi_{NN}\sim15\%$} for partonic scattering scales of \confirm{$5-10\GeV$}, which reduce to \confirm{$2-3\%$} in the central $\sqrt{\hat{s}}$ regime that we consider, before increasing to \confirm{$5\%$} for scales in the range $\sqrt{\hat{s}}\sim500\GeV-2\TeV$.

Finally, our results show that the $qg$ luminosity at $\sqrt{s_{NN}}=5\TeV$ is the largest over the entire kinematic regime considered. It converges to the $gg$ density for \confirm{$\sqrt{\hat{s}}\lesssim 30\GeV$} and then surpasses it up to a factor of \confirm{10 for $\sqrt{\hat{s}}\gtrsim 300\GeV$}. The related uncertainties correspond to a geometric average of the uncertainties associated with the $q\bar q$ and $gg$ luminosities. Hence, they vary from \confirm{$5\%-15\%$}. The largest uncertainties of \confirm{$10\% - 15\%$} are found for partonic center-of-mass energies below \confirm{$20\GeV$} and above \confirm{$800\GeV$}, whereas for more central energies the uncertainties are stable and reach a few percent. These properties stem from the fact that the $qg$ partonic configuration combines the large gluon density at small $x$ with the valence-quark domination of the $u$ and $d$ densities at large $x$.

In the second panel of Fig.~\ref{fig:ionLumi_MultiIon_Q_xx5TeV} we present the ratios of the four luminosities for $^{208}$Pb collisions relative to same luminosities for proton collisions. We observe that in the center of the kinematic regime under consideration, \confirm{$\sqrt{\hat{s}}\sim100-200\GeV\ (\tau\sim10^{-3})$}, the ratio of lead-lead and proton-proton $gg$ luminosities is compatible with unity (up to uncertainties), reflecting an accidental cancellation between the suppression of gluon densities at small momentum fractions $(x\lesssim10^{-3})$ and the enhancement of gluon densities at large $(x\gtrsim0.3)$. The below-unity value of the $gg$ parton luminosity ratio for smaller center-of-mass energies  then reflects the suppression of the gluon nPDF at smaller $x$ values, although this conclusion is subject to the large uncertainties in this kinematic limit. For larger energies the ratio of lead-lead to proton-proton luminosities is smaller than unity, and even suggests a stronger suppression than for lower collision energies. Once again, the uncertainties in this regime prevent too conclusive statements, although this behavior is consistent with a shallow suppression (up to uncertainties) of the gluon density for \confirm{$x\gtrsim 0.1-0.3$}.

\begin{figure}
\subfigure[]{\includegraphics[width=0.48\textwidth]{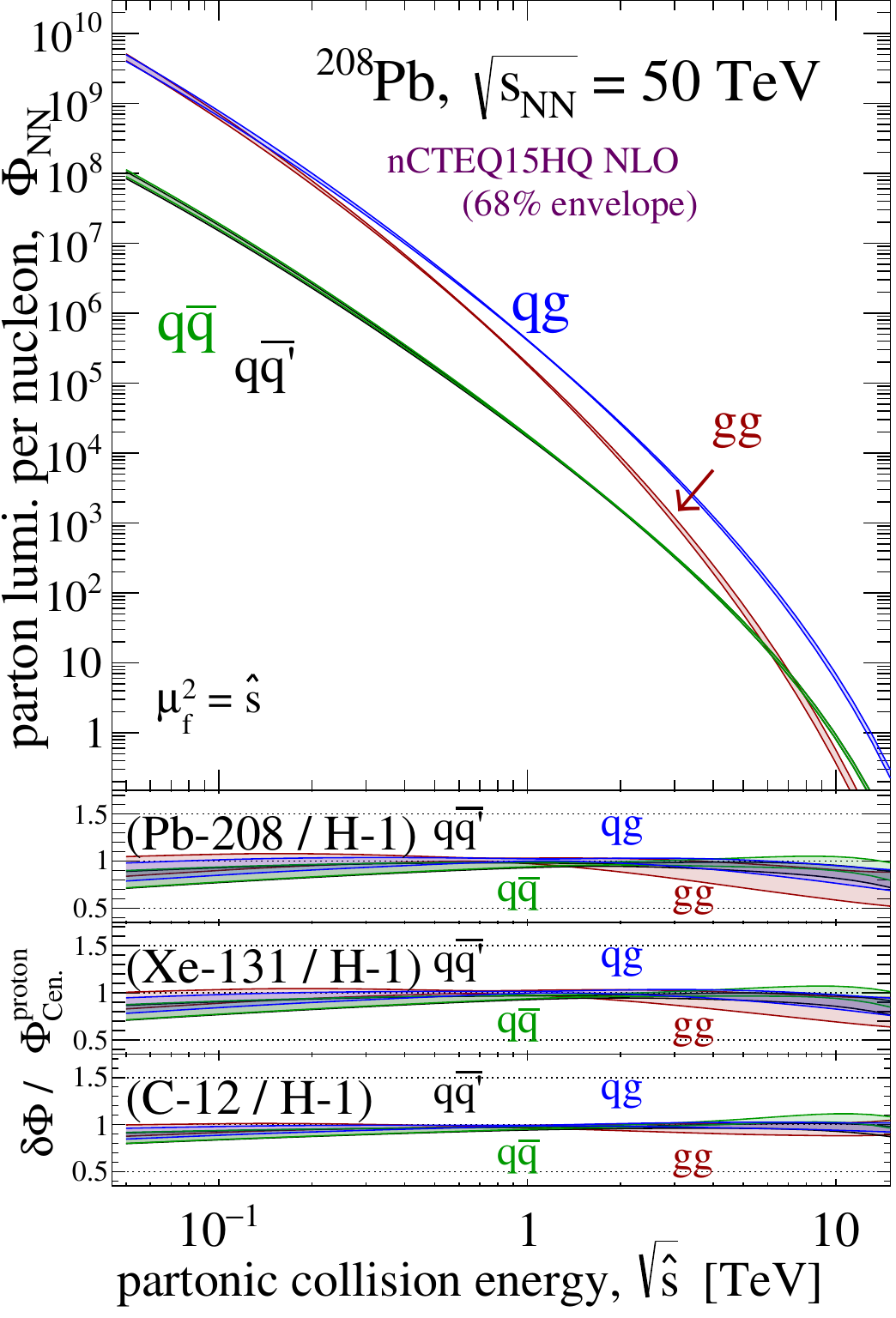}
\label{fig:ionLumi_MultiIon_Q_x50TeV}}\hfill
\subfigure[]{\includegraphics[width=0.48\textwidth]{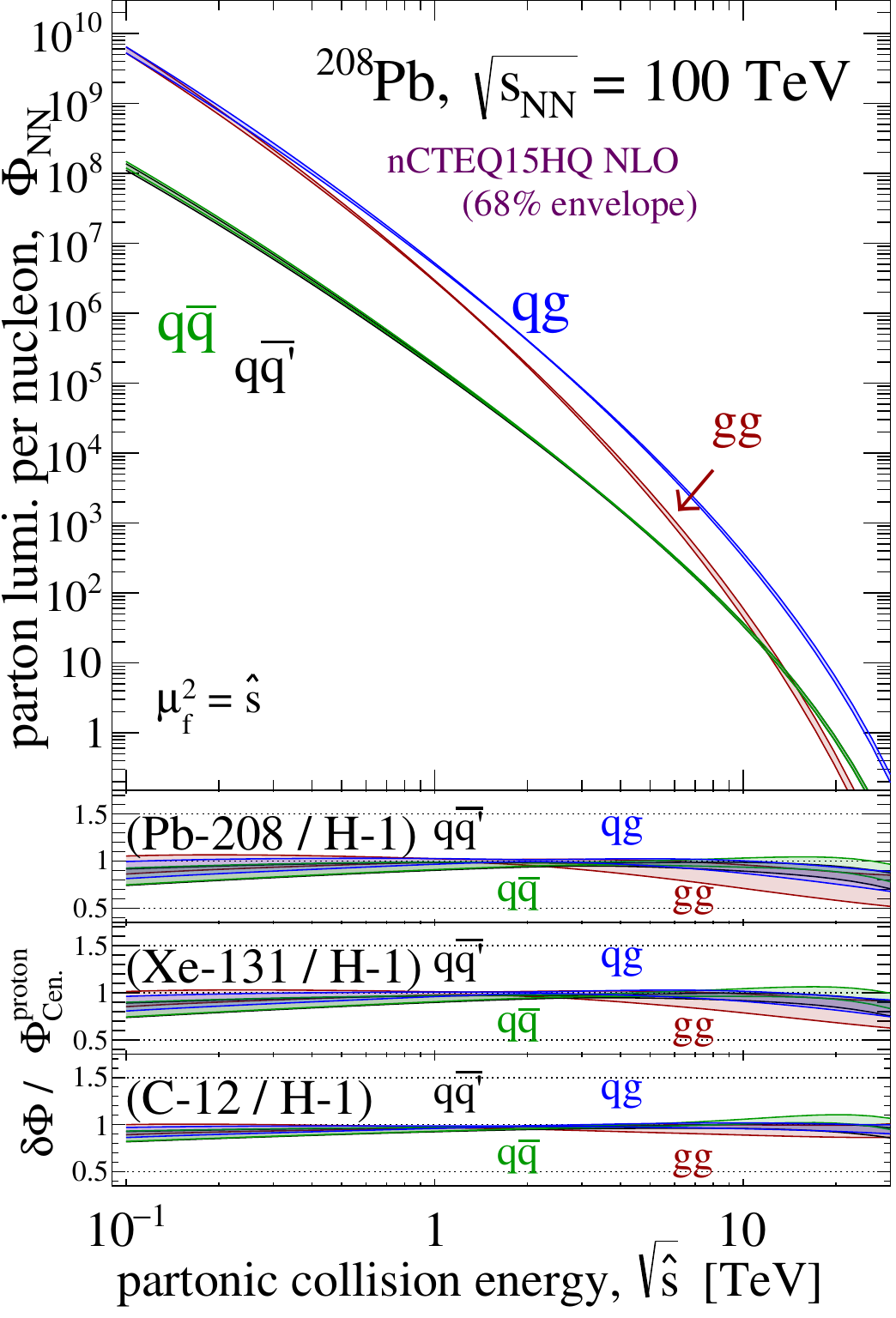}
\label{fig:ionLumi_MultiIon_Q_100TeV}}
  \caption{Same as Fig.~\ref{fig:ionLumi_MultiIon_loQ} but for (a) $\sqrt{s_{NN}}=50\TeV$ and (b) $\sqrt{s_{NN}}=100\TeV$.}
  \label{fig:ionLumi_MultiIon_hiQ}
\end{figure}

In the case of the ratio of $qg$ luminosities, and particularly for \confirm{$\sqrt{\hat{s}}\gtrsim 200\GeV\ (\tau\gtrsim 10^{-3})$}, the aforementioned suppression of the gluon density at moderate-to-high $x$ is compensated by the enlarged valence down quark content in $^{208}$Pb. (We reiterate that $^{208}$Pb is neutron rich and therefore the average nucleon is neutron-like.) These competing effects lead to a $qg$ luminosity that is comparable to the proton for \confirm{$\sqrt{\hat{s}}\gtrsim100\GeV\ (\tau\gtrsim4\times10^{-4})$.} However, for partonic center-of-mass energies below this threshold, the $qg$ luminosity of lead falls, reaching about \confirm{three-quarters} the size of the proton's at about \confirm{$\sqrt{\hat{s}}\sim 5\GeV\ (\tau\sim10^{-6})$}. Qualitatively, similar conclusions hold for the two ratios of quark-antiquark luminosities, although in some respects the effects are exacerbated. For example: the $q\overline{q}$ and $q\overline{q'}$ luminosity ratios are smaller than for $gg$ and $qg$ at the smallest partonic center-of-mass energies under consideration but are larger and closer to unity at the largest $\sqrt{\hat{s}}$.

In the third and fourth panels of Fig.~\ref{fig:ionLumi_MultiIon_Q_xx5TeV},  we present the same ratios in the case of symmetric xenon collisions and carbon collisions, respectively. As expected, the ratios approach unity as $A$ becomes smaller, \textit{i.e.}, more proton like. Uncertainties are additionally reduced, mirroring the reduced uncertainties originating from the nPDF fits. It is interesting to note that the dynamic behavior in nPDF at $x\gtrsim10^{-2}$, which leads to individual parton species having a much larger or much smaller density than the proton, is softened at the  level of luminosities. While nucleon-level luminosities in nuclei can exceed those of the proton, the excess is modest. This is because parton luminosities at a fixed $\sqrt{\hat{s}}$ typically sample a large range of possible momentum fractions, which leads to a compensation between enhancements and suppressions from opposing extremes. The notable exception is the extreme case of very small $\sqrt{\hat{s}}$; in this case, parton luminosities at the nucleon level tend to be \confirm{$10\%-50\%$} (up to uncertainties) smaller than in the proton.

In Figs.~\ref{fig:ionLumi_MultiIon_Q_x10TeV}, \ref{fig:ionLumi_MultiIon_Q_x50TeV}, and \ref{fig:ionLumi_MultiIon_Q_100TeV}, we present the same results as for Figs.~\ref{fig:ionLumi_MultiIon_Q_xx5TeV} but for nucleon-nucleon collision energies of $\sqrt{s_{NN}}=10$, 50 and $100\TeV$, respectively. The qualitative behavior of the results is similar to that featured at $5\TeV$, although the uncertainties are found to decrease with increasing evolution scale  since the bulk of the relevant dynamics becomes dictated by $g\to q\bar q$ splitting, and hence perturbative QCD, which is under good theoretical control. The  exception is the large partonic center-of-mass energy regime, which exhibits predictions with larger uncertainties; this follows from nPDF uncertainties in the large-$x$ regime (see, \textit{e.g.}, Fig.~\ref{fig:ionPDFxRatio_MultiSet_Q_100GeV}).

\section{Total and Fiducial Cross Sections}\label{sec:xsec}

In this section we present the main results of our work: a survey of production cross sections $(\sigma)$, at NLO in QCD, and their uncertainties for high-energy processes in symmetric ion collisions with  nucleon-nucleon collision energies spanning $\sqrt{s_{NN}}=1-100\TeV$. We focus on processes that at lowest order, \textit{i.e.}, the Born level, are described by tree-level partonic processes of the form
\begin{align}
    i\ +\ j\ \to\ \mathcal{F}\ .
\end{align}
Here, $i,j$ are any massless QCD partons $(n_f=5)$ and $\mathcal{F}$ is an $n_\mathcal{F}$-body final state with up to three SM particles from the collection $\{W^\pm, Z, \gamma, H, t,\overline{t}\}$.

For incoming ions $\mathcal{A}_1$ and $\mathcal{A}_2$, the inclusive, \textit{nucleus}-level cross section $(\sigma)$ is given by
\begin{align}
\label{eq:xsec_def_nucleon}
    \sigma(\mathcal{A}_1 \mathcal{A}_2 \to \mathcal{F}+X)\
    =\
    A_1\ \times\ A_2\
     \times\ \sigma_{NN}(N_1 N_2 \to \mathcal{F}+X)\ ,
\end{align}
where $X$ represents any and all outgoing hadronic activity associated with the beam remnants. In addition, $A_k$ is the atomic number of incoming nucleus $k\in\{1,2\}$, and $\sigma_{NN}$ is the inclusive \textit{nucleon}-level cross section for $N_1 N_2 \to \mathcal{F}$, with $N_k$ being an ``average'' nucleon of $\mathcal{A}_k$. In general, but especially in the large-$A$ limit, $N_k$ can be modeled\footnote{While the notion of modeling nuclei as a collection of bound nucleons is historical and justifiable, this formulation is not strictly necessary. It is possible to model nuclei directly in terms of partonic degrees of freedom~\cite{Ruiz:2023ozv}.} as a linear combination / admixture of a proton $\mathcal{P}$ and neutron $\mathcal{N}$, \textit{i.e.},
\begin{align}
    \vert N\rangle\ =\ \alpha \vert \mathcal{P} \rangle\ +\ \beta \vert \mathcal{N} \rangle\ ,
\label{eq:def_admix}
\end{align}
where $\vert\alpha\vert^2 = (Z/A)$,  $\vert\beta\vert^2=(A-Z)/A$, and $\vert\alpha\vert^2 + \vert\beta\vert^2= 1$. The kinematics of $N_k$ and $\mathcal{A}_k$ are then related by a rescaling of one or the other by $A_k$ or $1/A_k$~\cite{Schienbein:2007gr,Ruiz:2023ozv}. We focus on symmetric ion beams, meaning that $\mathcal{A}_1=\mathcal{A}_2$ and $N_1=N_2$. Furthermore, for $(A_k,Z_k)=(1,1)$, $\mathcal{A}_k$ and $N_k$ reduce to being each a proton, $\mathcal{A}_1 = \mathcal{P}$ and $N_1 = \mathcal{P}$.

To calculate nucleon-level cross sections, we apply the Collinear Factorization Theorem~\cite{Collins:1984kg,Collins:1985ue,Collins:2011zzd},
\begin{subequations}
\begin{align}
\sigma_{NN}\ &=\
\int dPS_{n_\mathcal{F}}\ \frac{d\sigma_{NN}}{dPS_{n_\mathcal{F}}},\quad \text{where}
\\
\frac{d\sigma_{NN}}{dPS_{n_\mathcal{F}}}\ &=\
\frac{1}{(1+\delta_{ij})}\
\sum_{i,j=u,g,\dots}\
\Delta_{ij}\
\otimes\ f_{i/\mathcal{A}_1}\
\otimes\ f_{j/\mathcal{A}_2}\
\otimes\
\frac{d\hat{\sigma}_{ij\to \mathcal{F}}}{dPS_{n_\mathcal{F}}}
\ ,
\end{align}
which has been derived for (select processes in) proton scattering and is assumed to hold for hard, inclusive nuclear collisions. In the above, $\Delta_{ij}$ is the Sudakov factor that accounts for renormalization-group (RG) evolution, \textit{e.g.}, parton showering; at the level of inclusive cross sections, it can be approximated as a Dirac-$\delta$ function and its convolution\footnote{The symbols $\otimes$ denote a convolution over a dimensionless variable $z$, $x_1$, or $x_2$, over the range $[0,1]$.} $\Delta_{ij}\otimes$ can be set to unity. The quantities $f_{i/A_{k}}(x_k,\mu_f)$ are the nPDFs of parton species $i$ for averaged nucleons in nucleus $\mathcal{A}_{k}$. Their two arguments are: (i) the momentum fraction $x_k=E_i/E_k$, which defines the momentum $p_i$ of parton $i$ relative to the momentum $P_k^N$ of its parent nucleon $N_k$ as
\begin{align}
p_i\ =\ x_k\ P_k^N\ =\ (x_k/A_k)\ P_k^\mathcal{A},\quad \text{with}\quad
P_k^N = \frac{\sqrt{s_{NN}}}{2}\ (1,0,0,\pm1)\ ,
\end{align}
where $P_k^\mathcal{A}=\sqrt{s}(1,0,0,\pm1)/2$ is the incoming ion's momentum, and (ii) the collinear factorization scale, below which  initial-state QCD radiation is resummed and nPDFs are RG-evolved via the DGLAP evolution equations.

Finally, $d\hat{\sigma}_{ij\to \mathcal{F}}$ is the totally differentiated \textit{parton}-level $ij\to \mathcal{F}$ cross section given by
\begin{align}
    \frac{d\hat{\sigma}_{ij\to \mathcal{F}}}{dPS_{n_\mathcal{F}}}\
    =\
    \frac{1}{2\hat{s}}\
    \frac{1}{\mathcal{S}_i\mathcal{S}_jN_c^iN_c^j}\
    \sum_{\rm dof}
    \vert \mathcal{M}\vert^2 \ .
\end{align}
Here, $\hat{s} = (p_i + p_j)^2$ is the (squared) hard scattering scale, $\mathcal{S}_k$ and $N_c^k$ are respectively the spin- and color-averaging factors, $\mathcal{M}$ is the $ij\to \mathcal{F}$ matrix element,
and the summation is over all discrete degrees of freedom (dof), \textit{e.g.}, helicity and color.
The $n_\mathcal{F}$-body phase space volume element is
\begin{align}
    dPS_{n_\mathcal{F}}\ =\ \frac{1}{\Omega}\,
    (2\pi)^4\
    \delta^4\left(p_i + p_j - \sum_{l=1}^{n_\mathcal{F}}p_l\right)\
    \times\
\prod_{l=1}^{n_\mathcal{F}}
\frac{d^3 p_l}{(2\pi)^3 2E_l}\ ,
\end{align}
\end{subequations}
where our definition includes a symmetry factor $\Omega$ relevant for the production of identical particles. Formally, $\Omega$ is the product of $(n_l!)$ factors, where $n_l$ is the number indistinguishable final-state particles of species $l$. All cross sections are computed numerically according to the methodology described in Sec.~\ref{sec:setup}.

For processes without infrared divergences in their Born-level matrix element, \textit{e.g.}, inclusive $W$ or $Z$ production via the Drell-Yan process ($q\overline{q}$ annihilation), we compute the total cross section, \textit{i.e.}, we integrate over all available phase space.
In general,
processes with final-state photons,
\textit{e.g.}, $W\gamma$ production,
contain infrared singularities in their matrix elements at the Born level.
For such processes,
we compute fiducial cross sections (instead of total cross sections) by imposing the following phase space restrictions on photons $(\gamma)$:
\begin{align}
\label{eq:cuts_photon}
  p_T^\gamma > 150\GeV \quad\text{and}\quad   \vert\eta^\gamma\vert<2.4\ .
\end{align}
For a final-state photon, $p_T^\gamma$ is the magnitude of its transverse momentum and $\eta^\gamma$ is its pseudorapidity.
We note that these restrictions
far exceed reconstruction and identification
thresholds for the LHC
experiments~\cite{ALICE:2008ngc,ATLASCollaboration:2012ilu,Contardo:2015bmq}.
We impose such stringent requirements for
two reasons:
(i) To ensure photon production originates from a hard-scattering process and not quark-gluon plasma~\cite{Bhattacharya:2015ada,Berges:2017eom}.
(ii) To minimize the numerical impact of soft/collinear logarithms in photon production at ultra high collider energies.
Throughout this work
we use the photon isolation criteria
as defined in Ref.~\cite{Frixione:1998jh}
and implemented in {\mgamc}, with
the angular separation  $(R_\gamma)$,
energy fraction $(\epsilon_\gamma)$,
and angular suppression factor $(n)$ parameters set to
the following values:
\begin{align}
\label{eq:cuts_photon_iso}
 R_{\gamma}=0.4,\
 \epsilon_\gamma=1.0,\
 n=1\ .
\end{align}

In Sec.~\ref{sec:xsec_xxv} we focus on single boson production, while in Sec.~\ref{sec:xsec_xvv} we consider diboson production without the Higgs or a photon. Associated photon $(V\gamma)$ production is discussed in Sec.~\ref{sec:xsec_xva}. Associated Higgs $(VH)$ production is covered in Sec.~\ref{sec:xsec_xvh}. Triboson $(VV'V'')$ processes are covered in Secs.~\ref{sec:xsec_vvv} and \ref{sec:xsec_wzv}. We address photon $(\gamma+X)$ processes in Sec.~\ref{sec:xsec_axx} before finishing with top quark pair $(t\overline{t}X)$ processes
in Sec.~\ref{sec:xsec_ttx}.
In Sec.~\ref{sec:xsec_yields}, we briefly discuss
the prospect for observing diboson and triboson processes
in ion collisions at the HL-LHC and FCC.
We give a summary of our results in Sec.~\ref{sec:xsec_summary}.

\subsection{Inclusive single boson production}\label{sec:xsec_xxv}

\begin{figure}[!t]
\includegraphics[width=.95\textwidth]{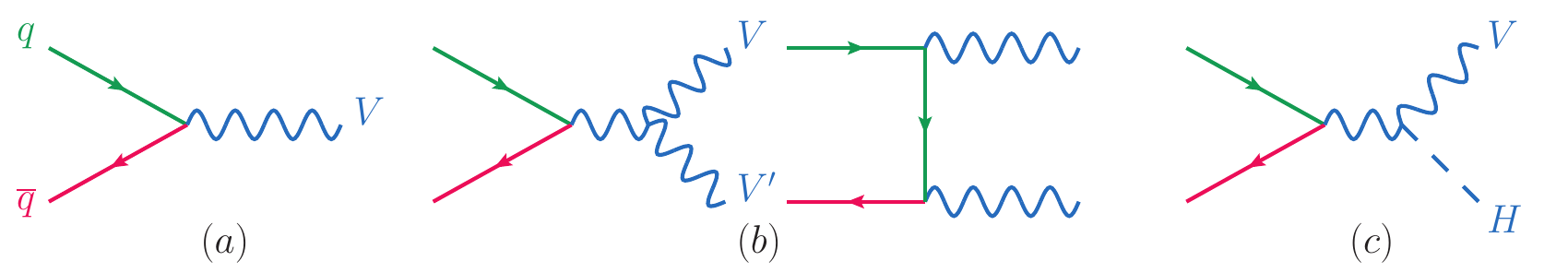}
\caption{Representative Feynman diagrams at the Born level depicting partonic production of (a) a weak boson $V\in\{W^\pm,Z\}$, (b) pair production of two weak bosons, and (c) associated $V$-Higgs (H) production.
}
\label{fig:diagram_MultiBoson_V_VV_VH}
\end{figure}

We begin our survey with the inclusive production of a single weak boson, and specifically with processes facilitated at LO by quark-antiquark annihilations, illustrated in Fig.~\ref{fig:diagram_MultiBoson_V_VV_VH}(a), and given by
\begin{align}
    q\overline{q},\ q\overline{q'}\
    \to\ W^\pm,\ W^-,\ W^+,\ Z\ .
\end{align}
NLO QCD predictions for weak boson production were computed first in Refs.~\cite{Altarelli:1978id,Altarelli:1979ub},
and at higher orders in Refs.~\cite{Hamberg:1990np,Harlander:2002wh,Duhr:2020seh,Duhr:2020sdp}.
Predictions for inclusive $W^\pm$ and $Z$
production specifically in ion collisions at
next-to-next-to-leading order (NNLO) QCD
are reported in Ref.~\cite{Helenius:2021tof}.
Inclusive $W$ and $Z$ production have been measured
by LHC experiments
in Pb-Pb collisions at
$\sqrt{s_{NN}}=2.76\TeV$~\cite{ATLAS:2012qdj,CMS:2011zfr,CMS:2012fgk,CMS:2014dyj,ATLAS:2014sic}
and
$\sqrt{s_{NN}}=5.02\TeV$~\cite{CMS:2021otx,CMS:2021kvd},
as well as in p-Pb collisions at
$\sqrt{s_{NN}}=5.02\TeV$~\cite{ATLAS:2015mwq,CMS:2015ehw,CMS:2015zlj,ALICE:2016rzo}
and $\sqrt{s_{NN}}=8.16\TeV$~\cite{CMS:2019leu}.

\begin{figure}[!t]
\subfigure[]{
\includegraphics[width=.47\textwidth]{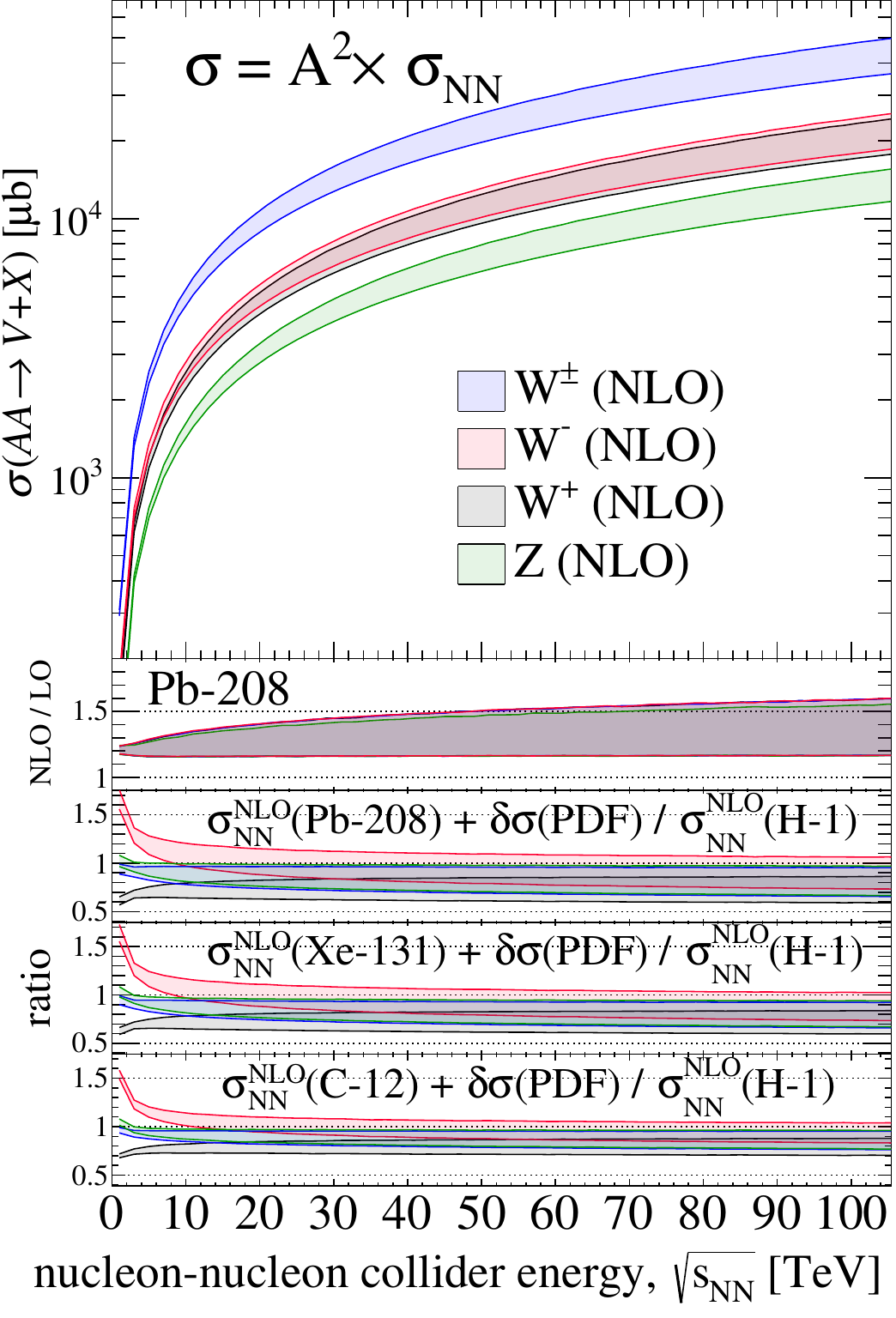}
\label{fig:ionsNLO_XSec_vXXX_vs_Beam}}
\subfigure[]{\includegraphics[width=.47\textwidth]{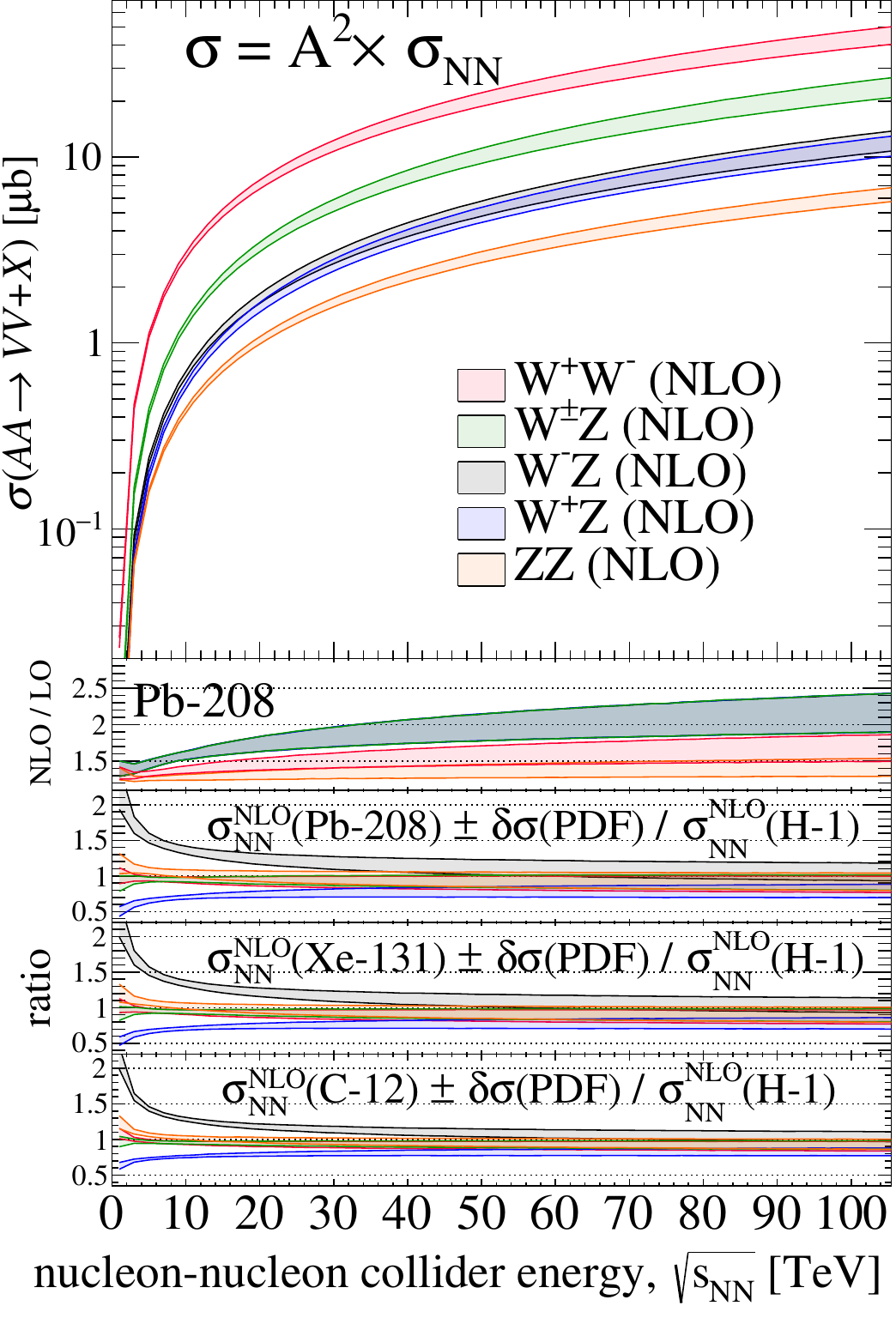}
\label{fig:ionsNLO_XSec_vvXX_vs_Beam}}
\caption{(a)
Top panel: the nucleus-level total cross section  at NLO in QCD for $W^\pm$ (blue), $W^-$ (red), $W^+$ (black), and $Z$ (green) production, shown in descending order, in symmetric $^{208}$Pb-$^{208}$Pb collisions as a function of nucleon-nucleon collision energy $\sqrt{s_{NN}}$. Band thickness corresponds to the residual scale uncertainty at this perturbative order. Second panel: The ratio of NLO cross sections and their scale uncertainties relative to the central LO cross section for $^{208}$Pb collisions. Third panel: The ratio of the nucleon-level NLO cross sections $(\sigma_{NN})$ and their PDF uncertainties for $^{208}$Pb collisions relative to the central NLO rate for protons. Fourth and bottom panel: same as third panel but for $^{131}$Xe and $^{12}$C, respectively. (b) Same as Fig.~\ref{fig:ionsNLO_XSec_vXXX_vs_Beam} but for $W^+W^-$ (red), $W^\pm Z$ (green), $WW^+Z$ (blue), $W^-Z$ (black), and $ZZ$ (orange) production, again shown in descending order.
}
\end{figure}

We show in the top panel of Fig.~\ref{fig:ionsNLO_XSec_vXXX_vs_Beam} the nucleus-level total cross section at NLO in QCD for $W^\pm$ (blue), $W^-$ (red), $W^+$ (black),  and $Z$ (green) production, shown in descending order, in  $^{208}$Pb-$^{208}$Pb collisions as a function of nucleon-nucleon collision energy $\sqrt{s_{NN}}$. The band thickness corresponds to the residual scale uncertainty at this order. In the second panel we show the ratio of NLO cross sections and their scale uncertainties relative to the central LO cross section for $^{208}$Pb collisions.
This is the QCD $K$-factor at NLO and is conventionally defined by the relationship
\begin{align}
    K^{\rm NLO}\ \pm\ \delta K\
    =\
    \sigma^{\rm NLO}/ \sigma^{\rm LO}\ \pm\ \delta\sigma / \sigma^{\rm LO}\ ,
\end{align}
where $\delta K$ represents the uncertainty in $K^{\rm NLO}$ that stems from the uncertainty in the NLO cross section prediction, which we denote by $\delta\sigma$.
In the third panel we have the ratio of the nucleon-level NLO cross sections $(\sigma_{NN})$ and their PDF uncertainties for $^{208}$Pb collisions relative to the central NLO rate for protons. In the fourth and bottom panels we show the same ion-over-proton ratios but for $^{131}$Xe and $^{12}$C, respectively.

Focusing first on the top panel, we report that over the  range $\sqrt{s_{NN}}=1\TeV-100\TeV$, the nucleus-level scattering rates and their residual scale uncertainty at NLO for lead collisions span approximately
\begin{subequations}
\begin{align}
    \sigma^{\rm NLO}_{AA\to V} &\sim 10^2\ub - 5\times10^4\ub,
    \\
    \delta\sigma^{\rm NLO}/\sigma^{\rm NLO} & \sim \confirm{\pm10\% - \pm30\%}\ .
\end{align}
\end{subequations}
For all collider energies, the $W^\pm$ production rate is the largest, followed by $W^-$, $W^+$, and finally $Z$ production. This hierarchy reflects an interplay between (a) the different gauge charges of quarks and (b) the size of parton densities. For instance: the difference between the $W^\pm$ and $Z$ rates at a fixed collider energy is largely due to the differences in gauge couplings. The ratio of the $W-u-\overline{d}$ and $Z-d-\overline{d}$ couplings (squared) scales as $(\Gamma_{Wu\overline{d}}/\Gamma_{Zd\overline{d}})^2 \sim (\sqrt{2} \cos\theta_W)^2\sim 1.5$. This is approximately the difference observed between the  $W^+$ (or $W^-$) production rate and the $Z$ production rate.

In addition, since $M_W,M_Z\lesssim\mathcal{O}(100\GeV)$, inclusive $W/Z$ production at low (high) collider energies is driven by valence-sea (sea-sea) quark scattering. Since the hard scattering scale $\sqrt{\hat{s}}=M_V$ is fixed by the matrix element and phase space, the kinematic threshold $\tau=x_1x_2=\hat{s}/s_{NN}$ necessarily decreases with increasing collision energies; this corresponds to probing smaller $x_k$ and an increased importance of low-$x$ partons. As shown in Secs.~\ref{sec:xpdf} and \ref{sec:lumi}, $^{208}$Pb has a larger $d_v$ content than $u_v$ content due to its surplus of neutrons. This means $^{208}$Pb exhibits a larger $d_v \overline{u}_{\rm sea}$ luminosity than $u_v\overline{d}_{\rm sea}$ luminosity. In other words,  at lower $\sqrt{s_{NN}}$ we expect a $W^-$ production rate that is larger than the $W^+$  rate. At larger collider energies, the $u_{\rm sea} \overline{d}_{\rm sea}$ and $d_{\rm sea} \overline{u}_{\rm sea}$ luminosities are similar since their production is driven by $g\to q\overline{q}$ splitting in QCD, which is flavor symmetric. This leads to the comparable $W^+$ and $W^-$ production rates observed in the figure. We additionally refer to Tables~\ref{tab:summary_pb208}-\ref{tab:summary_hx1_bis} (rows 1-4) for a quantitative comparison of the predictions at $\sqrt{s_{NN}} = 5$~TeV, 5.52~TeV, 39~TeV, and 100~TeV, illustrating this behavior.

For the NLO $K$-factor, we observe that all four channels exhibit comparable NLO QCD corrections, with
\begin{align}
    K^{\rm NLO}_{AA\to V} \sim 1.2 - 1.4\ ,
\end{align}
ranging from low-to-high $\sqrt{s_{NN}}$. These are consistent with perturbative QCD (pQCD) corrections for proton-proton collisions. There is similarity because the $\mathcal{O}(\alpha_s)$ corrections to quark-antiquark annihilation (Drell-Yan) are driven by the finite parts of virtual and soft-real corrections, which are positive, universal,  and scale as $d\sigma^\mathrm{virt+soft}\vert_{\rm finite}\sim \mathcal{O}( \alpha_s C_F\pi/3)\sim\mathcal{O}(0.14)$~\cite{Altarelli:1978id,Altarelli:1979ub}. However, at larger collider energies, the increased importance of pQCD corrections, which reach $\mathcal{O}(+40\%)$,
is due the largeness of the gluon density and thus the importance of the $qg\to Vq$ and $\overline{q}g\to V\overline{q}$ channels that open at this order. This demonstrates that even for ``simple'' processes LO predictions can severely underestimate total cross sections at large collision energies.

Focusing on the lower three panels, we observe a number of differences and similarities between the proton and an average nucleon from our representative isotopes. Qualitatively, we observe that for \confirm{$\sqrt{s_{NN}} \gtrsim 20\TeV$} (per nucleon), or when \confirm{$(\sqrt{\hat{s}}/\sqrt{s_{NN}})\sim (M_V/\sqrt{s_{NN}})\lesssim 4\times10^{-4}$}, all cross section ratios are largely flat and differ only by process-dependent normalizations. For instance: the per nucleon production rates of $Z$ bosons at large collider energies are about \confirm{$\mathcal{O}(80\%)$} compared to proton-proton collisions, while the $W^-$ production rates are about \confirm{$\mathcal{O}(90\%-95\%)$}. For the charge-symmetric channels, that is $W^\pm$ and $Z$ production, this ``convergence''-like behavior occurs even earlier at  \confirm{$\sqrt{s_{NN}} \gtrsim 5\TeV$}. This indicates that the \textit{rate} at which cross sections grow with collider energy is the same for the proton and the three ions, suggesting that the growth is dominated by partonic dynamics, \textit{i.e.}, by DGLAP evolution, and not by hadronic or nuclear dynamics.

Extracting the precise dependence of these global normalizations on the $(A,Z)$ numbers is obfuscated by two features: (i) the sizable uncertainties of nPDFs, which can exceed \confirm{$\mathcal{O}(\pm20\%)$}, and (ii) the relative nucleon content of our isotopes, which are comparable and are approximately $Z:(A-Z)\sim 40:60\ (40:60)\ [50:50]$ for $^{208}$Pb ($^{131}$Xe) [$^{12}$C]. However, we note ongoing efforts to determine such scaling relationships across nPDFs~\cite{Denniston:2023dwd}.

\subsection{Diboson processes}\label{sec:xsec_xvv}

We continue our survey with the inclusive production of two weak bosons. We specifically focus on processes facilitated at LO by quark-antiquark annihilation, illustrated in Fig.~\ref{fig:diagram_MultiBoson_V_VV_VH}(b), and given by
\begin{align}
    q\overline{q},\ q\overline{q'}\
    \to\
    W^+W^-,\    W^\pm Z,\ W^-Z,\ W^+Z,\ ZZ\ .
\end{align}
NLO in QCD predictions were first computed in Refs.~\cite{Ohnemus:1990za,Mele:1990bq,Ohnemus:1991kk,Ohnemus:1991gb,Frixione:1992pj,Frixione:1993yp} and at higher orders in Refs.~\cite{Gehrmann:2014fva,Cascioli:2014yka,Grazzini:2016swo}.

We show in the top panel of Fig.~\ref{fig:ionsNLO_XSec_vvXX_vs_Beam} the nucleus-level total cross section at NLO in QCD for $W^+W^-$ (red), $W^\pm Z$ (green),  $W^- Z$ (black), $W^+ Z$ (blue), and $ZZ$ (orange) production in symmetric $^{208}$Pb-$^{208}$Pb collisions as a function of nucleon-nucleon collision energy $\sqrt{s_{NN}}$. The band thickness corresponds to the residual scale uncertainty at this order. In the lower panels we show the same ratios as in Fig.~\ref{fig:ionsNLO_XSec_vXXX_vs_Beam} for single-boson production.
In the top panel, we observe that over the  range $\sqrt{s_{NN}}=1\TeV-100\TeV$, the nucleus-level scattering rates and their residual scale uncertainty at NLO for lead-lead collisions span approximately
\begin{subequations}
\begin{align}
    \sigma^{\rm NLO}_{AA\to VV} &\sim 0.1 \ub - 40\ub,
    \\
    \delta\sigma^{\rm NLO}/\sigma^{\rm NLO} & \sim \confirm{\pm10\% - \pm50\%}\ .
\end{align}
\end{subequations}
For all collider energies, the $W^+W^-$ production rate is the largest, followed by those for $W^\pm Z$, $W^-Z$, $W^+Z$, and finally $ZZ$ production. This hierarchy again reflects the interplay between the gauge charges of quarks and the density of quarks and gluons. The systematic $\mathcal{O}(10^3)$ drop in cross sections for diboson production relative to the single-boson case in Sec.~\ref{sec:xsec_xxv} is also consistent with na\"ive power counting and phase-space suppression associated with adding one  final-state leg. That is, for $V, V'\in\{W^\pm,Z\}$ and a hard scale of $Q=2M_V$ one has\footnote{The double logarithm originates from phase space integration over the additional propagator in $VV'$ production. At high energies, these capture the leading contributions of the emission of ``soft'' and ``collinear'' weak bosons.}
\begin{align}
\label{eq:power_counting_vv}
\frac{\sigma(AA \to VV'+X)}{\sigma(AA \to V+X)}\
\sim\
\mathcal{O}\left[\frac{\alpha_W}{4\pi}\log^2\left(\frac{Q^2}{M_V^2}\right)\right]\
\sim\ 5\times10^{-3}\ .
\end{align}
Furthermore, as in the single-boson case, the $W^-Z$ rate is larger than the $W^+Z$ due to the $d$ content of $^{208}$Pb.

In the second panel of the figure, we show the NLO $K$-factors for diboson production. These are large $(\gtrsim1.5)$, grow monotonically with energy, and span from low-to-high $\sqrt{s_{NN}}$:
\begin{subequations}
\begin{align}
    K^{\rm NLO}_{AA\to WZ} &\sim \confirm{1.4 - 2.2}\ ,
    \\
    K^{\rm NLO}_{AA\to WW} &\sim \confirm{1.3 - 1.7}\ ,
    \\
    K^{\rm NLO}_{AA\to ZZ} &\sim \confirm{1.3 - 1.5}\ .
\end{align}
\end{subequations}
The largeness of the perturbative corrections is real and is the result of so-called ``radiation amplitude zeros'' in the Born amplitudes~\cite{Mikaelian:1977ux,Brown:1979ux,Mikaelian:1979nr,Zhu:1980sz,Brodsky:1982sh,Brown:1982xx}. In essence, the tree-level $q\overline{q'}\to WZ$ process has a large, destructive interference that suppresses the LO cross section. This interference is disrupted in the $qg$ and $\overline{q}g$ channels and leads to a net $\mathcal{O}(\alpha_s)$ correction that is comparable to or larger than the Born result. Moreover, as the collider energy increases, the gluon density becomes  enhanced (see Sec.~\ref{sec:xpdf}), leading to much larger $qg$ and $\overline{q}g$ luminosities (see Sec.~\ref{sec:lumi}), and ultimately to $\mathcal{O}(\alpha_s)$ corrections to diboson production that increase with $\sqrt{s_{NN}}$. For proton collisions, pQCD corrections are more modest at NNLO, modulo the opening of $gg$ fusion channels, suggesting that the perturbative expansion stabilizes~\cite{Cascioli:2014yka,Gehrmann:2014fva,Grazzini:2016swo}.

The three lower panels show the (per nucleon) cross section ratios with respect to the cross section for $pp$ collisions. For the $W^+W^-$, $W^\pm Z$, and $ZZ$ channels we report that the ratios sit \confirm{at or just below unity for $\sqrt{s_{NN}}\gtrsim2-4\TeV$}. This follows from the three channels being charge- and flavor-symmetric.\footnote{We briefly note that charge- and flavor-symmetric channels, which run over all quark species, do not imply flavor universality, which means being the same for all quark species. $ZZ$ production, for example, is flavor symmetric since all $q\overline{q}$ and $qg$ configurations are included  but it is not flavor universal since the $Z-q-\overline{q}$ coupling depends on flavor. This implies that flavor-summed channels can still be sensitive to the relative abundance of partons, and hence the proton-to-neutron ratio of a nucleus.} More specifically, the enhanced $d$-quark density (or $d\overline{u}$ luminosity) that drives the $W^-Z$ ratio well above unity for \confirm{$\sqrt{s_{NN}}\lesssim30\TeV$} is mostly, but not exactly, compensated by the suppressed $u$-quark density (or $u\overline{d}$ luminosity) that drives the $W^+Z$ ratio well below unity for \confirm{$\sqrt{s_{NN}}\lesssim20\TeV$}. So while the relative importance of the $u$ and $d$ densities in  $q\overline{q}\to VV'$ (or $qg\to VV'q$) scattering shifts as the $(A,Z)$ composition of an average nucleon becomes more proton-like or more neutron-like, we still sum over all charge and flavor configurations of initial-state parton pairs. As a consequence, differences in scattering rates between  a ``neutron-like'' nucleon and the proton (approximately) vanish due to (approximate) isospin symmetry. Remaining differences between the isotopes and the proton in $W^+W^-$, $W^\pm Z$, and $ZZ$ are then due to (i) violations of isospin symmetry between protons and neutrons at the hadronic level and (ii) intra-nuclear exchanges at the nuclear level. For instance: the ratios for these three channels reflect the fact that the $q\overline{q}$, $q\overline{q'}$, and $qg$  luminosities at $\sqrt{\hat{s}}=100\GeV$ (see  Sec.~\ref{sec:lumi}) sit just below those for the proton. Finally, we find that the ratios stabilize and become flat-like when collision energies exceed $\sqrt{s_{NN}}\sim 40-50\TeV$. This corresponds to the ratio \confirm{$(\sqrt{\hat{s}}/\sqrt{s_{NN}})\sim (2M_V/\sqrt{s_{NN}})\lesssim 4\times10^{-4}$}.
We refer to Tables~\ref{tab:summary_pb208}-\ref{tab:summary_hx1_bis} (rows 5-9) for a quantitative comparison of diboson predictions at $\sqrt{s_{NN}} = 5$~TeV, 5.52~TeV, 39~TeV, and 100~TeV.

\subsection{Photon associated processes}\label{sec:xsec_xva}

\begin{figure}[!t]
\subfigure[]{\includegraphics[width=.47\textwidth]{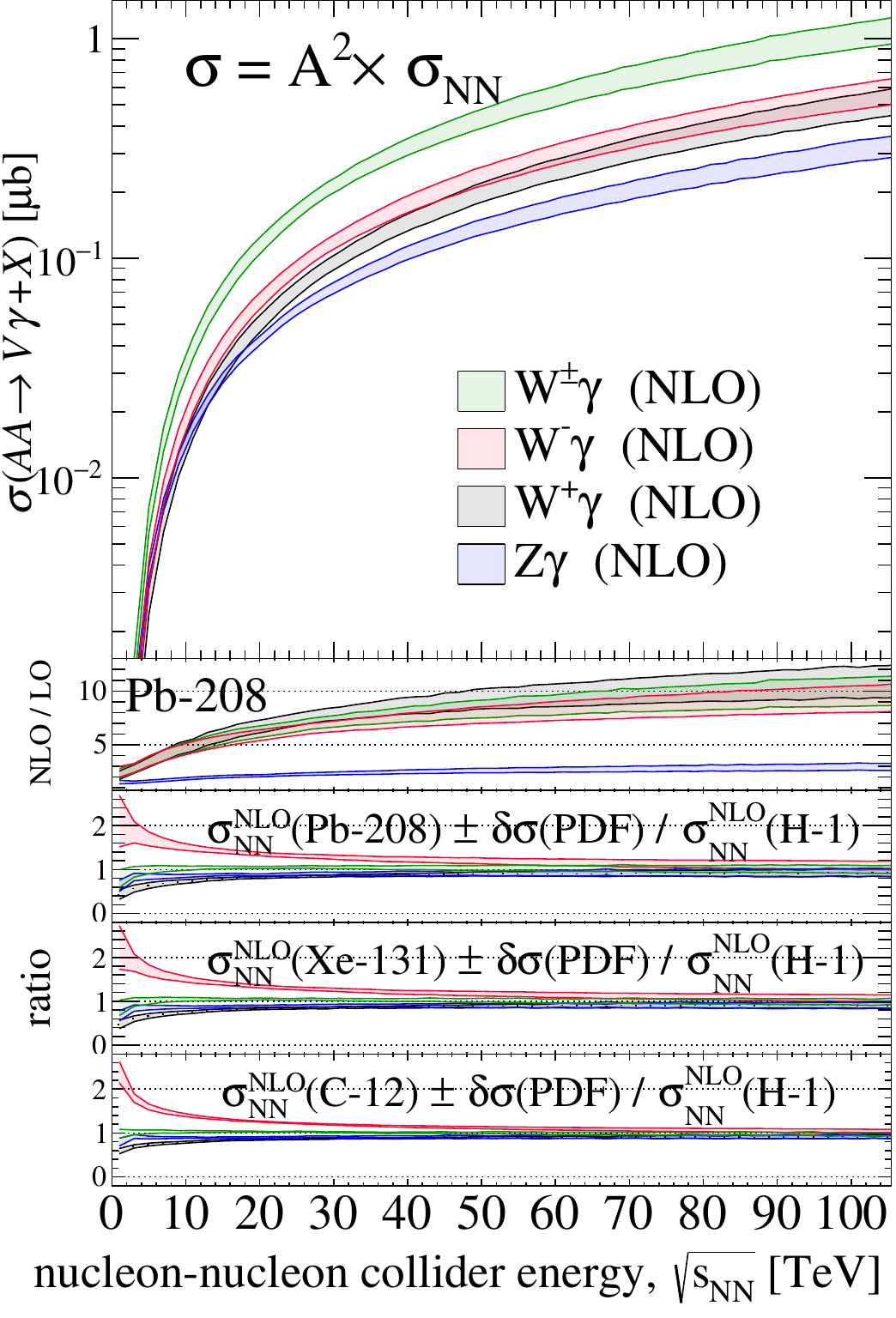}
\label{fig:ionsNLO_XSec_vaXX_vs_Beam}}
\subfigure[]{\includegraphics[width=.47\textwidth]{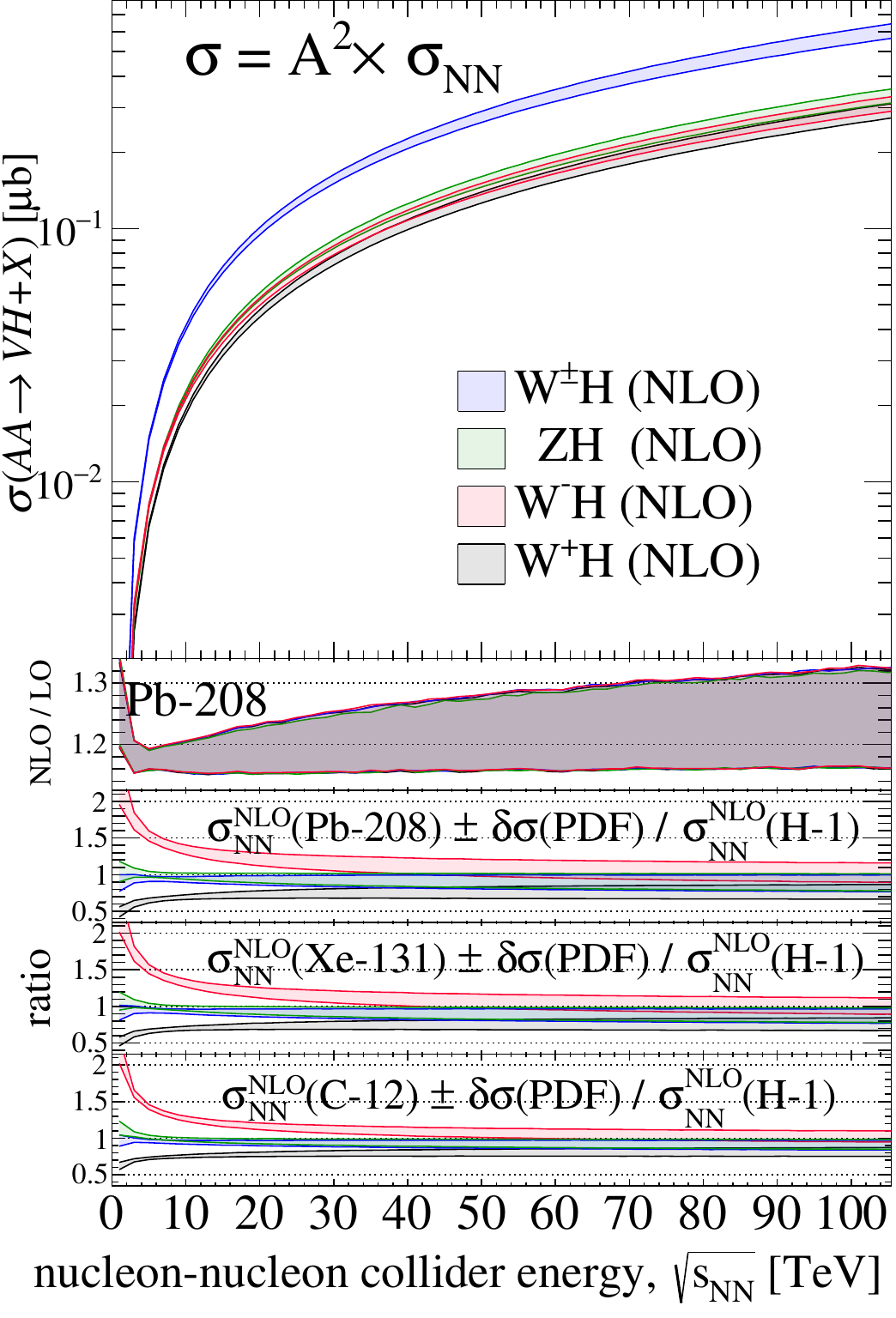}
\label{fig:ionsNLO_XSec_vhXX_vs_Beam}}
\caption{(a) Same as Fig.~\ref{fig:ionsNLO_XSec_vXXX_vs_Beam} but for $W^\pm \gamma$ (green), $W^- \gamma$ (red), $W^+\gamma$ (black), and $Z \gamma$ (blue) production with the phase space cuts of Eqs.~\eqref{eq:cuts_photon} and \eqref{eq:cuts_photon_iso}, shown in descending order. (b) Same as Fig.~\ref{fig:ionsNLO_XSec_vXXX_vs_Beam} but for $W^\pm H$ (blue), $Z H$ (green), $W^- H$ (red), and $W^+Z$ (black) production, also shown in descending order.}
\end{figure}

We now focus on the production of a high-energy photon $(\gamma)$ in association with one weak boson, as illustrated diagrammatically in Fig.~\ref{fig:diagram_MultiBoson_V_VV_VH}(b), and given at LO by
\begin{align}
    q\overline{q},\ q\overline{q'}\
    \to&\
    W^\pm \gamma,\
    W^- \gamma,\
    W^+ \gamma,\
    Z \gamma\ .
\end{align}
To regulate infrared divergences in matrix elements, we impose the phase space cuts of  Eq.~\eqref{eq:cuts_photon}.
Full NLO in QCD predictions for $V\gamma$ were first reported in Refs.~\cite{Ohnemus:1992jn,Baur:1997kz} and are now available at higher orders~\cite{Baur:1997kz}.

We show in the top panel of Fig.~\ref{fig:ionsNLO_XSec_vaXX_vs_Beam} the nucleus-level total cross section at NLO in QCD for $W^\pm \gamma$ (green), $W^- \gamma$ (red), $W^+ \gamma$ (black), and $Z\gamma$ (blue) production, shown in descending order, in $^{208}$Pb-$^{208}$Pb collisions and as a function of $\sqrt{s_{NN}}$. The band thickness corresponds to the residual scale uncertainty at this order. In the lower panels we show the same ratios as in Fig.~\ref{fig:ionsNLO_XSec_vXXX_vs_Beam} for single-boson production.
Focusing on the top panel, nucleus-level cross sections and uncertainties at NLO roughly span
\begin{subequations}
\begin{align}
    \sigma^{\rm NLO}_{AA\to V\gamma} &\sim 10^{-3}\ub - 1\ub,
    \\
    \delta\sigma^{\rm NLO}/\sigma^{\rm NLO} & \sim \confirm{\pm10\% - \pm20\%}\ .
\end{align}
\end{subequations}
For all collision energies, the $W^\pm \gamma$ rate is the largest, followed by the $W^-\gamma$, $W^+\gamma$, and finally $Z\gamma$ rates. As in the single boson case (Sec.~\ref{sec:xsec_xxv}), this hierarchy reflects gauge charges and parton densities.

We immediately focus on the $K$-factors, which are large and span with increasing energy:
\begin{subequations}
\begin{align}
    K^{\rm NLO}_{AA\to W\gamma} &\sim \confirm{2 - 10}\ ,
    \\
    K^{\rm NLO}_{AA\to Z\gamma} &\sim \confirm{1.5 - 3}\ .
\end{align}
\end{subequations}
As in the diboson case (Sec.~\ref{sec:xsec_xvv}),
the largeness of these corrections
can be attributed to the presence of radiation amplitude zeros at the Born level. The amplitude zero is more severe in $V\gamma$ production than in $VV'$ production due to (i) the Abelian nature of QED and (ii) the masslessness of the photon~\cite{Mikaelian:1977ux,Brown:1979ux,Mikaelian:1979nr,Zhu:1980sz,Brodsky:1982sh,Brown:1982xx}. However, potential enhancements from soft-photon emission are moderated by the phase space cuts of Eqs.~\eqref{eq:cuts_photon} and \eqref{eq:cuts_photon_iso}. The absence of a $\gamma-Z-Z$ and $\gamma-\gamma-Z$ vertices also implies that $Z\gamma$ production experiences a smaller destructive interference at the Born level than $W\gamma$ production, which implies a smaller amplitude zero and subsequently to a smaller QCD $K$-factor at NLO. And as in the $VV'$ case, corrections at NNLO suggest that the perturbative series stabilizes~\cite{Grazzini:2015nwa}.

The three lower panels show the (per nucleon) cross section ratio with respect to the $pp$ cross section. We report that the cross section ratios for charge- and flavor-symmetric channels ($W^\pm \gamma$ and $Z\gamma$) have mostly flat ratios \confirm{at or just below unity for $\sqrt{s_{NN}}\gtrsim 5\TeV$}, whereas the charge- and flavor-asymmetric channels ($W^-\gamma$ and $W^+\gamma$) deviate significantly from unity, particularly for \confirm{$\sqrt{s_{NN}}\lesssim30\TeV$}. For example: at $\sqrt{s_{NN}}\sim2\TeV$, the per nucleon rates for $W^-\gamma$ in lead, xenon, and carbon are nearly twice as large as the rate of for proton-proton collisions. Despite being a charge- and flavor-symmetric channel, $Z\gamma$ production can only occur through the emission of a photon off an initial-state quark. Hence, $Z\gamma$ is sensitive to the net weak and electric charges carried by \textit{all} partons in an average nucleon, and consequentially the nucleon's net nuclear isospin. For all three ion configurations, we find that the ratios stabilize, \textit{i.e.}, become flat, for collision energies that exceed $\sqrt{s_{NN}}\gtrsim60-70\TeV$, or when \confirm{$(\sqrt{\hat{s}}/\sqrt{s_{NN}}) \sim (M_V+p_T^\gamma)/\sqrt{s_{NN}}\sim 4\times10^{-3}$}. We refer to Tables~\ref{tab:summary_pb208}-\ref{tab:summary_hx1_bis} (rows 10-13) for a quantitative comparison of $V\gamma$ predictions at $\sqrt{s_{NN}} = 5$~TeV, 5.52~TeV, 39~TeV, and 100~TeV.

\subsection{Higgs associated processes}\label{sec:xsec_xvh}

We now focus on the production of the Higgs boson $(H)$ in association with one weak boson, as illustrated diagrammatically in Fig.~\ref{fig:diagram_MultiBoson_V_VV_VH}(c), and specifically on processes given at LO by
\begin{align}
    q\overline{q},\ q\overline{q'}\
    \to&\
    W^\pm H,\
    Z H,\
    W^- H,\
    W^+ H\ .
\end{align}
NLO in QCD predictions for $VH$ production were first reported in Ref.~\cite{Han:1991ia} for $pp$ collisions and are now available at higher orders~\cite{Brein:2003wg,Baglio:2022wzu}.
Prospects for observing Higgs production via gluon fusion  in hadronic and non-hadronic decay channels have been reported elsewhere for a variety of ion configurations and collision energies~\cite{Berger:2018mtg,dEnterria:2019cps}.

We show in the top panel of Fig.~\ref{fig:ionsNLO_XSec_vhXX_vs_Beam} the nucleus-level total cross section at NLO in QCD for $W^\pm H$ (blue), $ZH$ (green), $W^- H$ (red), and $W^+ H$ (black) production, shown in descending order, in $^{208}$Pb-$^{208}$Pb collisions versus $\sqrt{s_{NN}}$. As above, band thickness denotes residual scale uncertainty at this order. The lower panels show the same ratios as shown in Fig.~\ref{fig:ionsNLO_XSec_vXXX_vs_Beam} for single-boson production.
Nucleus-level cross sections and scale uncertainties at NLO for lead-lead collisions span approximately
\begin{subequations}
\begin{align}
    \sigma^{\rm NLO}_{AA\to VH} &\sim 10^{-3}\ub - 0.6\ub,
    \\
    \delta\sigma^{\rm NLO}/\sigma^{\rm NLO} & \sim \confirm{\pm10\% - \pm20\%}\ .
\end{align}
\end{subequations}
For all collision energies, the $W^\pm H$ rate is the largest, followed by the $ZH$, $W^-H$, and then $W^+H$ rates.

The most remarkable feature is the similarity of the individual $ZH$, $W^-H$, and $W^+H$ rates, which is qualitatively different than what was observed for single-boson production (Sec.~\ref{sec:xsec_xxv}) and diboson production (Sec.~\ref{sec:xsec_xvv}). This behavior is accidental: the smaller $Z-q-\overline{q}$ couplings that lead to a suppressed single-$Z$ production rate [see Fig.~\ref{fig:ionsNLO_XSec_vXXX_vs_Beam}] are balanced by the larger $H-Z-Z$ coupling. For example: at $\sqrt{s_{NN}}=100\TeV$ the $(ZH)$-over-$(W^-H)$ cross section ratio is approximately
\begin{align}
    \frac{\sigma_{AA\to ZH}^{\rm NLO}}{\sigma_{AA\to W^-H}^{\rm NLO}}\Bigg\vert_{\sqrt{s_{NN}}=100\TeV}
    ^{\rm ^{208}Pb-^{208}Pb}\
    \approx \confirm{1.071}\ ,
\end{align}
whereas the $Z$-over-$W^-$ cross section ratio weighted by the $(ZH)$-over-$(W^-H)$ coupling ratio is
\begin{align}
    \left(\frac{M_Z}{M_W\cos\theta_W}\right)^2\ \times\
    \frac{\sigma_{AA\to Z}^{\rm NLO}}{\sigma_{AA\to W^-}^{\rm NLO}}\Bigg\vert_{\sqrt{s_{NN}}=100\TeV}
    ^{\rm ^{208}Pb-^{208}Pb}
    \approx \left(1.286\right)^2\ \times\
    \confirm{\left(0.617\right)}\
    \approx \confirm{1.020}\ .
\end{align}
We attribute the small, \confirm{$\mathcal{O}(5\%)$} difference between these sets of ratios to the small kinematical differences in matrix elements, \textit{e.g.}, $\sqrt{\hat{s}}\sim M_V$ versus $\sqrt{\hat{s}}\sim (M_V+m_H)$, as well as to the small variations in PDFs, \textit{e.g.}, $x\sim M_V/\sqrt{s_{NN}}$ versus $x\sim (M_V+m_H)/\sqrt{s_{NN}}$.

The second panel shows the QCD $K$-factor at NLO, which grows monotonically and spans
\begin{align}
    K^{\rm NLO}_{AA\to VH} \sim \confirm{1.2 - 1.3}\
\end{align}
for all channels. The color structure of $VH$ production is identical to single weak boson production  (Sec.~\ref{sec:xsec_xxv}) and therefore each receives the same QCD corrections at NLO. However, in the present case, NLO corrections are more modest than in single boson production. This follows from the increase in hard scale for the present case, \textit{i.e.}, $\sqrt{\hat{s}}\sim (M_V+m_H)$ for associated $VH$ production versus $\sqrt{\hat{s}}\sim M_V$ for single $V$ production. A larger hard scale implies that the strong coupling constant $\alpha_s(\mu_r=\sqrt{\hat{s}})$ runs to smaller values in $VH$ production, and hence leads to smaller shifts in normalization.

The three lower panels show the (per nucleon) cross section ratio with respect to the cross section for $pp$ collisions. As in the diboson case (Sec.~\ref{sec:xsec_xvv}), we find that the cross section ratios for charge- and flavor-symmetric channels ($W^\pm H$ and $ZH$) have a ratio just below unity for collision scales above $\sqrt{s_{NN}}\gtrsim 5\TeV$. Contrary to this, the charge- and flavor-asymmetric channels ($W^-H$ and $W^+H$) deviate from unity for all $\sqrt{s_{NN}}$, with the most significant deviations below $\sqrt{s_{NN}}\lesssim30\TeV$. For all three configurations of symmetric ion collisions, we find that the ratios stabilize, \textit{i.e.}, become flat, for collision energies that exceed $\sqrt{s_{NN}}\gtrsim40-50\TeV$. This corresponds to the ratio \confirm{$(\sqrt{\hat{s}}/\sqrt{s_{NN}}) \sim (M_V+m_H)/\sqrt{s_{NN}}\sim 4-5\times10^{-3}$}. We refer to Tables~\ref{tab:summary_pb208}-\ref{tab:summary_hx1_bis} (rows 14-17) for a quantitative comparison of $VH$ predictions at $\sqrt{s_{NN}} = 5$~TeV, 5.52~TeV, 39~TeV, and 100~TeV.

\subsection[\texorpdfstring{Triboson I: $WWX$ and $ZZX$}{Triboson I: WWX and ZZX}]{Triboson Processes I: $WWX$ and $ZZX$}\label{sec:xsec_vvv}

\begin{figure}[!t]
\includegraphics[width=.95\textwidth]{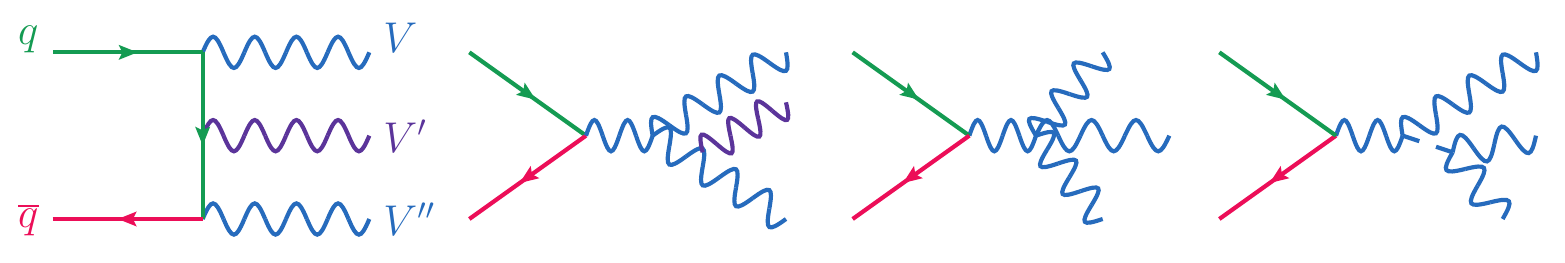}
\caption{Representative Born-level Feynman graphs
depicting the  production of three weak bosons.
}\label{fig:diagram_MultiBoson_VVV}
\end{figure}

\begin{figure}[!t]
\subfigure[]{\includegraphics[width=.47\textwidth]{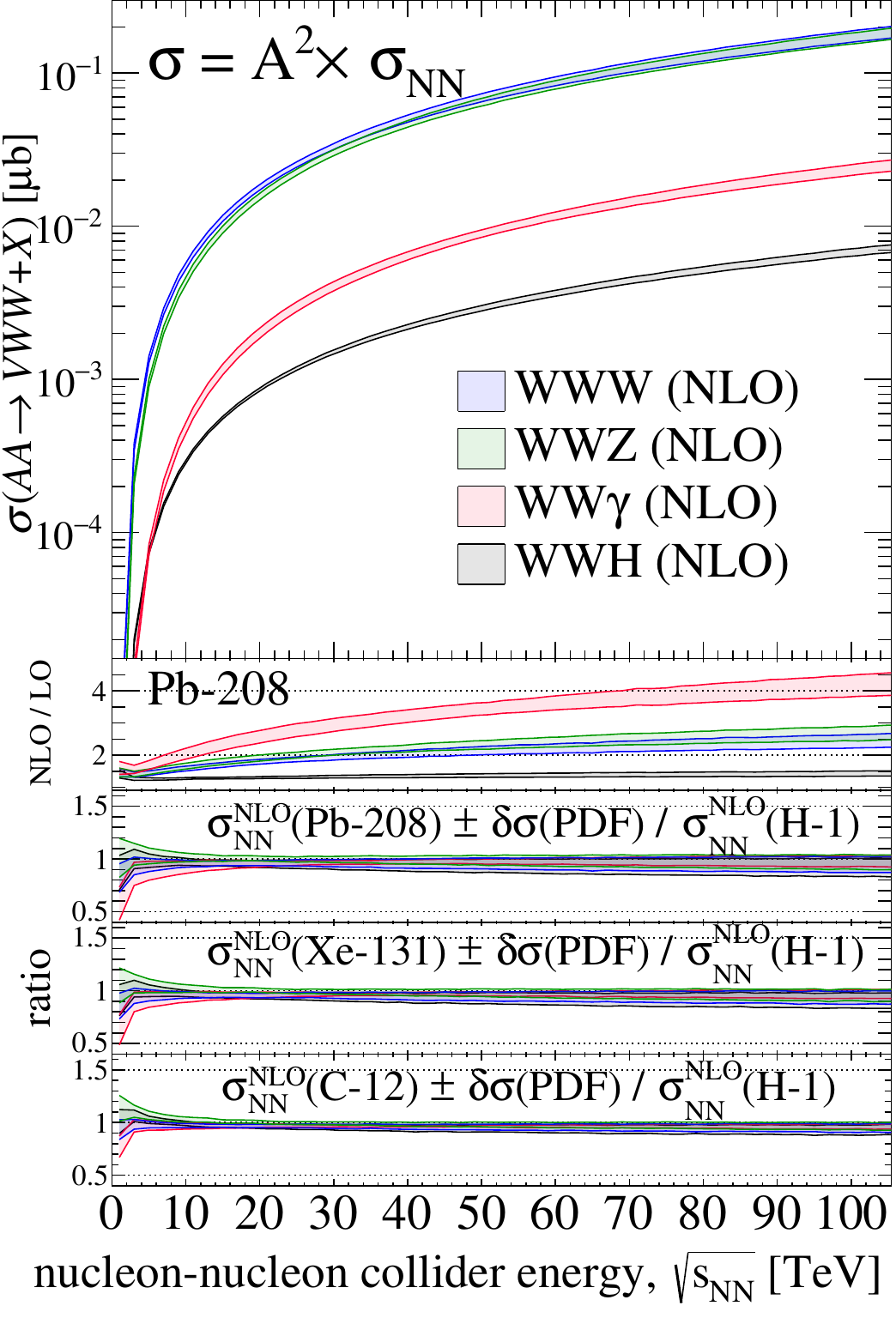}
\label{fig:ionsNLO_XSec_vwwX_vs_Beam}}
\subfigure[]{\includegraphics[width=.47\textwidth]{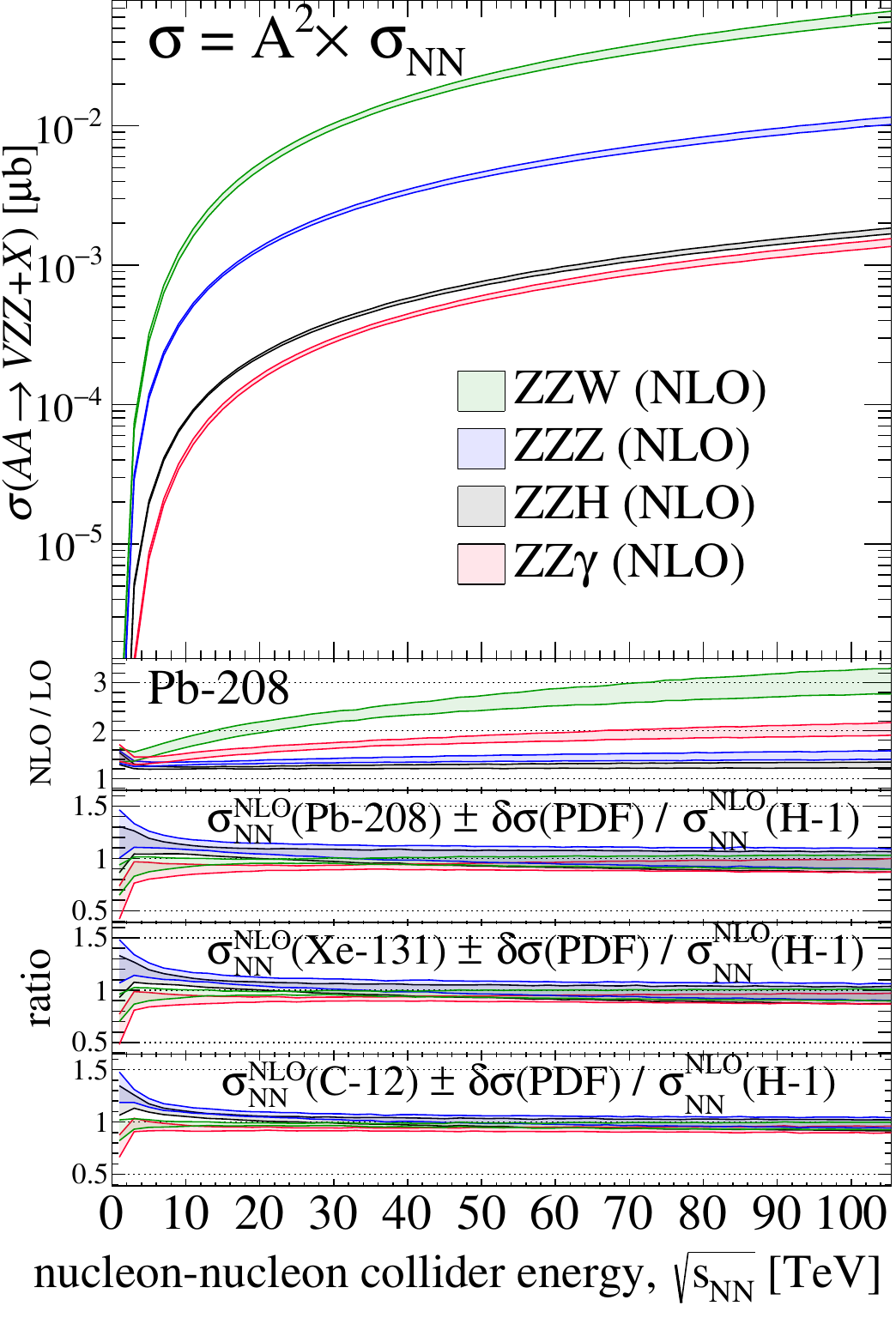}
\label{fig:ionsNLO_XSec_vzzX_vs_Beam}}
\caption{Same as Fig.~\ref{fig:ionsNLO_XSec_vXXX_vs_Beam} but for (a) $W^+W^-W^\pm$ (blue), $W^+W^-Z$ (green), $W^+W^-\gamma$ (red), and $W^+W^-H$ (black), as well as for (b) $ZZW^\pm$ (green), $ZZZ$ (blue), $ZZH$ (black) and $ZZ\gamma$ (red), all in descending order.}
\label{fig:ionsNLO_XSec_vvvX_vs_Beam}
\end{figure}

Turning to the first set of triboson processes, we consider in Fig.~\ref{fig:ionsNLO_XSec_vwwX_vs_Beam} those processes with at least two $W$ bosons and in Fig.~\ref{fig:ionsNLO_XSec_vzzX_vs_Beam} those processes with at least two $Z$ bosons, as schematically illustrated by the Feynman diagrams in Fig.~\ref{fig:diagram_MultiBoson_VVV}. Specifically, we consider the channels given at LO by the following partonic processes:
\begin{subequations}
\begin{align}
    q\overline{q},\ q\overline{q'}\
    &\to\
    W^+W^-W^\pm,\
    W^+W^-Z,\
    W^+W^-\gamma,\
    W^+W^-H,\quad\text{and}
    \\
    q\overline{q},\ q\overline{q'}\
    &\to\
    ZZW^\pm,\
    ZZZ,\
    ZZH,\
    ZZ\gamma\ .
\end{align}
\end{subequations}
NLO predictions for $W^+W^-X$ and $ZZX$ were first reported in Refs.~\cite{Lazopoulos:2007ix,Hankele:2007sb,Binoth:2008kt,Bozzi:2009ig,Baglio:2011juf,Alwall:2014hca}.
We remind the reader that for processes with a final-state photon, we impose the phase space cuts given in Eqs.~\eqref{eq:cuts_photon} and \eqref{eq:cuts_photon_iso}.

In the top panel of Fig.~\ref{fig:ionsNLO_XSec_vwwX_vs_Beam} we show the nucleus-level cross section at NLO in QCD for $W^+W^-W^\pm$ (blue), $W^+W^-Z$ (green), $W^+W^-\gamma$ (red), and $W^+W^-H$ (black) production in descending order for $^{208}$Pb-$^{208}$Pb collisions as a function of nucleon-nucleon collision energy. In the top panel of Fig.~\ref{fig:ionsNLO_XSec_vzzX_vs_Beam} we additionally show, also in descending order, predictions for $ZZW^\pm$ (green), $ZZZ$ (blue), $ZZH$ (black) and $ZZ\gamma$ (red) production. In all cases, band thickness represents the residual scale uncertainty. In the lower panels we show the same nucleon-level ratios as in Fig.~\ref{fig:ionsNLO_XSec_vXXX_vs_Beam} for single-boson production.

For $\sqrt{s_{NN}}=1-100\TeV$, nucleus-level cross sections and uncertainties at NLO roughly span
\begin{subequations}
\begin{align}
    \sigma^{\rm NLO}_{AA\to WWV} &\sim 10^{-5}\ub - 0.2\ub,
    \\
    \sigma^{\rm NLO}_{AA\to ZZV} &\sim 10^{-5}\ub - 6\times10^{-2}\ub,
    \\
    \delta\sigma^{\rm NLO}/\sigma^{\rm NLO} & \sim \confirm{\pm10\% - \pm20\%}\ .
\end{align}
\end{subequations}
Across all collision energies, the $W^+W^-W^\pm$ and $W^+W^-Z$ channels shown in Fig.~\ref{fig:ionsNLO_XSec_vwwX_vs_Beam} have the largest rates and are highly comparable, particularly for $\confirm{\sqrt{s_{NN}}\gtrsim30-40\TeV}$; below this energy range, the $W^+W^-W^\pm$ rate is slightly larger. Below these curves \confirm{by roughly a factor of $4$} lies the $ZZW^\pm$ channel in Fig.~\ref{fig:ionsNLO_XSec_vzzX_vs_Beam}. The $W^+W^-\gamma$, $ZZZ$, and $W^+W^-H$ channels are the next largest in descending order, reaching \confirm{$\sigma^{\rm NLO}\sim \mathcal{O}(10^{-4}\ub)\ [\mathcal{O}(10^{-2}\ub)]$ at $\sqrt{s_{NN}}\sim5\TeV\ [100\TeV]$.} The $ZZH$ and $ZZ\gamma$ rates in Fig.~\ref{fig:ionsNLO_XSec_vzzX_vs_Beam} have the smallest rates, with $ZZ\gamma$ being smallest.
This comparison is illustrated quantitatively in Tables~\ref{tab:summary_pb208}-\ref{tab:summary_hx1_bis} (rows 18-25) for $\sqrt{s_{NN}} = 5$~TeV, 5.52~TeV, 39~TeV, and 100~TeV.

As with single boson production (Sec.~\ref{sec:xsec_xxv}), diboson production (Sec.~\ref{sec:xsec_xvv}), and $VH$/$V\gamma$ associated production (Secs.~\ref{sec:xsec_xva} and \ref{sec:xsec_xvh}), the relative sizes of the triboson channels can be understood by power counting of coupling constants and phase space suppression. The latter manifests through the appearance of a $(1/4\pi)$ factor for each new final-state particle. For example: the $W^+W^-W^\pm$ rate at NLO in QCD is roughly \confirm{$\mathcal{O}\left(5\times10^{-6}\right)$} smaller than the inclusive $W^\pm$ rate [see Fig.~\ref{fig:ionsNLO_XSec_vXXX_vs_Beam}]. To some degree, the $WWW$ channel can be modeled as a real radiative correction to inclusive $W^\pm$ production via a $\gamma^* \to W^+W^-$ leg. By power counting and phase-space suppression alone, one can estimate a cross-section ratio of
\begin{align}
\label{eq:power_counting_vvv}
\frac{\sigma^{\rm NLO}_{AA \to WWW}}{\sigma^{\rm NLO}_{AA \to W}}\
 \sim\
 \mathcal{O}\left(
 \frac{\alpha}{4\pi}
 \frac{\alpha_W}{4\pi}
 \right)\
\sim\ \confirm{1.5\times10^{-6}}\ .
\end{align}
Interfering channels, such as those with $Z^*/H^*\to W^+W^-$ legs, and differences in QCD corrections (discussed next) can account for the difference between Eq.~\eqref{eq:power_counting_vvv} and $\mathcal{O}\left(5\times10^{-6}\right)$. For completeness, we note that a comparable ratio can also be obtained by instead treating $WWW$ production as a single-emission radiative correction to various diboson production channels.

In the second panels, we observe a variety of QCD $K$-factors at NLO. These span roughly
\begin{subequations}
\begin{align}
    K^{\rm NLO}_{AA\to WW\gamma} &\sim \confirm{1.5 - 4}\ ,
    \\
    K^{\rm NLO}_{AA\to ZZW} &\sim \confirm{1.5 - 3}\ ,
    \\
    K^{\rm NLO}_{AA\to WWW, WWZ} &\sim \confirm{1.5 - 2.5}\ ,
    \\
    K^{\rm NLO}_{AA\to ZZ\gamma} &\sim \confirm{1.4 - 2}\ ,
    \\
    K^{\rm NLO}_{AA\to WWH,ZZH,ZZZ} &\sim \confirm{1.4 - 1.5}\ .
\end{align}
\end{subequations}
While more modest than for $VV'$ processes (see Secs.~\ref{sec:xsec_xvv} and \ref{sec:xsec_xva}), several of the $K$-factors here are large $(\gtrsim2)$, particularly for channels with final-state photons. Overall, the sizable importance of these $K$-factors can be attributed to the combination of three ingredients~\cite{Binoth:2008kt}: (i) the disruption of radiation zeros, which can be expected since the gauge and color structures of many $VVV$ sub-channels are similar to diboson production; (ii) the opening of new partonic channels, \textit{i.e.}, $(gq)$-scattering; and (iii) sizable finite virtual+soft corrections, which are positive and can be larger than those in single-boson production.

In the three lower panels we again show the (per nucleon) cross section ratio with respect to $pp$ collisions. For $\confirm{\sqrt{s_{NN}}\gtrsim 10\TeV}$, we observe that for all channels the ratios are approximately flat and are consistent with being at or just below unity. Again, we attribute this similarity to $pp$ collisions to summing over all charge and flavor configurations for each process under consideration. This sum renders the nucleon-level cross section (approximately) isospin invariant, that is to say, it does not depend on the colliding nucleon being a pure proton state, pure neutron state, or some admixture of the two.

\subsection[\texorpdfstring{Triboson II: $WZX$, $\gamma\gamma V$, $HHV$, and other processes}{
Triboson II: WZX, AAV, HHV, and other processes}]{
Triboson II: $WZX$, $\gamma\gamma V$, $HHV$, and other processes} \label{sec:xsec_wzv}

\begin{figure}[!t]
\subfigure[]{\includegraphics[width=.47\textwidth]{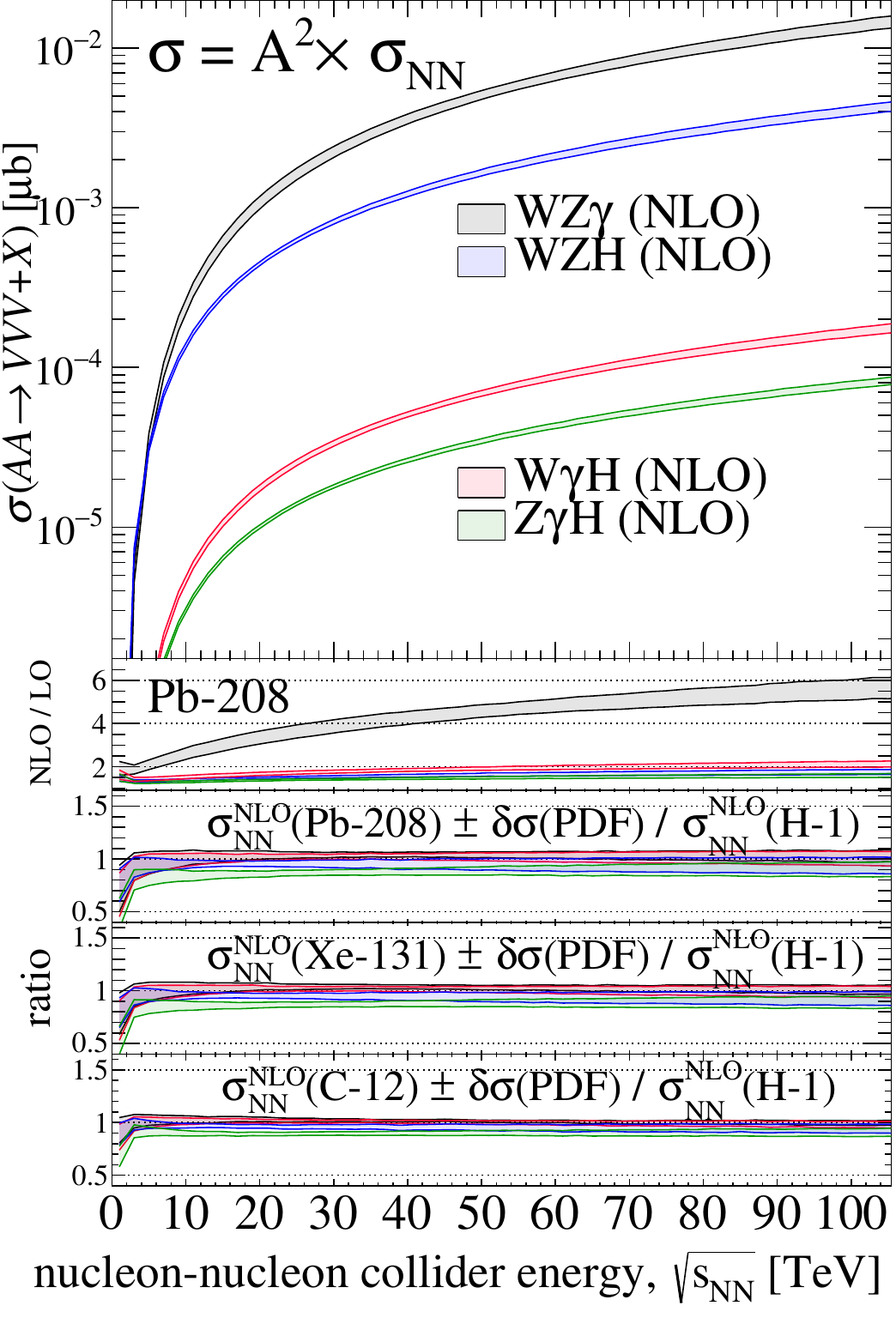}
\label{fig:ionsNLO_XSec_vwzX_vs_Beam}}
\subfigure[]{\includegraphics[width=.47\textwidth]{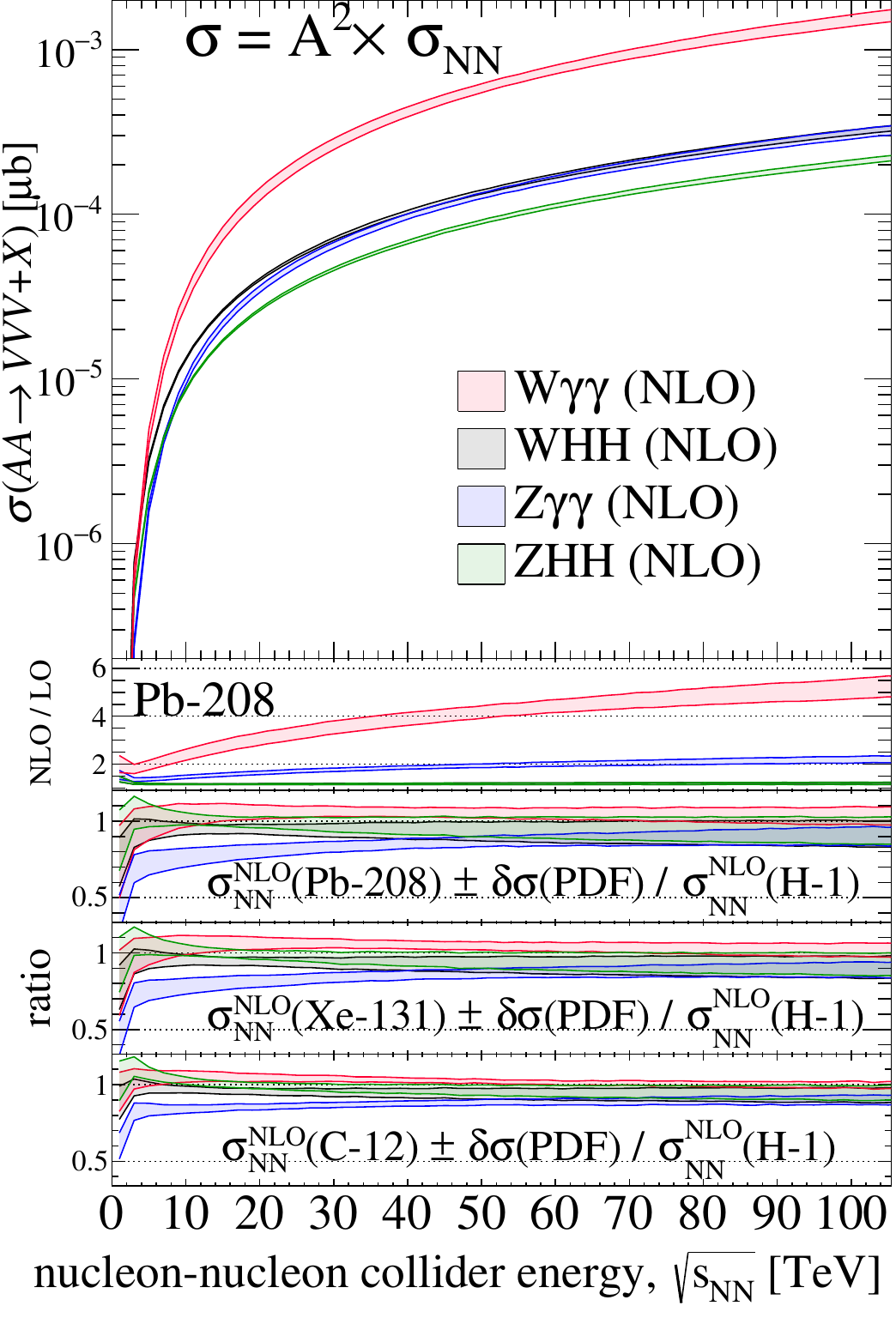}
\label{fig:ionsNLO_XSec_vaaX_vs_Beam}}
\caption{ Same as Fig.~\ref{fig:ionsNLO_XSec_vXXX_vs_Beam} but for (a) $W^\pm Z\gamma$ (black), $W^\pm ZH$ (blue), $W^\pm\gamma H$ (red), and $Z\gamma H$ (green), as well as (b) $W^\pm\gamma\gamma$ (red), $W^\pm HH$ (black), $Z\gamma\gamma$ (blue) and $ZHH$ (green), all in descending order.}
\label{fig:ionsNLO_XSec_vvv2_vs_Beam}
\end{figure}

We now continue in Fig.~\ref{fig:ionsNLO_XSec_vvv2_vs_Beam} with various other triboson processes that are given at LO by:
\begin{subequations}
\begin{align}
  q\overline{q},\ q\overline{q'}\
    &\to\
    W^\pm Z\gamma\ ,
    W^\pm ZH\ ,
    W^\pm\gamma H\ ,
    Z\gamma H\quad\text{and}
    \\
    q\overline{q},\ q\overline{q'}\
    &\to\
    W^\pm\gamma\gamma\ ,
    W^\pm HH\ ,
    Z\gamma\gamma\ ,
    ZHH\ .
\end{align}
\end{subequations}
The phase space cuts in Eqs.~\eqref{eq:cuts_photon} and \eqref{eq:cuts_photon_iso} are imposed on channels with final-state photons.
Predictions at NLO in QCD for these processes were first reported in Refs.~\cite{Baur:2010zf,Bozzi:2010sj,Bozzi:2011wwa,Bozzi:2011en,Song:2013sex,Frederix:2014hta,Alwall:2014hca}

In the top panel of Fig.~\ref{fig:ionsNLO_XSec_vwzX_vs_Beam} we show the nucleus-level cross section at NLO in QCD for $W^\pm Z\gamma$ (black), $W^\pm ZH$ (blue), $W^\pm\gamma H$ (red), and $Z\gamma H$ (green) production in descending order for symmetric $^{208}$Pb-$^{208}$Pb collisions as a function of the nucleon-nucleon collision energy $\sqrt{s_{NN}}$. Additionally, in the top panel of Fig.~\ref{fig:ionsNLO_XSec_vaaX_vs_Beam} we show, also in descending order, predictions for $W^\pm\gamma\gamma$ (red), $W^\pm HH$ (black),
$Z\gamma\gamma$ (blue) and $ZHH$ (green) production. In all cases, band thickness corresponds to the residual scale uncertainty at NLO in pQCD. In the lower panels we display the same nucleon-level ratios as in Fig.~\ref{fig:ionsNLO_XSec_vXXX_vs_Beam}.

For $\sqrt{s_{NN}}=1-100\TeV$, nucleus-level cross sections and uncertainties at NLO are roughly
 \begin{subequations}
\begin{align}
    \sigma^{\rm NLO}_{AA\to WZX} &\sim 10^{-6}\ub - 10^{-2}\ub,
    \\
    \sigma^{\rm NLO}_{AA\to W\gamma\gamma} &\sim 10^{-7}\ub - 10^{-3}\ub,
    \\
    \sigma^{\rm NLO}_{AA\to VHH, Z\gamma\gamma, V\gamma H} &\sim 10^{-6}\ub - 10^{-4}\ub,
    \\
    \delta\sigma^{\rm NLO}/\sigma^{\rm NLO} & \sim \confirm{\pm10\% - \pm30\%}\ .
\end{align}
\end{subequations}
For all $\sqrt{s_{NN}}$ values, the $W^\pm Z\gamma$ channel has the largest rate, followed by the $W^\pm ZH$ and $W\gamma\gamma$ channels by factors of a few to several.  About a factor of \confirm{10} below stand the remaining channels, which all have comparable rates. As with previous channels involving final-state photons, the hierarchy is  influenced by the  $p_T^\gamma> 150\GeV \sim 2M_V$ cut. Also playing a role at large $\sqrt{s_{NN}}$ is the emergence of collinear/soft $V^*\to V'V''$ splittings, for $V\in\{\gamma,W^\pm,Z\}$, which are logarithmically enhanced~\cite{Ciafaloni:1998xg,Beccaria:1998qe,Ciafaloni:2000df,Melles:2001ye,Bauer:2016kkv,Chen:2016wkt}.

In comparison to the triboson processes in Sec.~\ref{sec:xsec_vvv}, we observe that the $WZX$ and $W\gamma\gamma$ channels have comparable rates to those reported in Fig.~\ref{fig:ionsNLO_XSec_vvvX_vs_Beam}. This is follows from the presence of similar gauge couplings and large QCD corrections. All remaining channels have categorically smaller cross sections than those reported in Fig.~\ref{fig:ionsNLO_XSec_vvvX_vs_Beam} and follow from a mixture of relative coupling suppression, \textit{e.g.,} $(\alpha/\alpha_W) \sim 0.2$, more modest QCD corrections, and, most importantly, strong destructive interference. For example: the ratio of the $WZH$ and $W\gamma H$ cross sections at $\sqrt{s_{NN}}=100\TeV$ greatly exceeds the aforementioned coupling ratio and is approximately
\begin{align}
    \frac{\sigma_{AA\to W^\pm\gamma H}^{\rm NLO}}{\sigma_{AA\to W^\pm Z H}^{\rm NLO}}\Bigg\vert_{\sqrt{s_{NN}}=100\TeV}
    ^{\rm ^{208}Pb-^{208}Pb}\
    \sim\
    \confirm{\mathcal{O}(4\times10^{-2})}\ .
\end{align}

In the second panels, we observe much larger and much smaller QCD $K$-factors at NLO than reported for the $WWV$ and $ZZV$ channels in Sec.~\ref{sec:xsec_vvv}. Here, they roughly span
\begin{subequations}
\begin{align}
    K^{\rm NLO}_{AA\to WZ\gamma,\ W\gamma\gamma} &\sim \confirm{2 - 6}\ ,
    \\
    K^{\rm NLO}_{AA\to Z\gamma\gamma,\ V\gamma H,\ WZH} &\sim \confirm{1.5 - 2}\ ,
    \\
    K^{\rm NLO}_{AA\to VHH} &\sim \confirm{1.3-1.4}\ .
\end{align}
\end{subequations}
The largest corrections occur for $W\gamma\gamma$ and $WZ\gamma$ production, which feature large amplitude zeros\footnote{Strong ``zeros'' are anticipated in all processes of the form $AA\to W+n\gamma+mZ$~\cite{Zhu:1980sz,Brodsky:1982sh,Brown:1982xx}.}~\cite{Baur:1997bn,Bell:2009vh}. The rise in $K^{\rm NLO}$ in these two channels with increasing collision energy, and hence an increasing $g$ density, can be interpreted as the emerging dominance of $(qg)$-scattering, which opens at $\mathcal{O}(\alpha_s)$. NNLO predictions for these channels are not yet available; however, it is suspected that, as in NNLO predictions for diboson production, a modest rate increase will still occur due to the opening of the $(gg)$-scattering channels and NLO corrections to $(qg)$-scattering. This suggests that the scale uncertainties at NLO for triboson processes are not reliable estimates of missing higher-order terms.

We present in the three lower panels (per nucleon) cross section ratios with respect to $pp$ collisions. We observe that most all channels converge to flat values at or just below unity for $\confirm{\sqrt{s_{NN}}\gtrsim 5-10\TeV}$. This further supports our argument that charge- and flavor-summed channels are approximately isopsin symmetric. The only exception consists of the $Z\gamma\gamma$ channel, which as argued for $Z\gamma$ production (Sec.~\ref{sec:xsec_xvv}), is indirectly sensitive to the net nuclear isospin of the average nucleon because the $Z-q-\overline{q}$ and $\gamma-q-\overline{q}$ couplings are flavor-dependent. The cross section ratios for these two processes converge to flat values below unity for $\sqrt{s_{NN}}\gtrsim 60-70\TeV$. This corresponds to the fraction \confirm{$(\sqrt{\hat{s}}/\sqrt{s_{NN}}) \sim (M_Z+2p_T^\gamma)/\sqrt{s_{NN}}\lesssim6\times10^{-3}$}. We refer to Tables~\ref{tab:summary_pb208}-\ref{tab:summary_hx1_bis} (rows 26-33) for a quantitative comparison of predictions for this set of triboson processes at $\sqrt{s_{NN}} = 5$~TeV, 5.52~TeV, 39~TeV, and 100~TeV.

\subsection{Inclusive photon processes}\label{sec:xsec_axx}

\begin{figure}[!t]
\includegraphics[width=.95\textwidth]{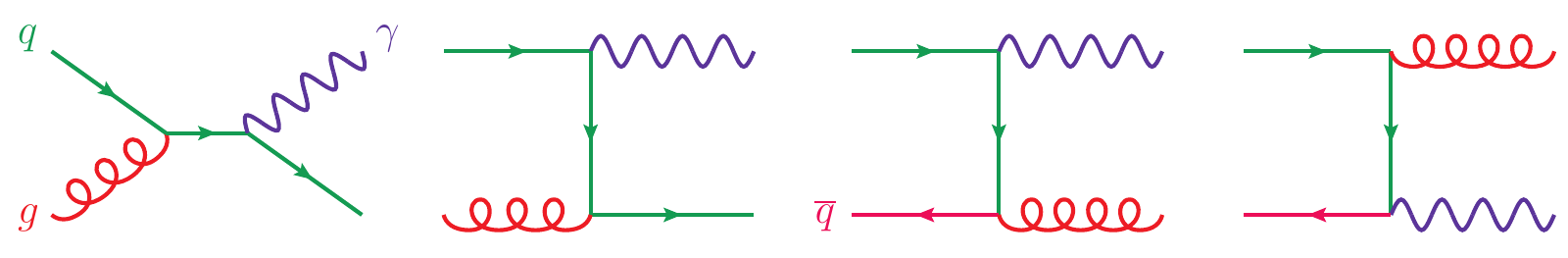}
\caption{Representative Feynman diagrams, at the Born level, depicting inclusive photon production.
}
\label{fig:diagram_MultiBoson_PhotonX}
\end{figure}

Moving beyond massive electroweak bosons, we now consider inclusive photon production processes. Specifically, we focus on prompt photon production, \textit{i.e.}, the production of a single, high-energy, central photon in association with anything, as illustrated diagrammatically in Fig.~\ref{fig:diagram_MultiBoson_PhotonX}, as well as on photon pair production (Fig.~\ref{fig:diagram_MultiBoson_V_VV_VH}) and triple photon production(Fig.~\ref{fig:diagram_MultiBoson_VVV}). At LO, these are described by
\begin{align}
    qg,\ \overline{q}g,\ q\overline{q}\to
    \gamma+X,\
    \text{where}\ X\in\{q,\overline{q},g\},
    \quad\text{and}\quad
    q\overline{q}\ \to\ \gamma\gamma,\
    \gamma\gamma\gamma\ .
\label{eq:proc_parton_axxx}
\end{align}
To regulate infrared divergencies in matrix elements, we impose the phase space cuts of  Eqs.~\eqref{eq:cuts_photon} and \eqref{eq:cuts_photon_iso}. Predictions at NLO in QCD for inclusive photon production processes were first reported in Refs.~\cite{Berger:1983yi,Field:1989uq,Baer:1990ra,Gordon:1994ut,Frixione:1998jh,Binoth:1999qq,Catani:2002ny,Mandal:2014vpa,Mandal:2014vpa}.
Direct/inclusive/prompt photon production
has been  measured by LHC experiments
in Pb-Pb collisions at
$\sqrt{s_{NN}}=2.76\TeV$~\cite{CMS:2012oiv,ATLAS:2015rlt}
and
$\sqrt{s_{NN}}=5.02\TeV$~\cite{CMS:2020oen},
as well as in p-Pb collisions at
$\sqrt{s_{NN}}=5.02\TeV$~\cite{ALICE:2023ode}
and $\sqrt{s_{NN}}=8.16\TeV$~\cite{ATLAS:2019ery}.

\begin{figure}[!t]
\subfigure[]{\includegraphics[width=.47\textwidth]{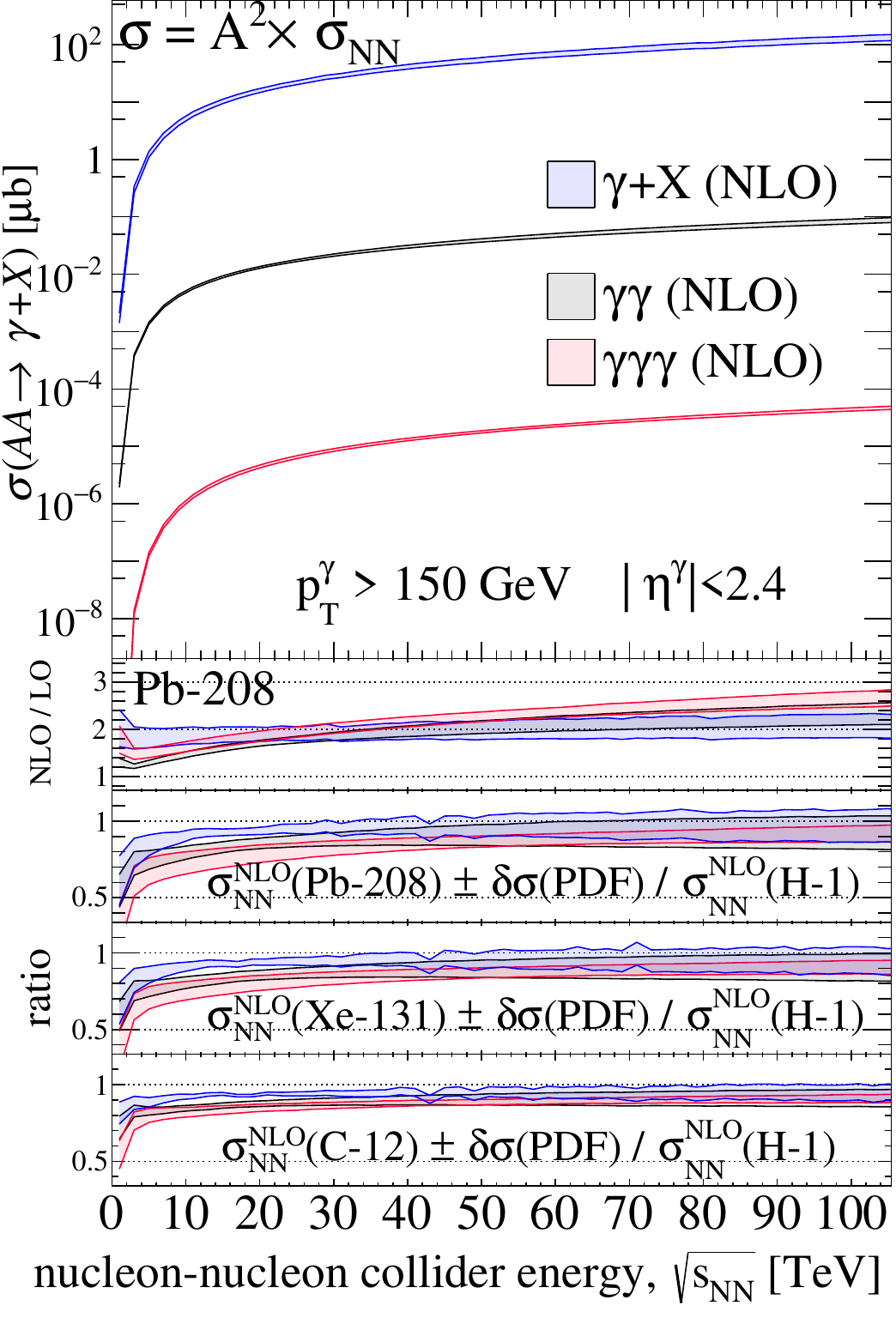}
\label{fig:ionsNLO_XSec_aXXX_vs_Beam}
}
\subfigure[]{\includegraphics[width=.47\textwidth]{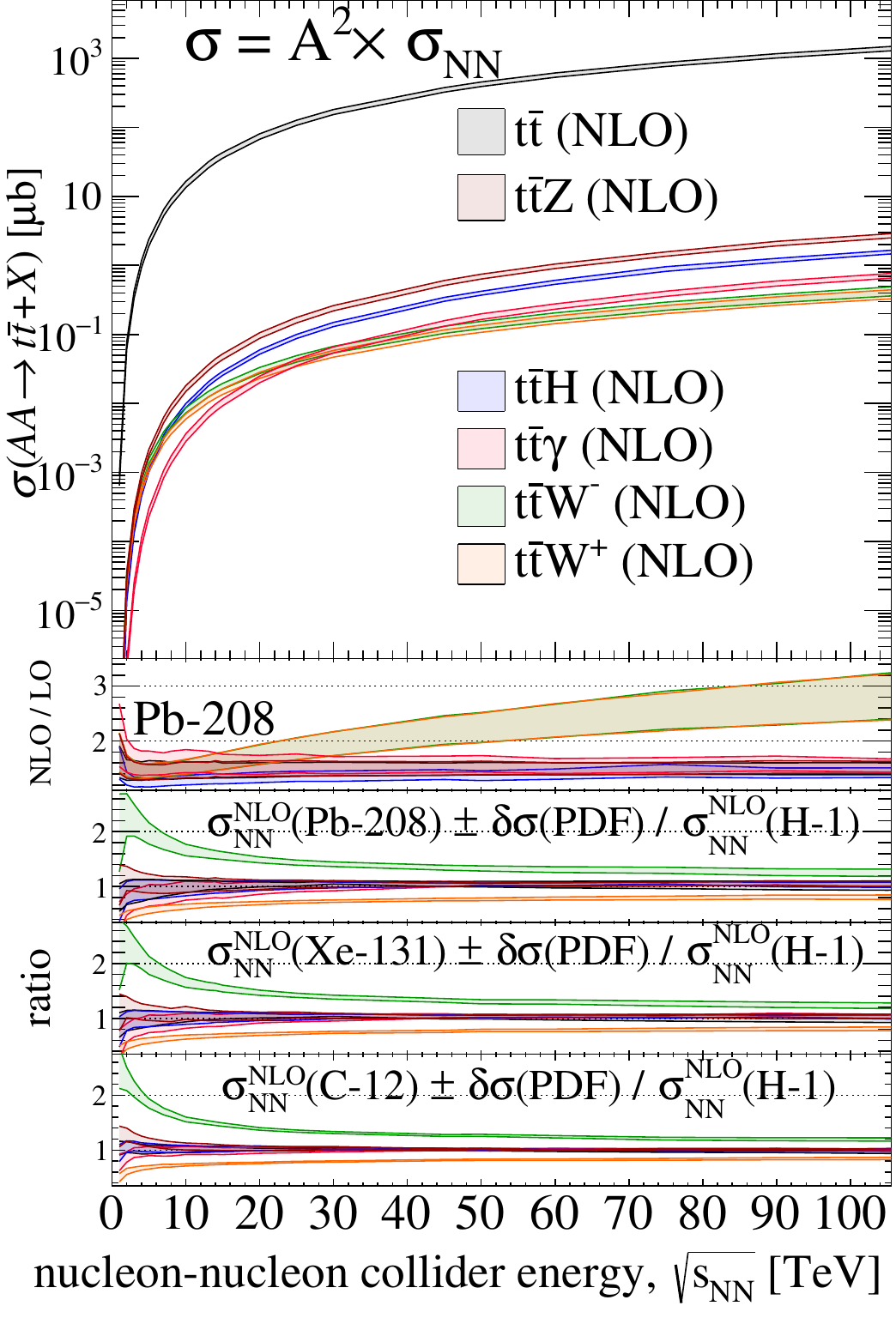}
\label{fig:ionsNLO_XSec_ttXX_vs_Beam}
}
\caption{Same as Fig.~\ref{fig:ionsNLO_XSec_vXXX_vs_Beam} but for (a) $\gamma+X$ (blue), $\gamma\gamma$ (black), and $\gamma\gamma\gamma$ (red) production, as well as for (b) $t\overline{t}$ (black), $t\overline{t}Z$ (purple), $t\overline{t}H$ (blue), $t\overline{t}\gamma$ (pink) $t\overline{t}W^-$ (green), and $t\overline{t}W^+$ (yellow) production, all in descending order.
}
\end{figure}

We show in the top panel of Fig.~\ref{fig:ionsNLO_XSec_aXXX_vs_Beam} the nucleus-level total cross section at NLO in QCD for $\gamma+X$ (blue), $\gamma\gamma$ (black), and $\gamma\gamma\gamma$ (red), shown in descending order, in $^{208}$Pb-$^{208}$Pb collisions and as a function of $\sqrt{s_{NN}}$. The band thickness corresponds to the residual scale uncertainty at this order. In the lower panels we show the same ratios as shown in Fig.~\ref{fig:ionsNLO_XSec_vXXX_vs_Beam}. Focusing on the top panel, cross sections and uncertainties at NLO roughly span
\begin{subequations}
\begin{align}
    \sigma^{\rm NLO}_{AA\to \gamma+X}\ &\sim\ 10^{-3}\ub - 10^{2}\ub,
    \\
    \sigma^{\rm NLO}_{AA\to \gamma\gamma}\ &\sim\ 10^{-5}\ub - 10^{-1}\ub,
    \\
    \sigma^{\rm NLO}_{AA\to \gamma\gamma\gamma}\ &\sim\ 10^{-9}\ub - 4\times10^{-5}\ub,
    \\
    \delta\sigma^{\rm NLO}/\sigma^{\rm NLO}\ & \sim\ \confirm{\pm10\% - \pm20\%}\ .
\end{align}
\end{subequations}
For all $\sqrt{s_{NN}}$, single prompt photon production has the largest rate, followed by pair and triple production. The strong hierarchy across all three channels reflects the relative coupling and phase space suppression for producing one, two, or three photons as well as the role of different parton densities. For example: already at LO, $\gamma+X$ is sensitive to the gluon nPDF though the $(qg)$ luminosity, which is the largest of all partonic configurations across all center-of-mass energies (see Sec.~\ref{sec:lumi}). At NLO, the $gg\to q\overline{q}\gamma$ channel opens, which contributes to large $\mathcal{O}(\alpha_s)$ corrections; for context, $gg\to q\overline{q}\gamma$ contributes to the dijet+photon process. On the other hand, the $\gamma\gamma$ and $\gamma\gamma\gamma$ processes are strictly initiated by $(q\overline{q})$ annihilation at LO and are only sensitive to the gluon distribution at $\mathcal{O}(\alpha_s)$, \textit{i.e.,} at NLO in QCD. The difference in parton luminosities alone accounts for an $\mathcal{O}(40)$ times difference between the prompt photon and multi-photon production rates. As shown by the following ratios, coupling and na\"ive phase space suppression can account for most of the differences at high energies:
\begin{align}
    \left(\frac{\alpha}{\alpha_s}\right)\
    \times\
    \left(\frac{\sum_{q=u}^b e_q^4}{\sum_{q=u}^b e_q^2}\right)
    \times\
    \left(\frac{\mathcal{L}_{q\overline{q}}}{\mathcal{L}_{qg}}
    \right)\
    \times\
    \frac{\sigma_{AA\to \gamma+X}^{\rm NLO}}{\sigma_{AA\to \gamma\gamma}^{\rm NLO}}\Bigg\vert_{\sqrt{s_{NN}}=100\TeV}
    ^{\rm ^{208}Pb-^{208}Pb}&
    \nonumber\\
    \approx\
    \left(0.075\right)\
    \times\
    \left(0.40\right)\
    \times\
    \left(\frac{1}{40}\right)\ \times\
    &\confirm{\left(1480\right)}\
    \approx\ \confirm{1.11}\ ,
    \\
    \left(\frac{\alpha}{4\pi}\right)\
    \times\
    \frac{\sigma_{AA\to \gamma\gamma}^{\rm NLO}}{\sigma_{AA\to \gamma\gamma\gamma}^{\rm NLO}}\Bigg\vert_{\sqrt{s_{NN}}=100\TeV}
    ^{\rm ^{208}Pb-^{208}Pb}\
    \approx\
    \left(6\times10^{-4}\right)\
    \times\
    &\confirm{\left(1900\right)}\
    \approx\
    \confirm{1.14}\ .
\end{align}
The closeness of these ratios to unity underscores how so few terms, namely gauge coupling, gauge charges, and luminosities, drive the global of size of photon processes. Exact differences in matrix elements, including color factors and sizable subleading channels,
account for the remainder.

Focusing on the QCD $K$-factors, which are all sizable for $\sqrt{s_{NN}}=1-100\TeV$, we report:
\begin{subequations}
\begin{align}
    K^{\rm NLO}_{AA\to \gamma+X}\ &\sim\ \confirm{1.75 - 2}\ ,
    \\
    K^{\rm NLO}_{AA\to \gamma\gamma}\ &\sim\ \confirm{1.5 - 2.4}\ ,
    \\
    K^{\rm NLO}_{AA\to \gamma\gamma\gamma}\ &\sim\ \confirm{1.5 - 2.7}\ .
\end{align}
\end{subequations}
Unlike many channels previously discussed, the NLO $K$-factor for prompt photon production remains relatively stable/flat with collider energy. Diphoton and triphoton production, on the other hand, exhibit QCD $K$-factors at NLO that grows with collider energy. We attribute this difference to the opening of new partonic channels in the $\gamma\gamma$ and $\gamma\gamma\gamma$ processes at $\mathcal{O}(\alpha_s)$, and particularly gluon-initiated channels, which introduce a reliance on the $(qg)$  parton luminosity and hence an increased dependence on $\sqrt{s_{NN}}$. As prompt photon production is already sensitive to $(qg)$-scattering at LO, such a dependence on $\sqrt{s_{NN}}$ is already present at LO. Instead, at NLO prompt photon production receives sizable real corrections from dijet+photon process $pp\to jj\gamma$, including the $gg\to q\overline{q}\gamma$ subchannel, as well as virtual corrections, which introduce logarithmic and color-factor enhancements~\cite{Baer:1990ra,Gordon:1994ut,Frixione:1998jh,Catani:2002ny}. We do not separately identify $q^*\to q\gamma$ fragmentation; this has some role in the largeness of QCD corrections found in  Fig.~\ref{fig:ionsNLO_XSec_aXXX_vs_Beam}. Corrections at NNLO in QCD to inclusive single and diphoton production remain large~\cite{Catani:2011qz,Campbell:2016yrh,Campbell:2016lzl,Chen:2019zmr}, again through a mix of new production topologies and virtual corrections. However, agreement with LHC $pp$  data suggest higher order perturbative corrections will not introduce additional large corrections~\cite{ATLAS:2017cvh,CMS:2018dqv,CMS:2019jlq,ATLAS:2019buk}.

The three lower panels show the (per nucleon) cross section ratios with respect to the $pp$ cross section. We report that all three cross section ratios sit significantly below unity for \confirm{$\sqrt{s_{NN}}\lesssim 10-20\TeV$}, and converge to flat values at or just below unity for \confirm{$\sqrt{s_{NN}}\gtrsim 50\TeV$}. (The ratios for the $\gamma+X$ channel appear to converge at lower collision energies, with \confirm{$\sqrt{s_{NN}}\gtrsim 20-30\TeV$}.) Overall, this corresponds to the ratio \confirm{$(\sqrt{\hat{s}}/\sqrt{s_{NN}}) \sim (2p_T^\gamma)/\sqrt{s_{NN}} - (3p_T^\gamma)/\sqrt{s_{NN}} \lesssim 0.01$}. We refer to Tables~\ref{tab:summary_pb208}-\ref{tab:summary_hx1_bis} (rows 34-36) for a quantitative comparison of predictions for single and multi-photon production at $\sqrt{s_{NN}} = 5$~TeV, 5.52~TeV, 39~TeV, and 100~TeV.

Despite \textit{appearing} charge and flavor symmetric, the three photon channels listed in  Eq.~\eqref{eq:proc_parton_axxx} are inherently sensitive to the net electric charge, and hence net nuclear isospin, of nucleons. More specifically, $^{208}$Pb, $^{131}$Xe, and $^{12}$C are neutron-rich, meaning that the \textit{average} nucleon is more neutron-like than proton-like. This implies that the average electric charge of an average nucleon is smaller than a proton. Hence, by electric charge conservation the average nucleons for our three isotopes have a larger number density of partons with negative electric charge than found in a proton. For instance: nucleon-level sum rules for valence $u$- and $d$-quarks can be written as~\cite{Ruiz:2023ozv}:
\begin{align}
 \int_0^A dx\ \left[u_v(x)-d_v(x)\right]\
 =\
 \left[2Z+(A-Z)\right] - \left[Z + 2(A-Z)\right]\
 =\
 2Z-A\ ,
\end{align}
which is negative for our three isotopes but positive for the proton. Assuming $\overline{u}(x)=\overline{d}(x)$, the relationship implies larger $(d_vg)$ and $(d_v\overline{d})$ luminosities and smaller $(u_vg)$ and $(u_v\overline{u})$ luminosities than found in a proton. Since inclusive photon production is proportional to the electric charges of quarks, when  $\gamma+X$, $\gamma\gamma$, and $\gamma\gamma\gamma$ production is driven by valence quarks, \textit{i.e.}, when $(p_T^\gamma/\sqrt{s_{NN}}) \sim \mathcal{O}(1)$, then one expects nucleon-level cross section ratios well below unity. Conversely, when production is driven by sea quarks, \textit{i.e.}, when $(p_T^\gamma/\sqrt{s_{NN}}) \ll \mathcal{O}(1)$, then one expects nucleon-level cross section ratios close unity.
(Since the contribution from $u_v$ and $d_v$ can never be removed, the ratio will always remain below unity.)

\subsection{Top quark processes}\label{sec:xsec_ttx}

\begin{figure}
\includegraphics[width=.95\textwidth]{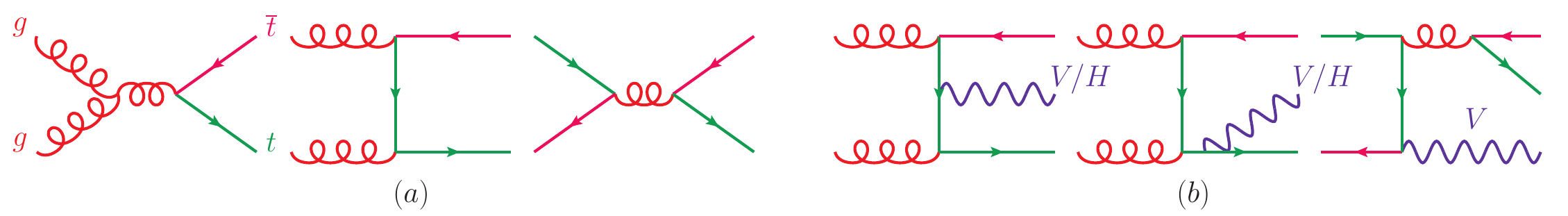}
\caption{Representative Feynman graphs, at the Born level, depicting production of (a) $t\overline{t}$ and (b) $t\overline{t}V$ and $t\overline{t}H$.
}
\label{fig:diagram_MultiBoson_tt_ttV}
\end{figure}

Finally, we consider the pair production of top quarks, together with the associated production of top quark pairs with an additional electroweak boson. Specifically, we focus on those processes described schematically by Fig.~\ref{fig:diagram_MultiBoson_tt_ttV} at LO, and by the partonic configurations
\begin{align}
    q\overline{q},\ q\overline{q'},\ gg\
    &\to\
    t\overline{t},\
    t\overline{t}Z,\
    t\overline{t}H\
    t\overline{t}\gamma,\
    t\overline{t}W^-,\
    t\overline{t}W^+\ .
\label{eq:proc_parton_ttx}
\end{align}
For processes with a final-state photon, we impose the phase space cuts given in Eqs.~\eqref{eq:cuts_photon} and \eqref{eq:cuts_photon_iso}.
Predictions for $t\overline{t}$ and $t\overline{t}X$ at NLO in QCD were first reported in Refs.~\cite{Nason:1987xz,Beenakker:1988bq,Beenakker:2002nc,Dawson:2002tg,Lazopoulos:2008de,Badger:2010mg,Melnikov:2011ta,Campbell:2012dh} and are available for some channels at higher precision~\cite{Czakon:2013goa,Buonocore:2023ljm}.
Predictions for  $t\overline{t}$
production in ion collisions
at NNLO QCD are reported in
Ref.~\cite{dEnterria:2015mgr,dEnterria:2017jyt}.
$t\overline{t}$  production
has been  measured by LHC experiments
in Pb-Pb collisions at
$\sqrt{s_{NN}}=5.02\TeV$~\cite{CMS:2020aem}
as well as in p-Pb collisions at
$\sqrt{s_{NN}}=8.16\TeV$~\cite{CMS:2017hnw,ATLAS:2024qdu}.

In the top panel of Fig.~\ref{fig:ionsNLO_XSec_ttXX_vs_Beam} we show the nucleus-level cross section at NLO in QCD for $t\overline{t}$ (black), $t\overline{t}Z$ (purple), $t\overline{t}H$ (blue), $t\overline{t}\gamma$ (pink) $t\overline{t}W^-$ (green), and $t\overline{t}W^+$ (yellow) production in descending order for symmetric $^{208}$Pb-$^{208}$Pb collisions as a function of $\sqrt{s_{NN}}$. As before, band thickness corresponds to the residual scale uncertainty at NLO in pQCD. In the lower panels we show the same nucleon-level ratios as in Fig.~\ref{fig:ionsNLO_XSec_vXXX_vs_Beam}.
For $\sqrt{s_{NN}}=1-100\TeV$, nuclear cross sections and uncertainties at NLO are approximately
 \begin{subequations}
\begin{align}
    \sigma^{\rm NLO}_{AA\to t\overline{t}} &\sim 10^{-2}\ub - 10^3\ub,
    \\
    \sigma^{\rm NLO}_{AA\to t\overline{t}V} &\sim 10^{-4}\ub - 1\ub,
    \\
    \delta\sigma^{\rm NLO}/\sigma^{\rm NLO} & \sim \confirm{\pm10\% - \pm20\%}\ .
\end{align}
\end{subequations}
For all collision energies, $t\overline{t}$ production is the dominant channel; this is to be expected  as $t\overline{t}$ is only a two-body final state, meaning it is less suppressed by (electroweak) coupling constants and phase-space constraints. Roughly two orders of magnitude below lie the rates for the associated production channels. This difference is consistent with na\"ive power counting and phase space suppression, \textit{e.g.}, $(\alpha_W/4\pi)\sim3\times10^{-3}$. Furthermore, at lower collision energies, there is an increased importance of phase space suppression associated with kinematic thresholds.

For all $\sqrt{s_{NN}}$ under consideration, the $t\overline{t}Z$ and $t\overline{t}H$ channels are respectively the leading and sub-leading associate production channels. The former is roughly \confirm{$1.5-2$ times larger} than the latter, with the difference slowly increasing with collision energy. In this case, the difference in scattering rates is a subtle interplay between kinematic factors  and the electroweak couplings of the $Z$ and $H$ boson to the top quark. More specifically, in the large $\sqrt{s_{NN}}$ limit, $t\overline{t}Z$ and $t\overline{t}H$ production in $AA$ collisions are driven by $(gg)$-scattering, implying that the final-state $Z$ and $H$ bosons are both sourced from top quarks. In this same limit, inclusive $t\overline{t}Z$ production
is enhanced by the emission of (relatively) soft and collinear $Z$ bosons~\cite{Ciafaloni:1998xg,Beccaria:1998qe,Ciafaloni:2000df,Melles:2001ye,Bauer:2016kkv,Chen:2016wkt}. This kinematical enhancement is offset by the $t-\overline{t}-Z$ coupling itself, which is much smaller than the $t-\overline{t}-H$ coupling. As a consequence, we see that at $\sqrt{s_{NN}}=100\TeV$ the $(t\overline{t}Z)$-over-$(t\overline{t}H)$ cross section ratio weighted by (a) the $(t\overline{t}H)$-over-$(t\overline{t}Z)$ coupling ratio and (b) electroweak logarithms for a representative hard scale of $Q=(2m_t+m_Z)$ roughly gives\footnote{This argument is complicated by the presence of relative helicity suppression in the $t\overline{t}H$ process and the $Z$ boson's polarization. For transversely polarized $Z$ bosons $(Z_T)$, the $t-\overline{t}-Z_T$ coupling is helicity conserving, meaning that squared matrix elements contain relative factors of $\vert\mathcal{M}_{t\overline{t}Z_T}\vert^2\sim [(\not\!\!q_t)^2/\hat{s}]\sim\mathcal{O}(1)$, where $q_t$ is the off-shell momentum of an internal top quark and is naturally the scale of the hard process, $\sqrt{\hat{s}}$. The $t-\overline{t}-H$ coupling, on the other hand, is helicity inverting, meaning that squared matrix elements contain relative factors of $\vert\mathcal{M}_{t\overline{t}H}\vert^2\sim (m_t^2/\hat{s})\sim(m_t^2)/(2m_t+m_H)^2\sim\mathcal{O}(10^{-1})$. However, in practice, the $Q^2$ argument in the double log  in Eq.~\eqref{eq:ttz_tth_scaling_ratio} is also much smaller. For longitudinally polarized $Z$ bosons $(Z_0)$, the $t-\overline{t}-Z_0$ coupling is helicity inverting, meaning that the $t\overline{t}Z_0$ rate  should converge to the $t\overline{t}H$ rate according to the Goldstone Equivalence Theorem~\cite{Lee:1977yc,Chanowitz:1985hj}.}
\begin{align}
\label{eq:ttz_tth_scaling_ratio}
    \cfrac{\left(\frac{gm_t}{2M_W}\right)^2}
    {\left(\frac{g}{\cos\theta_W}\right)^2
    \left[(Q_t\sin^2\theta_W)^2+(T_{3L}^t-Q_t\sin^2\theta_W)^2\right]\
    \times\
    \log^2\left(\frac{Q^2}{M_Z^2}\right)
    }\
    &
    \nonumber\\
    \times\
    \frac{\sigma_{AA\to t\overline{t}Z}^{\rm NLO}}{\sigma_{AA\to t\overline{t}H}^{\rm NLO}}\Bigg\vert_{\sqrt{s_{NN}}=100\TeV}
    ^{\rm ^{208}Pb-^{208}Pb}\
    \approx\
    \confirm{\left(0.627\right)}\
    &\times\
    \confirm{\left(1.715\right)}\
    \approx\
    \confirm{1.075}\ .
\end{align}

For lower collision energies with $\sqrt{s_{NN}}\lesssim30\TeV$, the $t\overline{t}W^-$ and $t\overline{t}W^+$ channels are the next largest scattering channels with $t\overline{t}\gamma$ exhibiting the smallest rate of all channels  listed in Eq.~\eqref{eq:proc_parton_ttx}. For larger $\sqrt{s_{NN}}$ values, the situation is reverse: the $t\overline{t}\gamma$ production rate exceeds both $t\overline{t}W^\pm$ cross sections. Over the entire range, $t\overline{t}W^-$ has a slightly larger cross section than $t\overline{t}W^+$. This is due to the fact that $t\overline{t}W$ is driven by $(q\overline{q'})$- and $(qg)$-scattering coupled with the the neutron-rich nature of $^{208}$Pb.

Focusing on $t\overline{t}\gamma$ production, note that the (relatively) dynamical dependence of $t\overline{t}\gamma$ on collision energy indicates a dependence on phase space. (At LO in the electroweak theory, electroweak couplings are scale invariant and therefore are constant as $\sqrt{s_{NN}}$ changes.) We attribute the behavior observed in Fig.~\ref{fig:ionsNLO_XSec_ttXX_vs_Beam} to the transverse momentum requirement on the photon, which scales as $p_T^\gamma \sim 2M_V$. As a consequence, the kinematic threshold for   $t\overline{t}\gamma$ production is about $\min(\sqrt{\hat{s}})\sim (2m_t+p_T^\gamma)\gtrsim500\GeV$. Evidently, the process is significantly more phase-space suppressed than $t\overline{t}W^\pm$ for $\sqrt{s_{NN}}\ll 10-20\TeV$ and relatively unsuppressed for $\sqrt{s_{NN}}\gg 20-30\TeV$.

In the second panels, we observe a variety of QCD $K$-factors at NLO. These span roughly
\begin{subequations}
\begin{align}
    K^{\rm NLO}_{AA\to t\overline{t}, t\overline{t}Z/H/\gamma} &\sim \confirm{1.6 - 1.5}\ ,
    \\
    K^{\rm NLO}_{AA\to t\overline{t}W^\pm} &\sim \confirm{1.4 - 2.8}\ .
\end{align}
\end{subequations}
While corrections at NLO in QCD for $t\overline{t}$ and $t\overline{t}Z/H/\gamma$ production seem modest, corrections beyond this order are still sizable~\cite{Czakon:2013goa,vonBuddenbrock:2020ter,Frederix:2021agh,Buonocore:2023ljm}. This is particularly true for $t\overline{t}W^\pm$ production~\cite{vonBuddenbrock:2020ter,Frederix:2021agh,Buonocore:2023ljm}, which proceeds at LO by $q\overline{q'}$ annihilation and only at NLO in QCD does the $(qg)$-scattering channel open. The large and comparable $K$-factors for $t\overline{t}W^-$ and $t\overline{t}W^+$ essentially reflect the opening of this partonic channel and a gluon distribution that grows with increasing $\sqrt{s_{NN}}$ (for a fixed scale). In summary, QCD corrections at NLO and beyond for inclusive and associated $t\overline{t}$ production in $AA$ collisions are necessary for reliable theoretical predictions. However, further exploration in this direction requires two-loop computations and  is beyond the scope the present work.

In the three lower panels we show (per nucleon) cross section ratios with respect to $pp$ collisions. For $\confirm{\sqrt{s_{NN}}\gtrsim 5-15\TeV}$, we observe that  all charge-symmetric and flavor-symmetric channels, \textit{i.e.,} the $t\overline{t}$, $t\overline{t}Z$, $t\overline{t}H$, and $t\overline{t}\gamma$ channels, converge to a flat value close to unity. This similarity reflects the fact that these channels are driven by $(gg)$-scattering, which exhibits a parton luminosity (see Sec.~\ref{sec:lumi}) that is largely insensitive to the precise $(A,Z)$ configuration, \textit{i.e.}, is approximately isospin symmetric. On the other hand, for $t\overline{t}W^-$ and $t\overline{t}W^+$ production, we find large pulls above and below unity respectively. As in inclusive $W^-$ and $W^+$ production in Sec.~\ref{sec:xsec_xxv}, these pulls are due to the processes $q\overline{q'}\to t\overline{t}W^-$ and $q\overline{q'}\to t\overline{t}W^+$ being charge-/flavor-asymmetric partonic scattering processes, that are thus sensitive to the $(A,Z)$ configuration of a nucleus. As for other processes, the cross section ratios for these two processes converge to flat values. Interestingly though, since the PDF uncertainties for $t\overline{t}W^-$ and $t\overline{t}W^+$ production are \confirm{sufficiently small and sufficiently away from unity}, it is possible to estimate the limits of the ratios. For asymptotically large collision energies, we find at the nucleon level
\begin{subequations}
\begin{align}
 \mathcal{R}_{NN}(t\overline{t}W^-)\ =\
 \lim_{\sqrt{s_{NN}}\to\infty}\
 \left(
 \frac{\sigma^{\rm NLO}_{NN\to t\overline{t}W^-}}
    {\sigma^{\rm NLO}_{pp\to t\overline{t}W^-}}
    \right)\
 &\ \approx\ 1.2-1.3\ ,
 \\
 \mathcal{R}_{NN}(t\overline{t}W^+)\ =\
 \lim_{\sqrt{s_{NN}}\to\infty}\
 \left(
 \frac{\sigma^{\rm NLO}_{NN\to t\overline{t}W^+}}
    {\sigma^{\rm NLO}_{pp\to t\overline{t}W^+}}
    \right)\
 &\ \approx\ 0.8-0.9\ ,
\end{align}
\end{subequations}
for our representative ion configurations. The processes reach these plateaus when $\sqrt{s_{NN}}\gtrsim 70-80\TeV$. This corresponds to the fraction \confirm{$(\sqrt{\hat{s}}/\sqrt{s_{NN}}) \sim (2m_t+M_W)/\sqrt{s_{NN}}\lesssim5-6\times10^{-3}$}. We refer to Tables~\ref{tab:summary_pb208}-\ref{tab:summary_hx1} (rows 37-42) for a quantitative comparison of predictions for top quark production at $\sqrt{s_{NN}} = 5$~TeV and 100~TeV.

\subsection{Prospects for observing multiboson processes in ion-ion collisions}\label{sec:xsec_yields}

Throughout Secs.~\ref{sec:xsec_xxv}-\ref{sec:xsec_ttx}
we presented numerical predictions for total or fiducial
cross sections for various resonant and
high-$p_T$ processes in the SM,
as a function of $\sqrt{s_{NN}}$ for various ion-ion
collision configurations.
While not exhaustive\footnote{Higgs production via gluon fusion, for example,
can be found elsewhere~\cite{Berger:2018mtg,dEnterria:2019cps}.},
our catalog is nonetheless representative for
single-boson, diboson, triple-boson,
and associated-single-boson processes.
We also showed how the scattering rates for these
processes can be estimated analytically.
Importantly,
na\"ive power counting
is insufficient to estimate all scattering rates
due to the emergence of electroweak logarithms
and the opening of numerically important partonic
channels at NLO in QCD.
These corrections become increasingly important
as the collider collision energies increase.
To our knowledge, our work
represents the first reliable catalog
of scattering rates for so many processes.

Given the integrated luminosity benchmarks for the
HL-LHC $[\mathcal{O}(1\ {\rm nb}^{-1})]$ and
the FCC $[\mathcal{O}(33\ {\rm nb}^{-1})]$,
we certainly do not expect all processes to be observable,
even with $A^2$ enhancements to hadronic cross sections.
However, given the various decay modes of electroweak bosons,
and different detector acceptance rates of different
particle species,
it is not obvious which specific channels
are observable without a more detailed study.
Such an expansive task is outside the scope
of our work.
Nevertheless, we give some guidance in this direction.

To carry out such an exercise, 
we consider the following $2\to 3$ and $2\to4$ processes  at NLO in QCD, but without matching fixed-order predictions with parton showering:
\begin{align}
\text{Proc.\ I}:\
\mathcal{A}\mathcal{A}\to e^\pm\mu^\mp\nu\overline{\nu}+X,\
\quad
\text{Proc.\ II}:\
\mathcal{A}\mathcal{A}\to \ell\nu_\ell \gamma+X,\
\quad
\text{Proc.\ III}:\
\mathcal{A}\mathcal{A}\to \ell\nu_\ell \gamma\gamma+X.
\end{align}
For $\ell\in\{e,\mu\}$, we consider all interfering diagrams, including non-resonant diagrams as well as the resonant contributions with intermediate $W^+W^-$ (Proc.~I),
$W^\pm\gamma$ (Proc.~II), and $W^\pm\gamma\gamma$ (Proc.~III) systems. We impose the following 
analysis-style kinematical cuts, typical of
LHC searches and measurements of
multiboson systems:
\begin{align}
 \vert\eta^{\ell,\gamma}\vert < 2.4,\quad
 p_T^\ell > 15\GeV,\quad
 p_T^\gamma > 150\GeV,\quad
 \Delta R(\ell,\ell')>0.4,\quad
 m_{\ell,\ell'} > 50\GeV ,
 \label{eq:cuts_lepton_analysis}
\end{align}
in addition to the photon isolation criteria listed
in Eq.~\eqref{eq:cuts_photon_iso}.
Matrix elements are generated and evaluated at NLO
(and LO) in QCD with the aforementioned phase space cuts
using {\mgamc}, as described in Sec.~\ref{sec:setup}.
As a benchmark, we consider $^{208}$Pb-$^{208}$Pb
collisions at $\sqrt{s}=5.52\TeV$ (HL-LHC)
and $\sqrt{s}=39\TeV$ (FCC),
assuming integrated luminosities of
$\mathcal{L}=1$ and $33$ nb$^{-1}$.
We report our results in Table~\ref{tab:events_yields_pb208}.

In row 2 of Table~\ref{tab:events_yields_pb208},
we report the generator-level, nucleon-level
cross section at NLO in QCD
$(\sigma^{\rm NLO}_{NN})$
after imposing the selection of Eqs.~\eqref{eq:cuts_lepton_analysis}
and~\eqref{eq:cuts_photon_iso} 
for Proc. I ($W^+W^-$/columns 2-3), Proc. II ($W^\pm\gamma$/columns 4-5), and Proc. III ($W^\pm\gamma\gamma$/columns 6-7).
At the nucleon level, \textit{i.e.}, without $A^2 = (208)^2\sim43\times10^3$ enhancements,
cross sections span, $\sigma_{NN}^{\rm NLO}\sim 10^{-12}\ub - 10^{-6}\ub$.
Some of these rates are indeed small.
However, the perceived smallness of all the rates
is partially due to the conventional choice
to report ion scatting rates in terms of$\ub$.
The small $W\gamma$ and $W\gamma\gamma$ rates also
reflect our choice in stringent phase space cuts.
For instance: the $p_T^\gamma$ threshold can be reduced by $2$ (so $p_T^\gamma > 75\GeV$),
and increase the $W\gamma(\gamma)$ rate by $2(4)$.

In rows 3 and 4, we show respectively
the residual scale uncertainty $(\delta_{\rm scale}$)
and PDF uncertainty $(\delta_{\rm PDF})$ on the
generator-level cross section.
Overall, scale uncertainties range from \confirm{$\delta_{\rm scale}\sim 2/3\%-15\%$},
while PDF uncertainties span \confirm{$\delta_{\rm PDF}\sim 4\%-11\%$}.
In row 5, we give the ratio
$K^{\rm NLO}=\sigma^{\rm NLO}_{NN}/\sigma^{\rm LO}_{NN}$, which spans $K^{\rm NLO} \sim 1.4-8.0$
and highlights the importance of formally subleading channels
at $\mathcal{O}(\alpha_s)$, as discussed in Sec.~\ref{sec:xsec_xva}.
In rows 6-7, we show the estimated event yields
for $\mathcal{L}=1$ nb$^{-1}$ (row 6) and
$\mathcal{L}=33$ nb$^{-1}$ (row 7) for each of the processes
and beam configurations. Given the kinematical
requirements of Eq.~\eqref{eq:cuts_lepton_analysis},
we assume $\varepsilon=100\%$ particle reconstruction
efficiency, which is not an unreasonable choice given the lower high-$p_T$ trigger demands for ion collisions and the real performance of ATLAS, CMS, and ALICE~\cite{CMS:2012oiv,ATLAS:2015rlt,CMS:2020oen,ALICE:2023ode,ATLAS:2019ery,CMS:2020aem,CMS:2017hnw,ATLAS:2024qdu}.
For instance: we could reduce lepton $p_T$ thresholds by 2-3 GeV, and impose instead  a $\varepsilon\gtrsim90\%$  reconstruction efficiency, resulting in comparable yields.

The takeaway of this exercise is the following:
With $\mathcal{L}=1-10$ nb$^{-1}$ of data,
one can expect enough events
to observe several of the possible diboson signal categories
at the HL-LHC, and therefore probe nuclei at
medium-to-large momentum fractions in the immediate future.
(We reiterate that halving the $p_T^\gamma$ requirement
can increase production rates twofold/fourfold for $W\gamma/W\gamma\gamma$.)
At the FCC and with $\mathcal{L}=10-30$ nb$^{-1}$ of data,
event yields allow one to establish a precision
program for diboson physics that probes
medium momentum fractions.
With more relaxed selection criteria,
or use of asymmetric $p-\mathcal{A}$ collisions,
which permit higher integrated luminosities
to be collected ~\cite{CMS:2017hnw,ATLAS:2024qdu},
it may be possible to carry out a triboson
program with ion collisions at the FCC.
Again, this is not a comprehensive collider analysis.
However, our predictions do provide guidance on how one can proceed further.

\begin{table*}[!t]\renewcommand{\tabcolsep}{4pt}\renewcommand{\arraystretch}{1.18}
\resizebox{\textwidth}{!}{
\begin{tabular}{ c | c | c | c | c | c | c }
\hline
 \multicolumn{7}{c}{$^{208}$Pb-$^{208}$Pb Collisions}
 \\
 \hline
  \multicolumn{1}{c}{}
& \multicolumn{2}{c}{$\mathcal{A}\mathcal{A}\to e^\pm\mu^\mp\nu\overline{\nu}+X$}
& \multicolumn{2}{c}{$\mathcal{A}\mathcal{A}\to \ell\nu_\ell \gamma+X$}
& \multicolumn{2}{c}{$\mathcal{A}\mathcal{A}\to \ell\nu_\ell \gamma\gamma+X$}
\\
\hline
$\sqrt{s_{NN}}$
& $5.52\TeV$ & $39\TeV$
& $5.52\TeV$ & $39\TeV$
& $5.52\TeV$ & $39\TeV$
\\
\makecell[c]{$\sigma^{\rm NLO}_{NN}$ [$\mu$b]
\\
after cuts [Eq.~\eqref{eq:cuts_lepton_analysis}]}
& $381\times10^{-9}$
& $2.95\times10^{-6}$
& $38.4\times10^{-9}$
& $1.18\times10^{-6}$
& $26.6\times10^{-12}$
& $1.44\times10^{-9}$
\\
$\pm\ \delta_{\rm scale}$
& $^{+3\%}_{-2\%}$
& $^{+7\%}_{-9\%}$
& $^{+15\%}_{-12\%}$
& $^{+9\%}_{-9\%}$
& $^{+12\%}_{-9\%}$
& $^{+8\%}_{-7\%}$
\\
$\pm\ \delta_{\rm PDF}$
& $\pm4\%$ & $\pm11\%$
& $\pm9\%$ & $\pm5\%$
& $\pm11\%$ & $\pm4\%$
\\
$K^{\rm NLO}\ =\ \sigma^{\rm NLO}_{NN}\ /\ \sigma^{\rm LO}_{NN}$
& $1.35$ & $1.70$ & $3.35$ & $8.03$ & $2.04$ & $4.00$
\\
Events / 1 nb$^{-1}$
& $16$
& $130$
& $1.7$
& $51$
& $1.1\times10^{-3}$
& $62\times10^{-3}$
\\
Events / 33 nb$^{-1}$
& $540$
& $4.2\times10^3$
& $55$
& $1.7\times10^3$
& $38\times10^{-3}$
& $2.1$
\\
\hline
\end{tabular}
}
\caption{For the processes
$\mathcal{A}\mathcal{A}\to e^\pm\mu^\mp\nu\overline{\nu}+X$
(columns 2-3),
$\mathcal{A}\mathcal{A}\to \ell\nu_\ell \gamma+X$
(columns 4-5),
and
$\mathcal{A}\mathcal{A}\to \ell\nu_\ell \gamma\gamma+X$
(columns 6-7)
in symmetric $^{208}$Pb-$^{208}$Pb collisions
at fixed nucleon-nucleon collision energies
$\sqrt{s_{NN}}$ (row 1),
the generator-level, nucleon-level
cross section at NLO in QCD
$\sigma^{\rm NLO}_{NN}$ [$\mu$b] (row 2)
after imposing kinematical cuts in Eq.~\eqref{eq:cuts_lepton_analysis} and
photon isolation of Eq.~\eqref{eq:cuts_photon_iso},
along with
the residual scale uncertainty $\delta_{\rm scale}$ [\%]
(row 3),
PDF uncertainty $\delta_{\rm PDF}$ [\%]
(row 4),
NLO in QCD $K$-factor (row 5),
and estimated event yields
for $\mathcal{L}=1$ nb$^{-1}$ (row 6) and
$\mathcal{L}=33$ nb$^{-1}$ (row 7). \label{tab:events_yields_pb208}}
\end{table*}

\subsection{Summary of cross sections}\label{sec:xsec_summary}

We summarize our catalog of $^{208}$Pb-$^{208}$Pb cross sections in Tables~\ref{tab:summary_pb208} and \ref{tab:summary_pb208_bis}. For each process that we considered (column~1), we give the per nucleon cross sections at NLO in pQCD [$\mu$b] for a representative collision energy of $\sqrt{s_{NN}}=5\TeV$ (column 3, Table~\ref{tab:summary_pb208}), the associated scale [$\%$] and PDF uncertainty [$\%$] (columns 4-5), and the QCD $K$-factor at NLO (column 6). We also give the analogous information for a representative collision energy of $\sqrt{s_{NN}}=5.52\TeV$ (column 3-6, Table~\ref{tab:summary_pb208_bis}), $39\TeV$ (column 7-10, Table~\ref{tab:summary_pb208_bis}), and $100\TeV$ (column 7-10, Table~\ref{tab:summary_pb208}). For reproducibility, we provide the {\mgamc} syntax used generate the matrix elements for each computation (column 2). For $^{131}$Xe, $^{12}$C, and $^{1}$H, we show the same information respectively in Tables~\ref{tab:summary_xe131} and \ref{tab:summary_xe131_bis},  \ref{tab:summary_cx12} and \ref{tab:summary_cx12_bis}, and \ref{tab:summary_hx1} and \ref{tab:summary_hx1_bis}.

\clearpage

\begin{table*}[!t]\renewcommand{\tabcolsep}{4pt}\renewcommand{\arraystretch}{1.18}
\resizebox{\textwidth}{!}{
\begin{tabular}{ l l ccc ccc }
\hline
\multicolumn{8}{c}{$^{208}$Pb-$^{208}$Pb Collisions}\\
\hline
\multicolumn{1}{c}{\multirow{2}{*}{Process}} & 
\multicolumn{1}{c}{\multirow{2}{*}{Syntax}} & 
\multicolumn{3}{c}{$\sqrt{s_{NN}}=5\TeV$} & 
\multicolumn{3}{c}{$\sqrt{s_{NN}}=100\TeV$}\\
& & $\sigma^\mathrm{NLO}$~[$\mu$b] & $\delta_\mathrm{scale}\quad \delta_\mathrm{PDF}$ & $K^\mathrm{NLO}$ & $\sigma^\mathrm{NLO}$~[$\mu$b] & $\delta_\mathrm{scale}\quad\delta_\mathrm{PDF}$ & $K^\mathrm{NLO}$\\
\hline
$N N \to W^\pm$ & \texttt{ p p > ww QED=1 QCD=0 [QCD]} & $5.714 \times 10^{-2}$ & $^{+4\%}_{-6\%}$ \quad $^{+8\%}_{-8\%}$ & 1.24 & $9.986 \times 10^{-1}$ & $^{+10\%}_{-19\%}$ \quad $^{+18\%}_{-18\%}$ & 1.45 \\
$N N \to W^-$ & \texttt{ p p > w- QED=1 QCD=0 [QCD]} & $3.022 \times 10^{-2}$ & $^{+4\%}_{-6\%}$ \quad $^{+8\%}_{-8\%}$ & 1.24 & $5.087 \times 10^{-1}$ & $^{+10\%}_{-19\%}$ \quad $^{+18\%}_{-18\%}$ & 1.44 \\
$N N \to W^+$ & \texttt{ p p > w+ QED=1 QCD=0 [QCD]} & $2.706 \times 10^{-2}$ & $^{+4\%}_{-6\%}$ \quad $^{+8\%}_{-8\%}$ & 1.24 & $4.858 \times 10^{-1}$ & $^{+10\%}_{-19\%}$ \quad $^{+18\%}_{-18\%}$ & 1.44 \\
$N N \to Z$ & \texttt{ p p > z QED=1 QCD=0 [QCD]} & $1.716 \times 10^{-2}$ & $^{+4\%}_{-5\%}$ \quad $^{+7\%}_{-7\%}$ & 1.23 & $3.147 \times 10^{-1}$ & $^{+10\%}_{-18\%}$ \quad $^{+18\%}_{-18\%}$ & 1.42 \\
\hline
$N N \to W^+ W^-$ & \texttt{ p p > w+ w- QED=2 QCD=0 [QCD]} & $2.529 \times 10^{-5}$ & $^{+3\%}_{-2\%}$ \quad $^{+3\%}_{-3\%}$ & 1.33 & $1.005 \times 10^{-3}$ & $^{+9\%}_{-12\%}$ \quad $^{+13\%}_{-13\%}$ & 1.70 \\
\hline
\end{tabular}
}
\caption{Per nucleon NLO in QCD cross sections (in $\mu$b) for various processes, for $^{208}$Pb collisions, and the representative collision energies $\sqrt{s_{NN}}=5\TeV$ and $100\TeV$, the corresponding {\mgamc} simulation syntax (with \texttt{define ww = w+ w-}), the associated scale and PDF uncertainties [$\%$], and QCD $K$-factor. \label{tab:summary_pb208}}
\end{table*}

\begin{table*}[!t]\renewcommand{\tabcolsep}{4pt}\renewcommand{\arraystretch}{1.18}
\resizebox{\textwidth}{!}{
\begin{tabular}{ l l ccc ccc }
\hline
\multicolumn{8}{c}{$^{208}$Pb-$^{208}$Pb Collisions}\\
\hline
\multirow{1}{*}{Process} & 
\multirow{1}{*}{Syntax} & 
\multicolumn{3}{c}{$\sqrt{s_{NN}}=5.52\TeV$} & 
\multicolumn{3}{c}{$\sqrt{s_{NN}}=39\TeV$}\\
& & $\sigma^\mathrm{NLO}$~[$\mu$b] & $\delta_\mathrm{scale}\quad \delta_\mathrm{PDF}$ & $K^\mathrm{NLO}$ & $\sigma^\mathrm{NLO}$~[$\mu$b] & $\delta_\mathrm{scale}\quad\delta_\mathrm{PDF}$ & $K^\mathrm{NLO}$\\
\hline
$N N \to W^\pm$ & \texttt{ p p > ww QED=1 QCD=0 [QCD]} & $6.390 \times 10^{-2}$ & $^{+4\%}_{-7\%}$ \quad $^{+8\%}_{-8\%}$ & 1.24 & $4.346 \times 10^{-1}$ & $^{+7\%}_{-15\%}$ \quad $^{+16\%}_{-16\%}$ & 1.36 \\
$N N \to W^-$ & \texttt{ p p > w- QED=1 QCD=0 [QCD]} & $3.348 \times 10^{-2}$ & $^{+4\%}_{-7\%}$ \quad $^{+8\%}_{-8\%}$ & 1.24 & $2.243 \times 10^{-1}$ & $^{+8\%}_{-15\%}$ \quad $^{+15\%}_{-15\%}$ & 1.37 \\
$N N \to W^+$ & \texttt{ p p > w+ QED=1 QCD=0 [QCD]} & $3.015 \times 10^{-2}$ & $^{+4\%}_{-7\%}$ \quad $^{+8\%}_{-8\%}$ & 1.24 & $2.113 \times 10^{-1}$ & $^{+8\%}_{-15\%}$ \quad $^{+16\%}_{-16\%}$ & 1.36 \\
$N N \to Z$ & \texttt{ p p > z QED=1 QCD=0 [QCD]} & $1.906 \times 10^{-2}$ & $^{+4\%}_{-6\%}$ \quad $^{+8\%}_{-8\%}$ & 1.22 & $1.351 \times 10^{-1}$ & $^{+6\%}_{-14\%}$ \quad $^{+15\%}_{-15\%}$ & 1.34 \\
\hline
$N N \to W^+ W^-$ & \texttt{ p p > w+ w- QED=2 QCD=0 [QCD]} & $2.945 \times 10^{-5}$ & $^{+3\%}_{-2\%}$ \quad $^{+3\%}_{-3\%}$ & 1.33 & $3.624 \times 10^{-4}$ & $^{+6\%}_{-8\%}$ \quad $^{+10\%}_{-10\%}$ & 1.56 \\
\hline
\end{tabular}
}
\caption{Same as Table~\ref{tab:summary_pb208} for the collision energies $\sqrt{s_{NN}}=5.52\TeV$ and $39\TeV$.\label{tab:summary_pb208_bis}}
\end{table*}

\clearpage
\begin{table*}[!t]\renewcommand{\tabcolsep}{4pt}\renewcommand{\arraystretch}{1.18}
\resizebox{\textwidth}{!}{
\begin{tabular}{ l l ccc  ccc }
\hline
\multicolumn{8}{c}{$^{131}$Xe-$^{131}$Xe Collisions}\\
\hline
\multicolumn{1}{c}{\multirow{2}{*}{Process}} & 
\multicolumn{1}{c}{\multirow{2}{*}{Syntax}} & 
\multicolumn{3}{c}{$\sqrt{s_{NN}}=5\TeV$} & 
\multicolumn{3}{c}{$\sqrt{s_{NN}}=100\TeV$}\\
& & $\sigma^\mathrm{NLO}$~[$\mu$b] & $\delta_\mathrm{scale}\quad \delta_\mathrm{PDF}$ & $K^\mathrm{NLO}$ & $\sigma^\mathrm{NLO}$~[$\mu$b] & $\delta_\mathrm{scale}\quad\delta_\mathrm{PDF}$ & $K^\mathrm{NLO}$\\
\hline
$N N \to W^\pm$ & \texttt{ p p > ww QED=1 QCD=0 [QCD]} & $5.651 \times 10^{-2}$ & $^{+4\%}_{-7\%}$ \quad $^{+7\%}_{-7\%}$ & 1.24 & $9.765 \times 10^{-1}$ & $^{+10\%}_{-19\%}$ \quad $^{+16\%}_{-16\%}$ & 1.43 \\
$N N \to W^-$ & \texttt{ p p > w- QED=1 QCD=0 [QCD]} & $2.957 \times 10^{-2}$ & $^{+4\%}_{-6\%}$ \quad $^{+7\%}_{-7\%}$ & 1.24 & $5.004 \times 10^{-1}$ & $^{+10\%}_{-19\%}$ \quad $^{+16\%}_{-16\%}$ & 1.45 \\
$N N \to W^+$ & \texttt{ p p > w+ QED=1 QCD=0 [QCD]} & $2.699 \times 10^{-2}$ & $^{+4\%}_{-6\%}$ \quad $^{+7\%}_{-7\%}$ & 1.24 & $4.814 \times 10^{-1}$ & $^{+10\%}_{-19\%}$ \quad $^{+17\%}_{-17\%}$ & 1.44 \\
$N N \to Z$ & \texttt{ p p > z QED=1 QCD=0 [QCD]} & $1.700 \times 10^{-2}$ & $^{+4\%}_{-6\%}$ \quad $^{+6\%}_{-6\%}$ & 1.23 & $3.078 \times 10^{-1}$ & $^{+9\%}_{-18\%}$ \quad $^{+17\%}_{-17\%}$ & 1.40 \\
\hline
$N N \to W^+ W^-$ & \texttt{ p p > w+ w- QED=2 QCD=0 [QCD]} & $2.529 \times 10^{-5}$ & $^{+3\%}_{-2\%}$ \quad $^{+3\%}_{-3\%}$ & 1.33 & $9.872 \times 10^{-4}$ & $^{+9\%}_{-12\%}$ \quad $^{+11\%}_{-11\%}$ & 1.69 \\
\hline
\end{tabular}
}
\caption{Per nucleon NLO in QCD cross sections (in $\mu$b) for various processes, for $^{131}$Xe collisions, and the representative collision energies $\sqrt{s_{NN}}=5\TeV$ and $100\TeV$, the corresponding {\mgamc} simulation syntax (with \texttt{define ww = w+ w-}), the associated scale and PDF uncertainties [$\%$], and QCD $K$-factor. \label{tab:summary_xe131}}
\end{table*}

\begin{table*}[!t]\renewcommand{\tabcolsep}{4pt}\renewcommand{\arraystretch}{1.18}
\resizebox{\textwidth}{!}{
\begin{tabular}{ l l ccc ccc }
\hline
\multicolumn{8}{c}{$^{131}$Xe-$^{131}$Xe Collisions}\\
\hline
\multirow{1}{*}{Process} & 
\multirow{1}{*}{Syntax} & 
\multicolumn{3}{c}{$\sqrt{s_{NN}}=5.52\TeV$} & 
\multicolumn{3}{c}{$\sqrt{s_{NN}}=39\TeV$}\\
& & $\sigma^\mathrm{NLO}$~[$\mu$b] & $\delta_\mathrm{scale}\quad \delta_\mathrm{PDF}$ & $K^\mathrm{NLO}$ & $\sigma^\mathrm{NLO}$~[$\mu$b] & $\delta_\mathrm{scale}\quad\delta_\mathrm{PDF}$ & $K^\mathrm{NLO}$\\
\hline
$N N \to W^\pm$ & \texttt{ p p > ww QED=1 QCD=0 [QCD]} & $6.295 \times 10^{-2}$ & $^{+4\%}_{-7\%}$ \quad $^{+7\%}_{-7\%}$ & 1.25 & $4.292 \times 10^{-1}$ & $^{+8\%}_{-15\%}$ \quad $^{+14\%}_{-14\%}$ & 1.37 \\
$N N \to W^-$ & \texttt{ p p > w- QED=1 QCD=0 [QCD]} & $3.294 \times 10^{-2}$ & $^{+4\%}_{-7\%}$ \quad $^{+7\%}_{-7\%}$ & 1.25 & $2.212 \times 10^{-1}$ & $^{+8\%}_{-15\%}$ \quad $^{+14\%}_{-14\%}$ & 1.38 \\
$N N \to W^+$ & \texttt{ p p > w+ QED=1 QCD=0 [QCD]} & $3.012 \times 10^{-2}$ & $^{+4\%}_{-7\%}$ \quad $^{+7\%}_{-7\%}$ & 1.25 & $2.103 \times 10^{-1}$ & $^{+8\%}_{-15\%}$ \quad $^{+14\%}_{-14\%}$ & 1.38 \\
$N N \to Z$ & \texttt{ p p > z QED=1 QCD=0 [QCD]} & $1.905 \times 10^{-2}$ & $^{+4\%}_{-6\%}$ \quad $^{+6\%}_{-6\%}$ & 1.24 & $1.331 \times 10^{-1}$ & $^{+7\%}_{-14\%}$ \quad $^{+14\%}_{-14\%}$ & 1.34 \\
\hline
$N N \to W^+ W^-$ & \texttt{ p p > w+ w- QED=2 QCD=0 [QCD]} & $2.948 \times 10^{-5}$ & $^{+3\%}_{-2\%}$ \quad $^{+3\%}_{-3\%}$ & 1.34 & $3.586 \times 10^{-4}$ & $^{+6\%}_{-8\%}$ \quad $^{+8\%}_{-8\%}$ & 1.57 \\
\hline
\end{tabular}
}
\caption{Same as Table~\ref{tab:summary_xe131} for the collision energies $\sqrt{s_{NN}}=5.52\TeV$ and $39\TeV$.\label{tab:summary_xe131_bis}}
\end{table*}

\clearpage
\begin{table*}[!t]\renewcommand{\tabcolsep}{4pt}\renewcommand{\arraystretch}{1.18}
\resizebox{\textwidth}{!}{
\begin{tabular}{ l l ccc ccc }
\hline
\multicolumn{8}{c}{$^{12}$C-$^{12}$C Collisions}\\
\hline
\multicolumn{1}{c}{\multirow{2}{*}{Process}} & 
\multicolumn{1}{c}{\multirow{2}{*}{Syntax}} & 
\multicolumn{3}{c}{$\sqrt{s_{NN}}=5\TeV$} & 
\multicolumn{3}{c}{$\sqrt{s_{NN}}=100\TeV$}\\
& & $\sigma^\mathrm{NLO}$~[$\mu$b] & $\delta_\mathrm{scale}\quad \delta_\mathrm{PDF}$ & $K^\mathrm{NLO}$ & $\sigma^\mathrm{NLO}$~[$\mu$b] & $\delta_\mathrm{scale}\quad\delta_\mathrm{PDF}$ & $K^\mathrm{NLO}$\\
\hline
$N N \to W^\pm$ & \texttt{ p p > ww QED=1 QCD=0 [QCD]} & $5.861 \times 10^{-2}$ & $^{+4\%}_{-6\%}$ \quad $^{+5\%}_{-5\%}$ & 1.24 & $1.063 \times 10^{+0}$ & $^{+10\%}_{-18\%}$ \quad $^{+11\%}_{-11\%}$ & 1.42 \\
$N N \to W^-$ & \texttt{ p p > w- QED=1 QCD=0 [QCD]} & $2.929 \times 10^{-2}$ & $^{+4\%}_{-6\%}$ \quad $^{+5\%}_{-5\%}$ & 1.25 & $5.283 \times 10^{-1}$ & $^{+10\%}_{-18\%}$ \quad $^{+11\%}_{-11\%}$ & 1.41 \\
$N N \to W^+$ & \texttt{ p p > w+ QED=1 QCD=0 [QCD]} & $2.922 \times 10^{-2}$ & $^{+4\%}_{-6\%}$ \quad $^{+5\%}_{-5\%}$ & 1.24 & $5.289 \times 10^{-1}$ & $^{+10\%}_{-18\%}$ \quad $^{+11\%}_{-11\%}$ & 1.41 \\
$N N \to Z$ & \texttt{ p p > z QED=1 QCD=0 [QCD]} & $1.746 \times 10^{-2}$ & $^{+4\%}_{-5\%}$ \quad $^{+4\%}_{-4\%}$ & 1.23 & $3.330 \times 10^{-1}$ & $^{+9\%}_{-17\%}$ \quad $^{+11\%}_{-11\%}$ & 1.40 \\
\hline
$N N \to W^+ W^-$ & \texttt{ p p > w+ w- QED=2 QCD=0 [QCD]} & $2.566 \times 10^{-5}$ & $^{+3\%}_{-2\%}$ \quad $^{+2\%}_{-2\%}$ & 1.33 & $1.029 \times 10^{-3}$ & $^{+9\%}_{-12\%}$ \quad $^{+7\%}_{-7\%}$ & 1.68 \\
\hline
\end{tabular}
}
\caption{Per nucleon NLO in QCD  cross sections (in $\mu$b) for various processes, for $^{12}$C collisions, and the representative collision energies $\sqrt{s_{NN}}=5\TeV$ and $100\TeV$, the corresponding {\mgamc} simulation syntax (with \texttt{define ww = w+ w-}), the associated scale and PDF uncertainties [$\%$], and QCD $K$-factor. \label{tab:summary_cx12}}
\end{table*}

\begin{table*}[!t]\renewcommand{\tabcolsep}{4pt}\renewcommand{\arraystretch}{1.18}
\resizebox{\textwidth}{!}{
\begin{tabular}{ l l ccc ccc }
\hline
\multicolumn{8}{c}{$^{12}$C-$^{12}$C Collisions}\\
\hline
\multirow{1}{*}{Process} & 
\multirow{1}{*}{Syntax} & 
\multicolumn{3}{c}{$\sqrt{s_{NN}}=5.52\TeV$} & 
\multicolumn{3}{c}{$\sqrt{s_{NN}}=39\TeV$}\\
& & $\sigma^\mathrm{NLO}$~[$\mu$b] & $\delta_\mathrm{scale}\quad \delta_\mathrm{PDF}$ & $K^\mathrm{NLO}$ & $\sigma^\mathrm{NLO}$~[$\mu$b] & $\delta_\mathrm{scale}\quad\delta_\mathrm{PDF}$ & $K^\mathrm{NLO}$\\
\hline
$N N \to W^\pm$ & \texttt{ p p > ww QED=1 QCD=0 [QCD]} & $6.514 \times 10^{-2}$ & $^{+4\%}_{-6\%}$ \quad $^{+5\%}_{-5\%}$ & 1.24 & $4.560 \times 10^{-1}$ & $^{+7\%}_{-14\%}$ \quad $^{+9\%}_{-9\%}$ & 1.35 \\
$N N \to W^-$ & \texttt{ p p > w- QED=1 QCD=0 [QCD]} & $3.281 \times 10^{-2}$ & $^{+4\%}_{-7\%}$ \quad $^{+5\%}_{-5\%}$ & 1.25 & $2.285 \times 10^{-1}$ & $^{+7\%}_{-14\%}$ \quad $^{+9\%}_{-9\%}$ & 1.35 \\
$N N \to W^+$ & \texttt{ p p > w+ QED=1 QCD=0 [QCD]} & $3.282 \times 10^{-2}$ & $^{+4\%}_{-7\%}$ \quad $^{+5\%}_{-5\%}$ & 1.25 & $2.301 \times 10^{-1}$ & $^{+8\%}_{-14\%}$ \quad $^{+9\%}_{-9\%}$ & 1.36 \\
$N N \to Z$ & \texttt{ p p > z QED=1 QCD=0 [QCD]} & $1.944 \times 10^{-2}$ & $^{+4\%}_{-6\%}$ \quad $^{+4\%}_{-4\%}$ & 1.24 & $1.410 \times 10^{-1}$ & $^{+6\%}_{-13\%}$ \quad $^{+9\%}_{-9\%}$ & 1.34 \\
\hline
$N N \to W^+ W^-$ & \texttt{ p p > w+ w- QED=2 QCD=0 [QCD]} & $2.975 \times 10^{-5}$ & $^{+3\%}_{-2\%}$ \quad $^{+2\%}_{-2\%}$ & 1.33 & $3.682 \times 10^{-4}$ & $^{+6\%}_{-8\%}$ \quad $^{+5\%}_{-5\%}$ & 1.55 \\
\hline
\end{tabular}
}
\caption{Same as Table~\ref{tab:summary_cx12} for the collision energies $\sqrt{s_{NN}}=5.52\TeV$ and $39\TeV$.\label{tab:summary_cx12_bis}}
\end{table*}

\clearpage
\begin{table*}[!t]\renewcommand{\tabcolsep}{4pt}\renewcommand{\arraystretch}{1.18}
\resizebox{\textwidth}{!}{
\begin{tabular}{ l l ccc ccc }
\hline
\multicolumn{8}{c}{$pp$ Collisions}\\
\hline
\multicolumn{1}{c}{\multirow{2}{*}{Process}} &
\multicolumn{1}{c}{\multirow{2}{*}{Syntax}} & 
\multicolumn{3}{c}{$\sqrt{s_{NN}}=5\TeV$} & 
\multicolumn{3}{c}{$\sqrt{s_{NN}}=100\TeV$}\\
& & $\sigma^\mathrm{NLO}$~[$\mu$b] & $\delta_\mathrm{scale}\quad \delta_\mathrm{PDF}$ & $K^\mathrm{NLO}$ & $\sigma^\mathrm{NLO}$~[$\mu$b] & $\delta_\mathrm{scale}\quad\delta_\mathrm{PDF}$ & $K^\mathrm{NLO}$\\
\hline
$p p \to W^\pm$ & \texttt{ p p > ww QED=1 QCD=0 [QCD]} & $6.419 \times 10^{-2}$ & $^{+4\%}_{-6\%}$ \quad $^{+4\%}_{-5\%}$ & 1.24 & $1.238 \times 10^{+0}$ & $^{+9\%}_{-17\%}$ \quad $^{+10\%}_{-6\%}$ & 1.40 \\
$p p \to W^-$ & \texttt{ p p > w- QED=1 QCD=0 [QCD]} & $2.541 \times 10^{-2}$ & $^{+4\%}_{-5\%}$ \quad $^{+3\%}_{-5\%}$ & 1.23 & $5.673 \times 10^{-1}$ & $^{+9\%}_{-17\%}$ \quad $^{+9\%}_{-6\%}$ & 1.40 \\
$p p \to W^+$ & \texttt{ p p > w+ QED=1 QCD=0 [QCD]} & $3.862 \times 10^{-2}$ & $^{+4\%}_{-6\%}$ \quad $^{+4\%}_{-5\%}$ & 1.24 & $6.708 \times 10^{-1}$ & $^{+9\%}_{-17\%}$ \quad $^{+10\%}_{-6\%}$ & 1.40 \\
$p p \to Z$ & \texttt{ p p > z QED=1 QCD=0 [QCD]} & $1.834 \times 10^{-2}$ & $^{+3\%}_{-5\%}$ \quad $^{+3\%}_{-4\%}$ & 1.23 & $3.834 \times 10^{-1}$ & $^{+8\%}_{-15\%}$ \quad $^{+9\%}_{-6\%}$ & 1.38 \\
\hline
$p p \to W^+ W^-$ & \texttt{ p p > w+ w- QED=2 QCD=0 [QCD]} & $2.626 \times 10^{-5}$ & $^{+3\%}_{-2\%}$ \quad $^{+4\%}_{-5\%}$ & 1.32 & $1.131 \times 10^{-3}$ & $^{+9\%}_{-11\%}$ \quad $^{+4\%}_{-4\%}$ & 1.64 \\
\hline
\end{tabular}
}
\caption{Per nucleon NLO in QCD cross sections (in $\mu$b) for various processes, for proton collisions, and the representative collision energies $\sqrt{s_{NN}}=5\TeV$ and $100\TeV$, the corresponding {\mgamc} simulation syntax (with \texttt{define ww = w+ w-}), the associated scale and PDF uncertainties [$\%$], and QCD $K$-factor. \label{tab:summary_hx1}}
\end{table*}

\begin{table*}[!t]\renewcommand{\tabcolsep}{4pt}\renewcommand{\arraystretch}{1.18}
\resizebox{\textwidth}{!}{
\begin{tabular}{ l l ccc ccc }
\hline
\multicolumn{8}{c}{$pp$ Collisions}\\
\hline
\multirow{1}{*}{Process} & 
\multirow{1}{*}{Syntax} & 
\multicolumn{3}{c}{$\sqrt{s_{NN}}=5.52\TeV$} & 
\multicolumn{3}{c}{$\sqrt{s_{NN}}=39\TeV$}\\
& & $\sigma^\mathrm{NLO}$~[$\mu$b] & $\delta_\mathrm{scale}\quad \delta_\mathrm{PDF}$ & $K^\mathrm{NLO}$ & $\sigma^\mathrm{NLO}$~[$\mu$b] & $\delta_\mathrm{scale}\quad\delta_\mathrm{PDF}$ & $K^\mathrm{NLO}$\\
\hline
$p p \to W^\pm$ & \texttt{ p p > ww QED=1 QCD=0 [QCD]} & $7.128 \times 10^{-2}$ & $^{+4\%}_{-6\%}$ \quad $^{+4\%}_{-4\%}$ & 1.24 & $5.209 \times 10^{-1}$ & $^{+6\%}_{-13\%}$ \quad $^{+5\%}_{-5\%}$ & 1.34 \\
$p p \to W^-$ & \texttt{ p p > w- QED=1 QCD=0 [QCD]} & $2.878 \times 10^{-2}$ & $^{+4\%}_{-6\%}$ \quad $^{+3\%}_{-5\%}$ & 1.24 & $2.338 \times 10^{-1}$ & $^{+7\%}_{-13\%}$ \quad $^{+5\%}_{-5\%}$ & 1.34 \\
$p p \to W^+$ & \texttt{ p p > w+ QED=1 QCD=0 [QCD]} & $4.267 \times 10^{-2}$ & $^{+4\%}_{-6\%}$ \quad $^{+4\%}_{-5\%}$ & 1.23 & $2.908 \times 10^{-1}$ & $^{+7\%}_{-13\%}$ \quad $^{+6\%}_{-5\%}$ & 1.34 \\
$p p \to Z$ & \texttt{ p p > z QED=1 QCD=0 [QCD]} & $2.057 \times 10^{-2}$ & $^{+3\%}_{-5\%}$ \quad $^{+3\%}_{-4\%}$ & 1.23 & $1.594 \times 10^{-1}$ & $^{+6\%}_{-12\%}$ \quad $^{+6\%}_{-5\%}$ & 1.32 \\
\hline
$p p \to W^+ W^-$ & \texttt{ p p > w+ w- QED=2 QCD=0 [QCD]} & $3.063 \times 10^{-5}$ & $^{+3\%}_{-2\%}$ \quad $^{+4\%}_{-5\%}$ & 1.33 & $3.986 \times 10^{-4}$ & $^{+6\%}_{-8\%}$ \quad $^{+3\%}_{-3\%}$ & 1.52 \\
\hline
\end{tabular}
}
\caption{Same as Table~\ref{tab:summary_hx1} for the collision energies $\sqrt{s_{NN}}=5.52\TeV$ and $39\TeV$.\label{tab:summary_hx1_bis}}
\end{table*}

\section{Extrapolating Cross Sections from One Nucleus to Another}
\label{sec:extrapolation}

In Sec.~\ref{sec:xsec} we surveyed the inclusive and fiducial cross sections for dozens of high-energy processes as a function of collider energy. One observation (of many) was that the ratios of the nucleon-level cross section for symmetric $^{208}$Pb, $^{131}$Xe, and $^{12}$C collisions with respect to symmetric proton collisions tended to converge to some number for sufficiently high collider energies. In many cases, these were near but not equal to unity. Despite different nuclei exhibiting qualitatively different parton densities (Sec.~\ref{sec:xpdf}) and parton luminosities (Sec.~\ref{sec:lumi}), findings suggest that it may be possible to extrapolate reliably a particular cross section  for one pair of colliding nuclei from a second pair at sufficiently high collider energies. In this section, we explore quantitatively this possibility.

For the inclusive production of a particular final state $\mathcal{F}$, possibly with phase space restrictions, we consider the cross section estimate (est.) obtained by multiply a true nuclear-level cross section (true) by the appropriate ratio of atomic numbers. Symbolically, this is given by
\begin{subequations}
\label{eq:extrapolation_def}
\begin{align}
    \sigma^{\rm est.}(\mathcal{A}_1\mathcal{A}_2 \to \mathcal{F}+X) &=\
    \left(\frac{A_1A_2}{A_3A_4}\right)\ \times\
    \sigma^{\rm true}(\mathcal{A}_3\mathcal{A}_4 \to \mathcal{F}+X)
    \\
    &=\
    \left(A_1A_2\right)\ \times\
    \sigma^{\rm true}_{NN}(N_3 N_4 \to \mathcal{F}+X)\ .
\end{align}
\end{subequations}
According to the definition in Eq.~\eqref{eq:xsec_def_nucleon},
this reduces to the product of (a) the target atomic numbers $(A_1A_2)$ and (b) the starting nucleon-level cross section $\sigma_{N_3N_4}$. One can then define the following as the  relative difference [$\%$] between the estimated and true nucleus-level cross sections:
\begin{subequations}
\label{eq:extrapolation_unc}
\begin{align}
\frac{\delta\sigma^{\mathcal{A}_1\mathcal{A}_2}_{\mathcal{A}_3\mathcal{A}_4}}
{\sigma^{\rm true}_{\mathcal{A}_1\mathcal{A}_2}}
&=\
\frac{\sigma^{\rm est.}(\mathcal{A}_1\mathcal{A}_2 \to \mathcal{F}+X)\
-\
\sigma^{\rm true}(\mathcal{A}_1\mathcal{A}_2 \to \mathcal{F}+X)}
{\sigma^{\rm true}_{\mathcal{A}_1\mathcal{A}_2}(\mathcal{A}_1\mathcal{A}_2 \to \mathcal{F}+X)}
\\
&=\
\frac{
\left(A_1A_2\right) \times
    \sigma^{\rm true}_{NN}(N_3 N_4 \to \mathcal{F}+X)\
    -\
\left(A_1A_2\right) \times
    \sigma^{\rm true}_{NN}(N_1 N_2 \to \mathcal{F}+X)\
}{\left(A_1A_2\right) \times
    \sigma^{\rm true}_{NN}(N_1 N_2 \to \mathcal{F}+X)}
\\
&=\
\frac{\sigma^{\rm true}_{NN}(N_3 N_4 \to \mathcal{F}+X)\
    -\
    \sigma^{\rm true}_{NN}(N_1 N_2 \to \mathcal{F}+X)\
}{\sigma^{\rm true}_{NN}(N_1 N_2 \to \mathcal{F}+X)}\ .
\end{align}
\end{subequations}

For the case of symmetric beams, we have $\mathcal{A}_4=\mathcal{A}_3$ and $\mathcal{A}_2=\mathcal{A}_1$ as well as  $N_4=N_3$ and $N_2=N_1$, but otherwise the expression remains the same. Two observations can be drawn from Eq.~\eqref{eq:extrapolation_unc}: First and most importantly, the expression quantifies the error in using Eq.~\eqref{eq:extrapolation_def} to extrapolate cross sections across nuclei. Second but more interesting, we find that an uncertainty at the \textit{nuclear level} can be expressed in terms of \textit{nucleon-level} quantities. To guide further discussion, we note that intuitively the closer this error is to zero, the better the extrapolation. However, a nonzero number is also useful as one can extract an ``offset'' factor $\Delta_{\rm offset}$. One can then, in principle, improve the extrapolation formula of Eq.~\eqref{eq:extrapolation_def} by the following modification
\begin{align}
\label{eq:extrapolation_offset}
    \sigma^{\rm est.}(\mathcal{A}_1\mathcal{A}_2 \to \mathcal{F}+X) &=\
    \left(A_1A_2\right)\ \times\
    \sigma^{\rm true}_{NN}(N_3 N_4 \to \mathcal{F}+X)\
    +\
    \Delta_{\rm offset}\ .
\end{align}

\begin{figure}[!th]
\begin{center}
\subfigure[]{
\includegraphics[width=.4\textwidth]{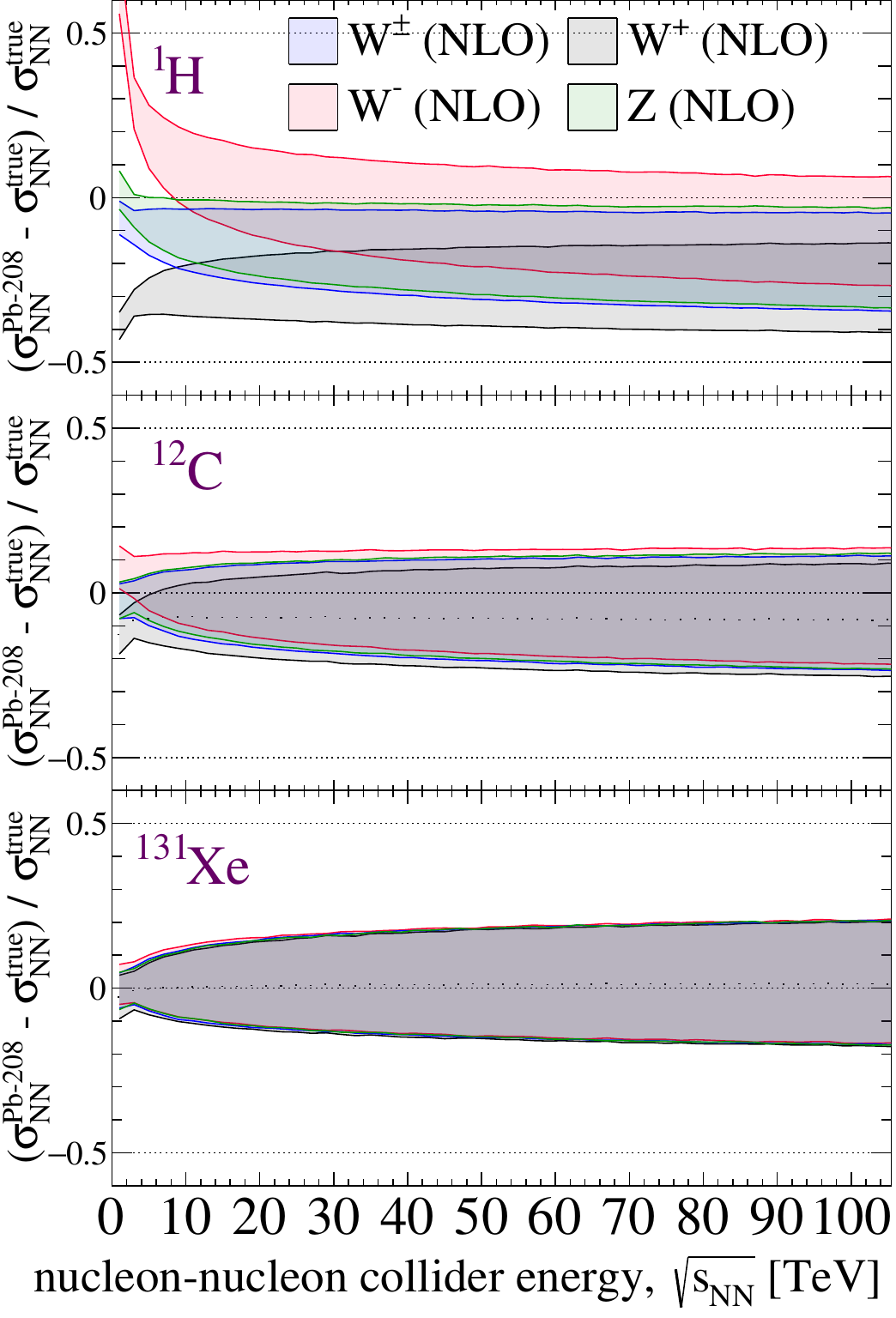}
\label{fig:ionsNLO_Extrap_vXXX_vs_Beam}}
\subfigure[]{
\includegraphics[width=.4\textwidth]{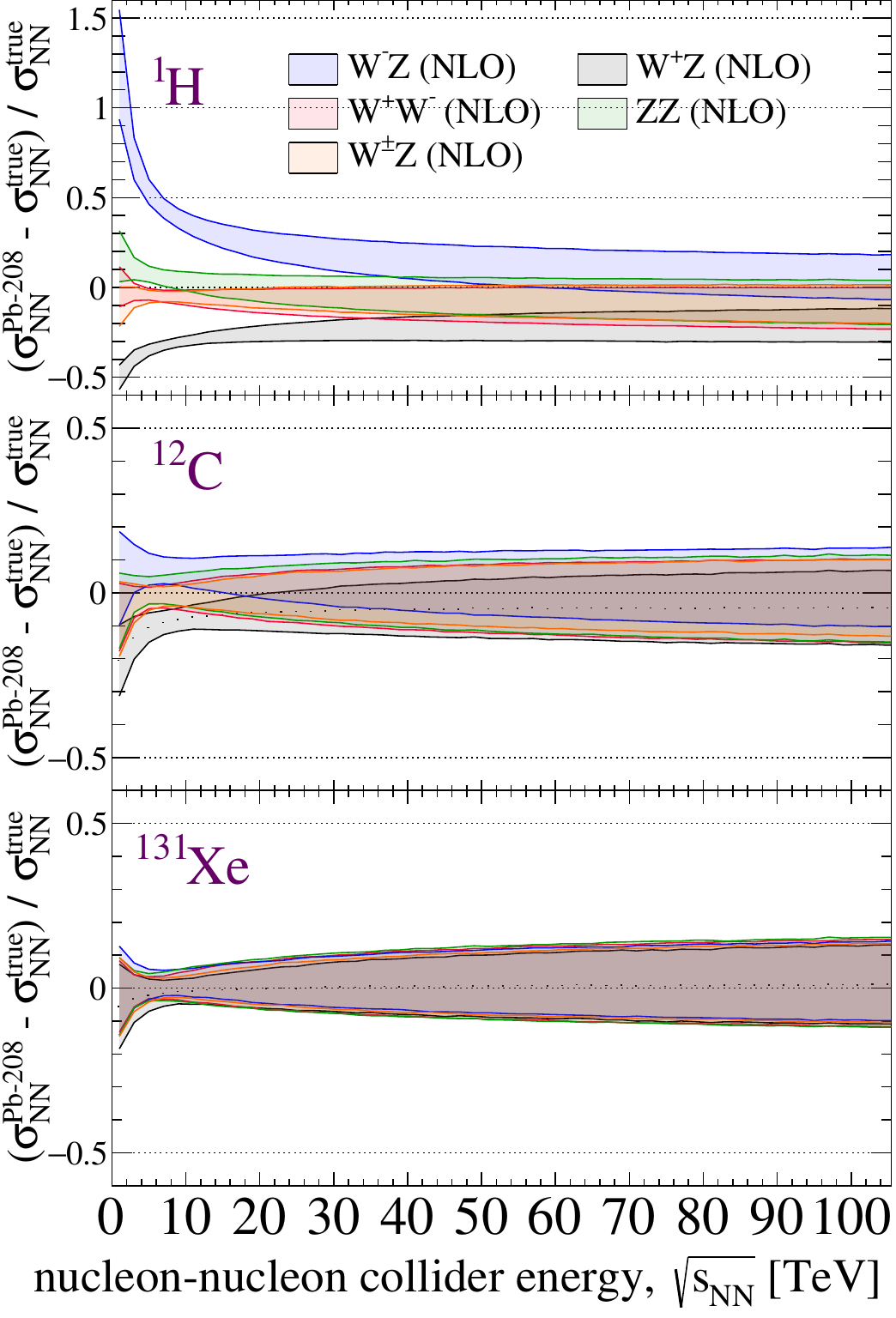}
\label{fig:ionsNLO_Extrap_vvXX_vs_Beam}}
\\
\subfigure[]{
\includegraphics[width=.4\textwidth]{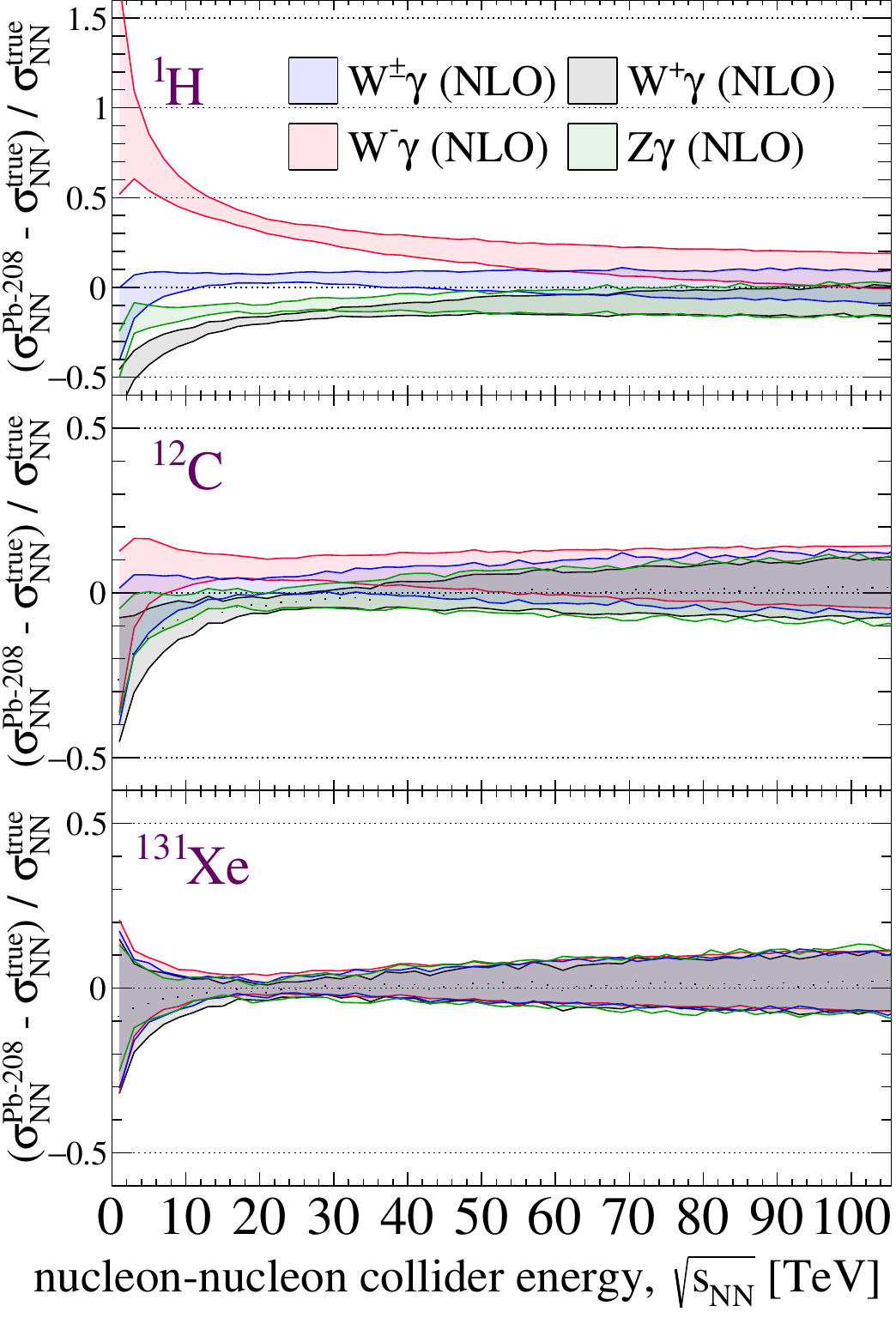}
\label{fig:ionsNLO_Extrap_vaXX_vs_Beam}}
\subfigure[]{
\includegraphics[width=.4\textwidth]{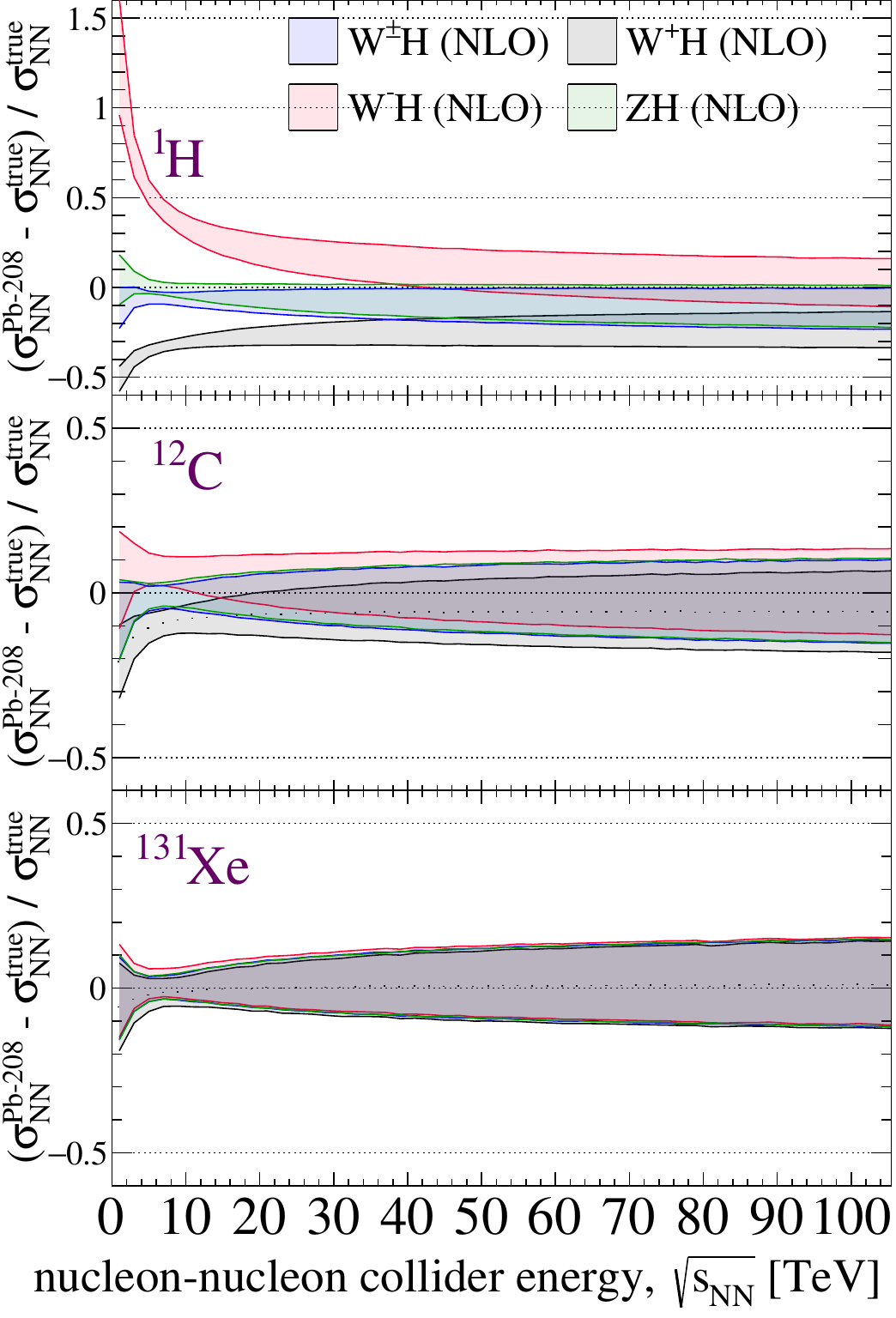}
\label{fig:ionsNLO_Extrap_vhXX_vs_Beam}}
\end{center}
\caption{As a function of $\sqrt{s_{NN}}$, the extrapolation error [Eq.~\eqref{eq:extrapolation_unc}] in estimating the NLO in QCD cross section,  with PDF uncertainty, for (a) single weak boson, (b) diboson, (c) associated photon, and (d) associated Higgs production in $^{1}$H (top panel), $^{12}$C (middle), and $^{131}$Xe (bottom) collisions from $^{208}$Pb collisions.
}
\label{fig:ionsNLO_Extrap_v_vv}
\end{figure}

\begin{figure}[!th]
\subfigure[]{
\includegraphics[width=.32\textwidth]{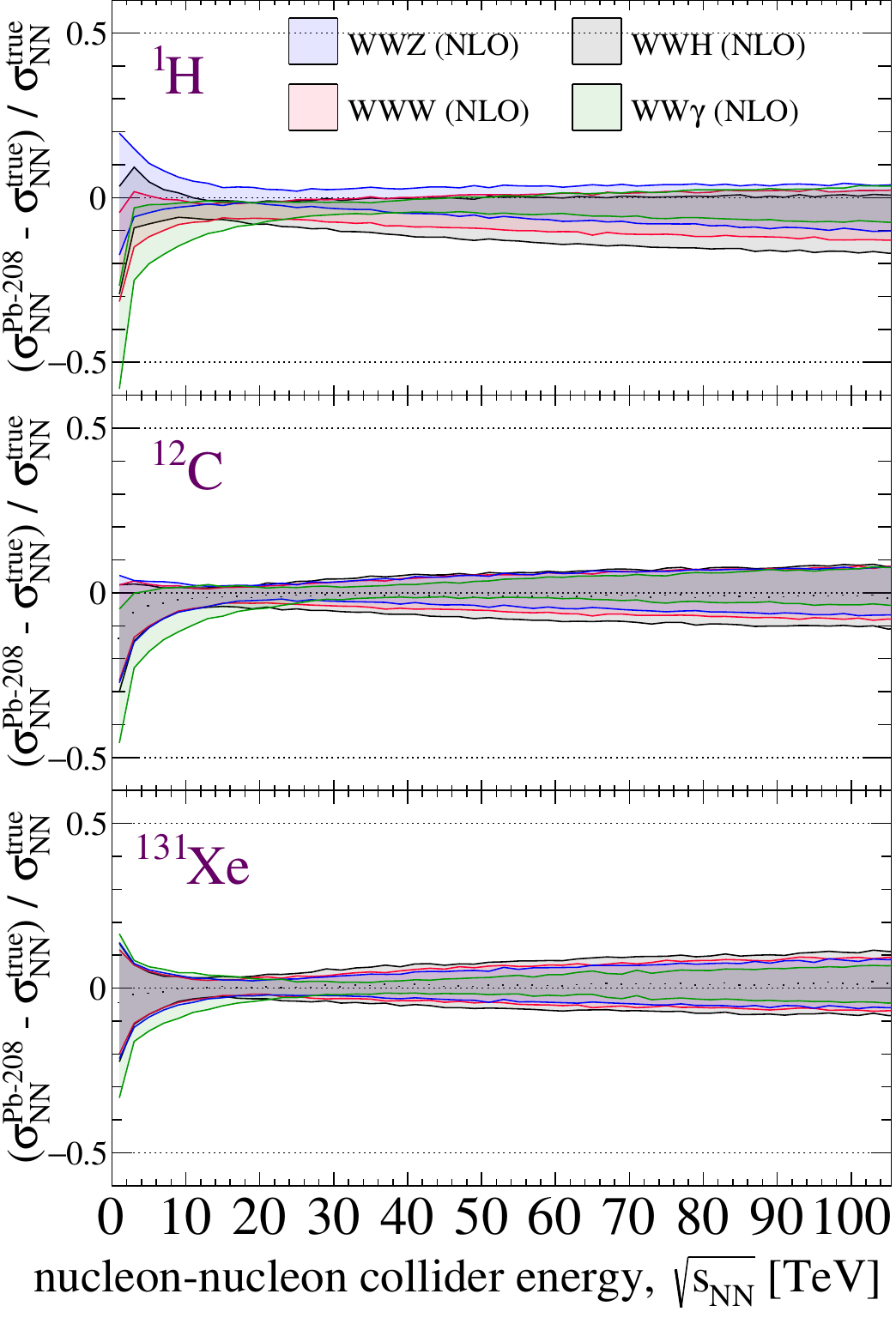}
\label{fig:ionsNLO_Extrap_vwwX_vs_Beam}}
\subfigure[]{
\includegraphics[width=.32\textwidth]{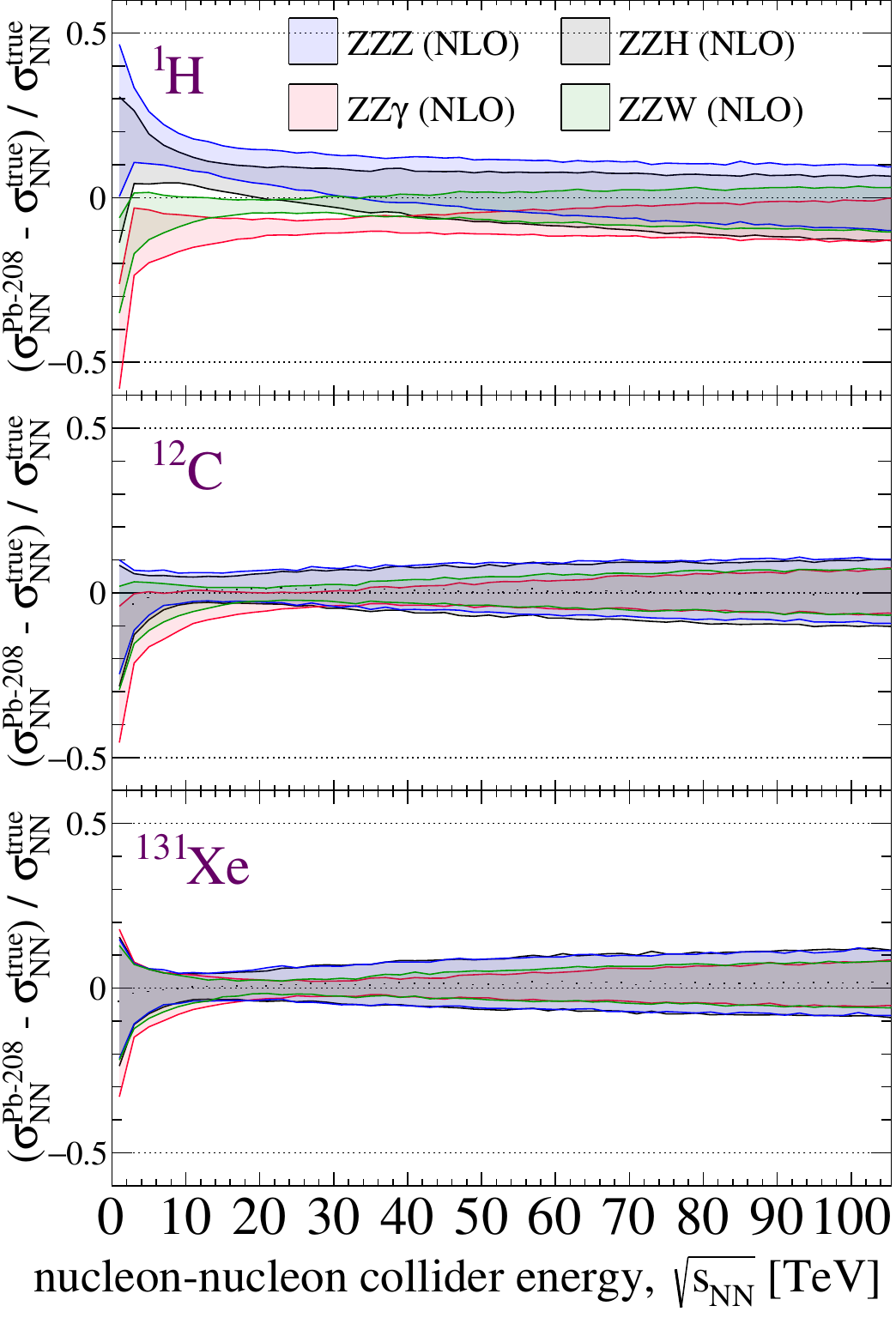}
\label{fig:ionsNLO_Extrap_vzzX_vs_Beam}}
\subfigure[]{
\includegraphics[width=.32\textwidth]{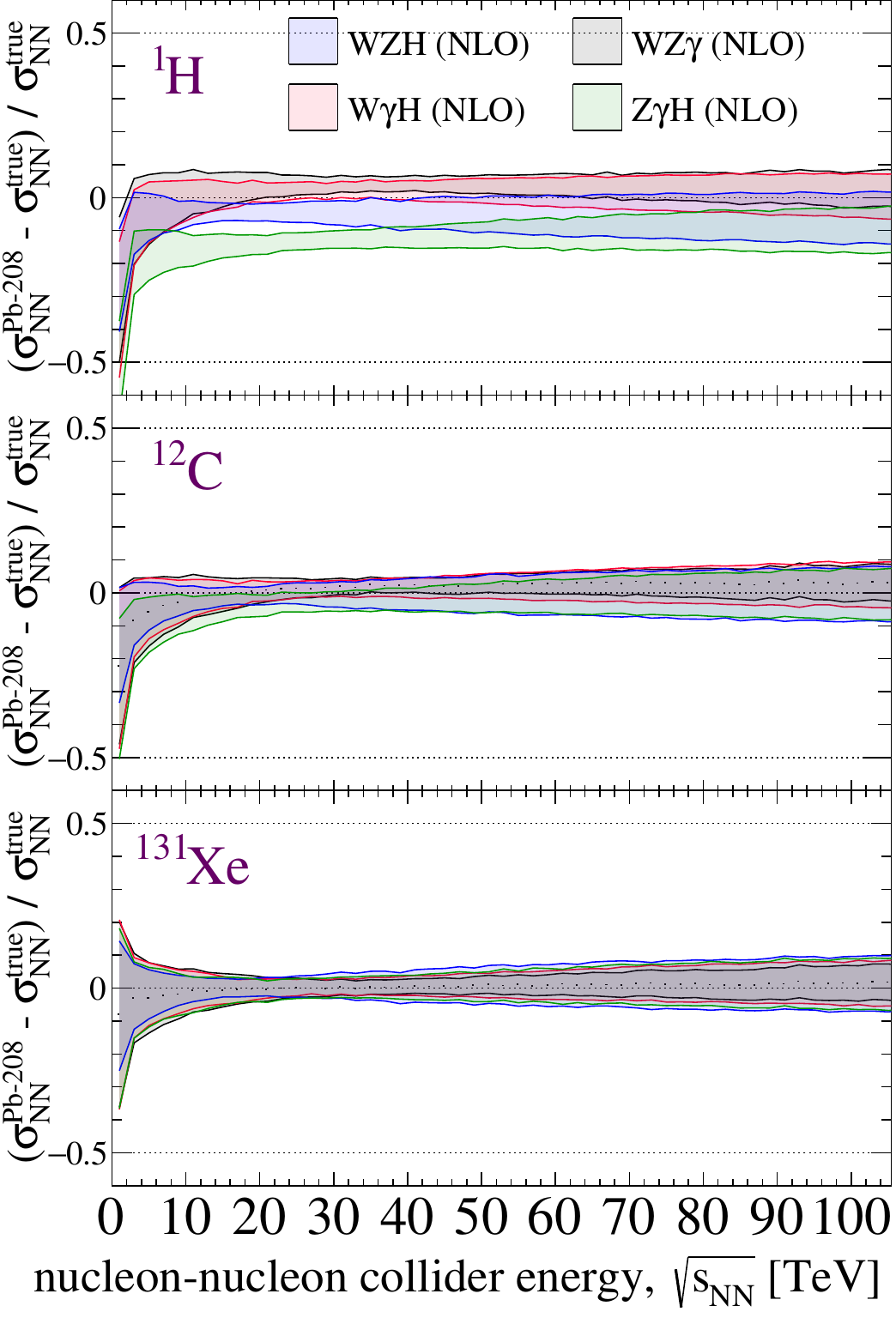}
\label{fig:ionsNLO_Extrap_vwzX_vs_Beam}}
\\
\subfigure[]{
\includegraphics[width=.32\textwidth]{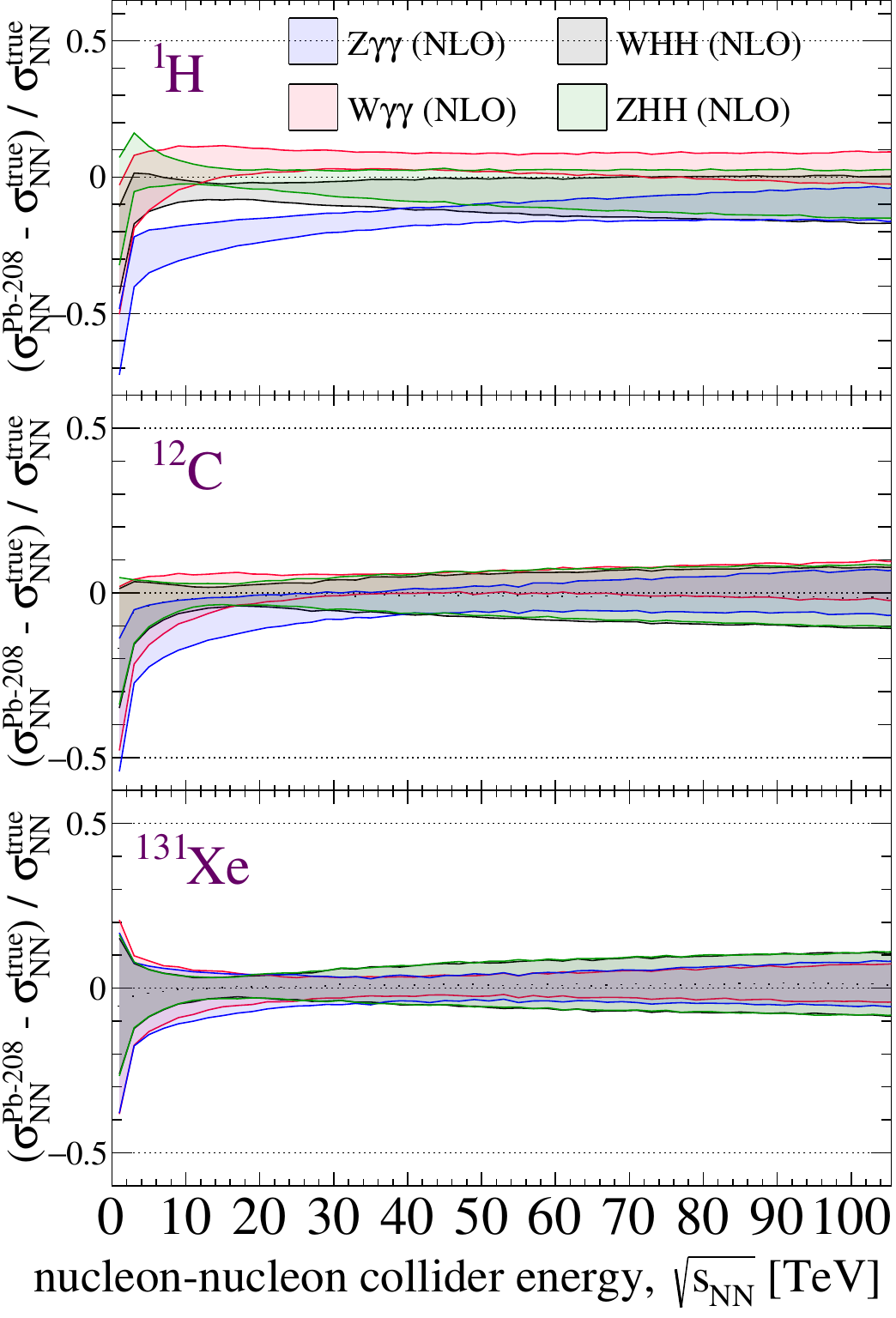}
\label{fig:ionsNLO_Extrap_vaaX_vs_Beam}}
\subfigure[]{
\includegraphics[width=.32\textwidth]{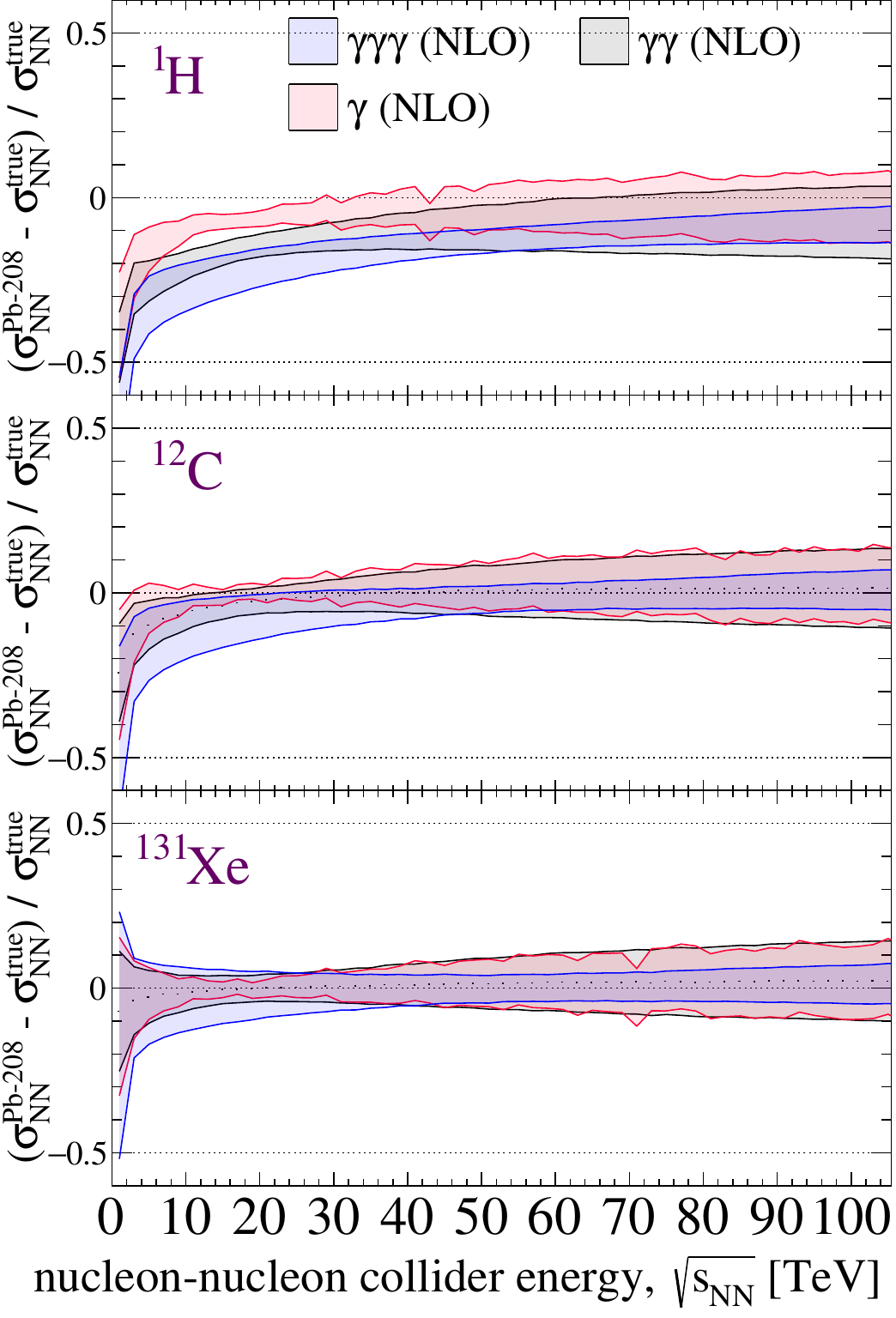}
\label{fig:ionsNLO_Extrap_aXXX_vs_Beam}}
\subfigure[]{
\includegraphics[width=.32\textwidth]{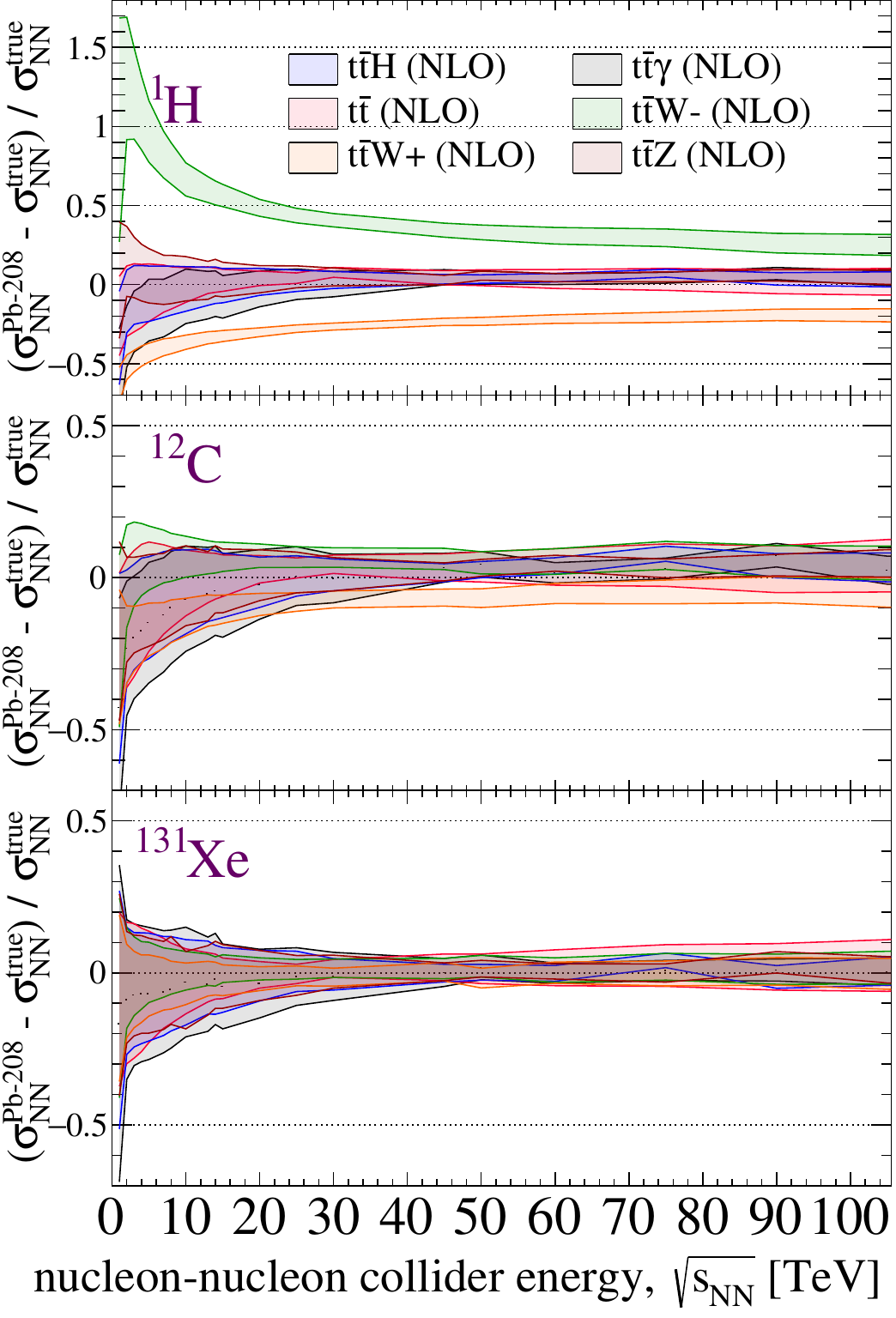}
\label{fig:ionsNLO_Extrap_ttXX_vs_Beam}}
\caption{Same as Fig.~\ref{fig:ionsNLO_Extrap_v_vv} but for (a) $W^+W^-X$, (b) $ZZX$, (c) $W^\pm Z X$ and $\gamma HX$, (d) $\gamma\gamma X$ and $HHX$, (e) inclusive photon, and (e) top quark processes.
}
\label{fig:ionsNLO_Extrap_other}
\end{figure}

In Figs.~\ref{fig:ionsNLO_Extrap_v_vv} and \ref{fig:ionsNLO_Extrap_other}, we show as a function of nucleon-nucleon collision energy $\sqrt{s_{NN}}$ the extrapolation error, as defined in Eq.~\eqref{eq:extrapolation_unc}, in estimating the scattering cross sections at NLO in QCD for our various processes and isotopes from the true predictions for $^{208}$Pb. Also shown is the uncertainty band stemming from the PDF uncertainty of $^{208}$Pb. In all plots, we consider (top panel) $^{1}$H, (middle panel) $^{12}$C, and (bottom panel) $^{131}$Xe.

We start our assessment with Fig.~\ref{fig:ionsNLO_Extrap_v_vv}, where we have: (a) single weak boson production, (b) diboson production, (c) associated photon-weak boson production, and (d) associated Higgs-weak boson production. The first remarkable finding is the extrapolation to $^{131}$Xe. For all processes and collision energies, the extrapolation error sits at or just around zero, with a PDF uncertainty in the range of \confirm{$\mathcal{O}(\pm5\%)-\mathcal{O}(\pm20\%)$.} For $\sqrt{s_{NN}}\gtrsim5-10\TeV$, a similar trend holds for $^{12}$C. While the error for carbon converges to \confirm{$\delta\sigma/\sigma\vert_{\rm est.}\sim0\%$ to $-10\%$}, this lies within the \confirm{$\mathcal{O}(\pm5\%)-\mathcal{O}(\pm20\%)$ PDF uncertainty.} For low collision energies, \textit{i.e.}, $\sqrt{s_{NN}}\lesssim5-10\TeV$, the error can grow to about \confirm{$\delta\sigma/\sigma\vert_{\rm est.}\sim-25\%$.} We broadly attribute these similarities and differences to two ingredients: (i) the relative nucleon content of the isotopes, which are comparable and are approximately $Z:(A-Z)\sim 40:60\ (40:60)\ [50:50]$ for $^{208}$Pb ($^{131}$Xe) [$^{12}$C], and (ii) the growing importance of valence quarks (gluons) at the lowest (highest) collision energies that we consider. For example: As discussed in Sec.~\ref{sec:xsec_xva}, $V\gamma$ production is sensitive to the nucleus' isospin since this is related to the net electric charge of its constituents.

When extrapolating predictions for proton collisions from lead collisions (in practice, one will more likely do the reverse), we observe much larger extrapolation errors and hence much worse agreement. At the largest collider energies, cross sections can be off by as much as \confirm{$\delta\sigma/\sigma\vert_{\rm est.}\sim-25\%$}, and over \confirm{$\delta\sigma/\sigma\vert_{\rm est.}\sim+100\%$} at the lowest energies. The largest disagreements are driven by the $W^-X$ channels, and ultimately reflect the difference in valance $d$ quark distributions in lead and the proton. However, in charge-symmetric channels such as $W^\pm+\gamma/Z/H$, $W^+W^-$, and $ZZ$ production, we find much better agreement. With the exception of $V\gamma$, extrapolation errors for these channels are small and comparable to those for $^{12}$C and $^{131}$Xe; differences remain below the \confirm{$\mathcal{O}(10\%)$ level.}

Moving onto Fig.~\ref{fig:ionsNLO_Extrap_other}, we show the same information as in Fig.~\ref{fig:ionsNLO_Extrap_v_vv} but for (a) $W^+W^-X$, (b) $ZZX$, (c) $W^\pm Z X$ and $\gamma HX$, (d) $\gamma\gamma X$ and $HHX$, (e) inclusive photon, and (e) top quark processes. Globally, we observe qualitatively and quantitatively similar behavior: For channels that are charge- and flavor-symmetric, scattering rates for symmetric $^{208}$Pb collisions can be used to reliably estimate scattering rates for  symmetric $^{1}$H, $^{12}$C, and $^{131}$Xe rates to within \confirm{$\mathcal{O}(10\%)$, particularly for collision energies $\sqrt{s_{NN}}\gtrsim5\TeV-10\TeV$.} The exception is channels with one or more high-energy photons in the final state; these retain sizable sensitivity to the net isospin of the nucleus. For charge- and flavor-asymmetric channels,  \textit{e.g.}, $t\overline{t}W^-$ and $t\overline{t}W^+$, we find that the extrapolation errors can quickly converse to nonzero values that are process and isotope dependent, suggesting that the offset correction in Eq.~\eqref{eq:extrapolation_offset} is a viable remedy.

\section{Discussion and Conclusions}\label{sec:summary}

The density and distribution of partons in nuclei differ both qualitatively and quantitatively from protons. This stems from an interplay of nuclear, hadronic, as well as partonic dynamics, and has a direct impact on parton luminosities. (See Fig.~\ref{fig:diagram_QCDexchanges} for illustrations of typical dynamics at these levels.) Subsequently, scattering cross sections of high-mass particle physics processes in high-energy ion collisions differ both qualitatively and quantitatively from those in proton collisions. We explore a variety of such high-energy processes, ranging from single $W^\pm$ and $Z$ production and associated $W^\pm H$ and $ZH$ production to triple $W^+W^- H$ and $ZZH$ production as well as associated $t\overline{t}W^\pm$ and $t\overline{t}Z$ production.

In this work, we have attempted to map out the ion and collider energy dependence by systematically cataloging the scattering rates and  theoretical uncertainties of \confirm{42} high-$p_T$ processes in $^{208}$Pb-$^{208}$Pb, $^{131}$Xe-$^{131}$Xe, and $^{12}$C-$^{12}$C collisions over the energy range $\sqrt{s_{NN}}=1-100$ TeV. Our work takes advantage of publicly available precision Monte Carlo simulation tools developed initially for proton-proton collisions~\cite{Stelzer:1994ta,Alwall:2014hca}. Our work also relies on advances in nuclear PDF fits, which have seen significant improvements over the past decade~\cite{Klasen:2023uqj}. We carry out this work up to NLO in QCD in order to properly account for additional partonic and kinematic sub-channels, which can be numerically important due to the presence of accidental cancellations at lowest order in perturbation theory, large virtual corrections at one loop, the largeness of the strong coupling constant, \textit{etc}.

While our main results are shown in Sec.~\ref{sec:xsec} (cross sections), Secs.~\ref{sec:xpdf} and \ref{sec:lumi} feature, respectively, a study of nPDFs and luminosities (and their ratios) for various partons, nuclei, scales, and kinematic ranges. When taken altogether, we are able to build an interesting narrative on how the interplay among nuclear dynamics (\textit{e.g.}, meson exchange), hadronic dynamics (\textit{e.g.}, long-range gluon exchanges), and partonic dynamics (\textit{e.g.}, DGLAP evolution) propagate into cross sections. In many cases, this interplay leads to sufficiently large differences in cross sections with sufficiently small uncertainties that they can be observed experimentally in ongoing and future high-energy ion collisions. However, at the same time, we find consistently the increasing role of partonic dynamics as the dominant driver of phenomenology at increasing collider energies for many processes. Motivated by this, we explore in Sec.~\ref{sec:extrapolation} the error made when trying to extrapolate cross sections for one nucleus from predictions for a second nucleus.
For illustrative purposes,
we briefly examine in Sec.~\ref{sec:xsec_yields}
the prospect for observing diboson and triboson processes
in symmetric ion collisions at the HL-LHC and FCC by means of a simple event analysis including conservative selection cuts.
Finally, in Sec.~\ref{sec:xsec_summary},
we catalog in
Tables~\ref{tab:summary_pb208}-\ref{tab:summary_hx1_bis}
various cross sections and uncertainties
for representative collision scales
$\sqrt{s_{NN}}=5\TeV,\ 5.52\TeV,\ 39\TeV,$ and $100\TeV$,
for all four ion-ion configurations under consideration.

For the benchmark collider scenario~\cite{FCC-ionsstudygroup:2017glf} of $\sqrt{s_{NN}}=39$ TeV, we report that
$\mathcal{O}(10^2-10^{8})$ resonant and high-$p_T$ Standard Model  events can be produced with  $\mathcal{L}=33$ nb$^{-1}$ of data. For the HL-LHC, one can expect $\mathcal{O}(10-10^6)$ single, multiboson, and top events
per $1$ nb$^{-1}$ at $\sqrt{s_{NN}}=5.52$ TeV.
While total production rates are large, decay rates and experimental selection/acceptance rates will impact final event yields,	and merit further study.

In the context of the inclusive production of high-energy systems in multi-TeV collisions, interesting findings include the following:
\begin{itemize}
 \item While nPDFs are known to exhibit large $A$ dependence for $x\gtrsim0.1$, much of this variation is  softened at the level of parton luminosities. We attribute this to an increased role of the gluon density and $g\to q\overline{q}$ splittings (that are  driven by perturbative dynamics), which are flavor symmetric (for massless quarks) and therefore exhibit approximate nuclear isospin symmetry, \textit{i.e.,} are approximately independent of $(A,Z)$ configurations. This is particularly true for kinematic thresholds below $(\hat{s}/s_{NN})\lesssim10^{-3}$.

 \item Categorically, for processes that sum over all charge and flavor configurations, we find that the ratio of  nucleon-averaged cross sections $\sigma_{NN}$ [as defined in Eq.~\eqref{eq:xsec_def_nucleon}] with respect to the proton cross section $\sigma_{pp}$ converges a flat value at or just below unity once the nucleon-nucleon collision energy surpasses $\sqrt{s_{NN}}\gtrsim5-10\TeV$. Not only does this include simpler single- and diboson processes, \textit{e.g.,} $Z$ and $W^\pm H$ production, but more complex  triboson and top quark processes, \textit{e.g.,} $W^+W^-Z$ and $t\overline{t}H$ production. This follows from the increased importance of sea quarks and sea antiquarks (and hence gluons) in the production of sub-TeV systems in multi-TeV collisions, and the smaller sensitivity of these partons to an averaged nucleon's $(A,Z)$ configuration. The exception are channels with multiple $Z$ bosons and photons since these are indirectly sensitive to the net isospin of nucleons, but the impact is small.

 \item Categorically, for processes that are charge or flavor asymmetric, \textit{e.g.,} $W^-$, $W^-H$, $W^-\gamma$, and $t\overline{t}W^-$ production, the aforementioned convergence still occurs when collision energies are sufficiently large such that \confirm{$(\sqrt{\hat{s}}/\sqrt{s_{NN}}) \lesssim  4-6\times10^{-3}$}. This universal-like behavior across so many processes and ion configurations suggests the onset of a common phenomenon, such as an increased importance / dominance of sea quarks and hence gluons in many-TeV collisions.

 \item In \textit{many} cases, we found large QCD corrections at NLO (with $K^{\rm NLO}>1.5-2$). Such large corrections, which are well-documented for proton collisions, have various origins. These include radiation amplitude zeros (accidental cancellations at lowest order), the opening of new partonic channels, DGLAP evolution, and sizable virtual corrections. It is important to stress that any  ion collision prediction of high-energy systems, albeit a precision measurement or not, should include QCD corrections if and when available. Importantly, nPDF uncertainties in cross sections are comparable or smaller than scale NLO in QCD uncertainties.

 \item In practice (see  Sec.~\ref{sec:extrapolation}), many nucleus-level cross sections and uncertainties for many processes can be reliably extrapolated, even at NLO in QCD, by scaling proton cross sections by the appropriate $A$ factor(s). For channels that are charge- and flavor-symmetric, scattering rates for symmetric $^{208}$Pb collisions can be used to reliably estimate scattering rates for symmetric $^{1}$H, $^{12}$C, and $^{131}$Xe collisions to within \confirm{$\mathcal{O}(10\%)$, particularly for collision energies $\sqrt{s_{NN}}\gtrsim5\TeV-10\TeV$.} Exceptions consist of channels with one or more high-energy photons in the final state; these retain sizable sensitivity to the net isospin of the nucleus.

 \item  With $\mathcal{L}=1-10$ nb$^{-1}$ of data,
one can expect enough events after reconstruction
and  selection to observe several diboson signal categories
at the HL-LHC. At the FCC and with $\mathcal{L}=10-30$ nb$^{-1}$, one can establish a precision program for diboson physics. Moreover, with more relaxed selection criteria and/or the use of asymmetric $p\mathcal{A}$ collisions, it may be possible to observe triboson processes in ion collisions at the FCC.

\end{itemize}

In conclusion, with the start of Run III activities at the LHC and ongoing preparations for its high luminosity era, we look forward to the upcoming transition to a precision particle physics program using high-energy ion-ion and ion-proton collisions. And subsequently, through the combined use of precision matrix elements, precision event generators / simulation tools, and precision detector systems, a precise determination of the nuclear structure could be achieved through the extraction of nuclear PDFs from measurements of hard, electroweak-scale processes in multi-TeV ion collisions.

\section*{Acknowledgements}
\addcontentsline{toc}{section}{Acknowledgements}

The authors thank  Aleksander Kusina, Krzysztof Kutak, and Fred Olness for enlightening discussions. BF acknowledges support from Grant ANR-21-CE31-0013, Project DMwithLLPatLHC, from the \emph{Agence Nationale de la Recherche} (ANR), France. RR acknowledges the support of Narodowe Centrum Nauki under Grant Nos.~2019/34/E/ST2/00186 and 2023/49/B/ST2/04330 (SNAIL). The authors acknowledge support from the COMETA COST Action CA22130. This work was partially funded by the Polish Academy of Sciences under the ``Funding for the dissemination and promotion of scientific activities'' program  (PAN.BFB.S.KA.140.022.2023).

\bibliography{ionsNLO_refs}

\providecommand{\href}[2]{#2}\begingroup\raggedright\begin{thebibliography}{100}

\bibitem{PHENIX:2003nhg}
{\scshape PHENIX} collaboration, K.~Adcox et~al., \textit{{PHENIX detector
  overview}},
  \href{https://doi.org/10.1016/S0168-9002(02)01950-2}{\textit{Nucl. Instrum.
  Meth. A} {\bfseries 499} (2003) 469--479}.

\bibitem{ALICE:2008ngc}
{\scshape ALICE} collaboration, K.~Aamodt et~al., \textit{{The ALICE experiment
  at the CERN LHC}},
  \href{https://doi.org/10.1088/1748-0221/3/08/S08002}{\textit{JINST}
  {\bfseries 3} (2008) S08002}.

\bibitem{Albacete:2016veq}
J.~L. Albacete et~al., \textit{{Predictions for $p+$Pb Collisions at
  $\sqrt{s_{NN}} = 5$ TeV: Comparison with Data}},
  \href{https://doi.org/10.1142/S0218301316300058}{\textit{Int. J. Mod. Phys.
  E} {\bfseries 25} (2016) 1630005},
  [\href{https://arxiv.org/abs/1605.09479}{{\ttfamily 1605.09479}}].

\bibitem{Dainese:2016gch}
A.~Dainese et~al., \textit{{Heavy ions at the Future Circular Collider}},
  \href{https://arxiv.org/abs/1605.01389}{{\ttfamily 1605.01389}}.

\bibitem{FCC-ionsstudygroup:2017glf}
{\scshape FCC-ions study group} collaboration, D.~d'Enterria et~al.,
  \textit{{Physics with ions at the Future Circular Collider}},
  \href{https://doi.org/10.1016/j.nuclphysa.2017.06.029}{\textit{Nucl. Phys. A}
  {\bfseries 967} (2017) 888--891},
  [\href{https://arxiv.org/abs/1704.05891}{{\ttfamily 1704.05891}}].

\bibitem{Citron:2018lsq}
Z.~Citron et~al., \textit{{Report from Working Group 5}: {Future physics
  opportunities for high-density QCD at the LHC with heavy-ion and proton
  beams}}, \href{https://doi.org/10.23731/CYRM-2019-007.1159}{\textit{CERN
  Yellow Rep. Monogr.} {\bfseries 7} (2019) 1159--1410},
  [\href{https://arxiv.org/abs/1812.06772}{{\ttfamily 1812.06772}}].

\bibitem{Brewer:2021kiv}
J.~Brewer, A.~Mazeliauskas and W.~van~der Schee, \textit{{Opportunities of OO
  and $p$O collisions at the LHC}},  in \textit{{Opportunities of OO and pO
  collisions at the LHC}}, 3, 2021,
  \href{https://arxiv.org/abs/2103.01939}{{\ttfamily 2103.01939}}.

\bibitem{Kovarik:2010uv}
K.~Kovarik, I.~Schienbein, F.~I. Olness, J.~Y. Yu, C.~Keppel, J.~G. Morfin
  et~al., \textit{{Nuclear Corrections in Neutrino-Nucleus DIS and Their
  Compatibility with Global NPDF Analyses}},
  \href{https://doi.org/10.1103/PhysRevLett.106.122301}{\textit{Phys. Rev.
  Lett.} {\bfseries 106} (2011) 122301},
  [\href{https://arxiv.org/abs/1012.0286}{{\ttfamily 1012.0286}}].

\bibitem{Kovarik:2015cma}
K.~Kovarik et~al., \textit{{nCTEQ15 - Global analysis of nuclear parton
  distributions with uncertainties in the CTEQ framework}},
  \href{https://doi.org/10.1103/PhysRevD.93.085037}{\textit{Phys. Rev. D}
  {\bfseries 93} (2016) 085037},
  [\href{https://arxiv.org/abs/1509.00792}{{\ttfamily 1509.00792}}].

\bibitem{Eskola:2016oht}
K.~J. Eskola, P.~Paakkinen, H.~Paukkunen and C.~A. Salgado, \textit{{EPPS16:
  Nuclear parton distributions with LHC data}},
  \href{https://doi.org/10.1140/epjc/s10052-017-4725-9}{\textit{Eur. Phys. J.
  C} {\bfseries 77} (2017) 163},
  [\href{https://arxiv.org/abs/1612.05741}{{\ttfamily 1612.05741}}].

\bibitem{Walt:2019slu}
M.~Walt, I.~Helenius and W.~Vogelsang, \textit{{Open-source QCD analysis of
  nuclear parton distribution functions at NLO and NNLO}},
  \href{https://doi.org/10.1103/PhysRevD.100.096015}{\textit{Phys. Rev. D}
  {\bfseries 100} (2019) 096015},
  [\href{https://arxiv.org/abs/1908.03355}{{\ttfamily 1908.03355}}].

\bibitem{Khanpour:2020zyu}
H.~Khanpour, M.~Soleymaninia, S.~Atashbar~Tehrani, H.~Spiesberger and V.~Guzey,
  \textit{{Nuclear parton distribution functions with uncertainties in a
  general mass variable flavor number scheme}},
  \href{https://doi.org/10.1103/PhysRevD.104.034010}{\textit{Phys. Rev. D}
  {\bfseries 104} (2021) 034010},
  [\href{https://arxiv.org/abs/2010.00555}{{\ttfamily 2010.00555}}].

\bibitem{Helenius:2021tof}
I.~Helenius, M.~Walt and W.~Vogelsang, \textit{{NNLO nuclear parton
  distribution functions with electroweak-boson production data from the LHC}},
  \href{https://doi.org/10.1103/PhysRevD.105.094031}{\textit{Phys. Rev. D}
  {\bfseries 105} (2022) 094031},
  [\href{https://arxiv.org/abs/2112.11904}{{\ttfamily 2112.11904}}].

\bibitem{Eskola:2021nhw}
K.~J. Eskola, P.~Paakkinen, H.~Paukkunen and C.~A. Salgado, \textit{{EPPS21: a
  global QCD analysis of nuclear PDFs}},
  \href{https://doi.org/10.1140/epjc/s10052-022-10359-0}{\textit{Eur. Phys. J.
  C} {\bfseries 82} (2022) 413},
  [\href{https://arxiv.org/abs/2112.12462}{{\ttfamily 2112.12462}}].

\bibitem{Eskola:2022rlm}
K.~J. Eskola, P.~Paakkinen, H.~Paukkunen and C.~A. Salgado, \textit{{Proton-PDF
  uncertainties in extracting nuclear PDFs from $W^\pm $ production in p+Pb
  collisions}},
  \href{https://doi.org/10.1140/epjc/s10052-022-10179-2}{\textit{Eur. Phys. J.
  C} {\bfseries 82} (2022) 271},
  [\href{https://arxiv.org/abs/2202.01074}{{\ttfamily 2202.01074}}].

\bibitem{Duwentaster:2022kpv}
P.~Duwent\"aster, T.~Je\v{z}o, M.~Klasen, K.~Kova\v{r}\'\i{}k, A.~Kusina, K.~F.
  Muzakka et~al., \textit{{Impact of heavy quark and quarkonium data on nuclear
  gluon PDFs}},
  \href{https://doi.org/10.1103/PhysRevD.105.114043}{\textit{Phys. Rev. D}
  {\bfseries 105} (2022) 114043},
  [\href{https://arxiv.org/abs/2204.09982}{{\ttfamily 2204.09982}}].

\bibitem{AbdulKhalek:2022fyi}
R.~Abdul~Khalek, R.~Gauld, T.~Giani, E.~R. Nocera, T.~R. Rabemananjara and
  J.~Rojo, \textit{{nNNPDF3.0: evidence for a modified partonic structure in
  heavy nuclei}},
  \href{https://doi.org/10.1140/epjc/s10052-022-10417-7}{\textit{Eur. Phys. J.
  C} {\bfseries 82} (2022) 507},
  [\href{https://arxiv.org/abs/2201.12363}{{\ttfamily 2201.12363}}].

\bibitem{nCTEQ:2023cpo}
{\scshape nCTEQ} collaboration, A.~W. Denniston et~al., \textit{{Modification
  of Quark-Gluon Distributions in Nuclei by Correlated Nucleon Pairs}},
  \href{https://doi.org/10.1103/PhysRevLett.133.152502}{\textit{Phys. Rev.
  Lett.} {\bfseries 133} (2024) 152502},
  [\href{https://arxiv.org/abs/2312.16293}{{\ttfamily 2312.16293}}].

\bibitem{Muller:2012zq}
B.~Muller, J.~Schukraft and B.~Wyslouch, \textit{{First Results from Pb+Pb
  collisions at the LHC}},
  \href{https://doi.org/10.1146/annurev-nucl-102711-094910}{\textit{Ann. Rev.
  Nucl. Part. Sci.} {\bfseries 62} (2012) 361--386},
  [\href{https://arxiv.org/abs/1202.3233}{{\ttfamily 1202.3233}}].

\bibitem{Klasen:2023uqj}
M.~Klasen and H.~Paukkunen, \textit{{Nuclear PDFs After the First Decade of LHC
  Data}},  \href{https://arxiv.org/abs/2311.00450}{{\ttfamily 2311.00450}}.

\bibitem{Achenbach:2023pba}
P.~Achenbach et~al., \textit{{The present and future of QCD}},
  \href{https://doi.org/10.1016/j.nuclphysa.2024.122874}{\textit{Nucl. Phys. A}
  {\bfseries 1047} (2024) 122874},
  [\href{https://arxiv.org/abs/2303.02579}{{\ttfamily 2303.02579}}].

\bibitem{Arslandok:2023utm}
M.~Arslandok et~al., \textit{{Hot QCD White Paper}},
  \href{https://arxiv.org/abs/2303.17254}{{\ttfamily 2303.17254}}.

\bibitem{Gomez:1993ri}
J.~Gomez et~al., \textit{{Measurement of the A-dependence of deep inelastic
  electron scattering}},
  \href{https://doi.org/10.1103/PhysRevD.49.4348}{\textit{Phys. Rev. D}
  {\bfseries 49} (1994) 4348--4372}.

\bibitem{NewMuon:1995cua}
{\scshape New Muon} collaboration, P.~Amaudruz et~al., \textit{{A Reevaluation
  of the nuclear structure function ratios for D, He, Li-6, C and Ca}},
  \href{https://doi.org/10.1016/0550-3213(94)00023-9}{\textit{Nucl. Phys. B}
  {\bfseries 441} (1995) 3--11},
  [\href{https://arxiv.org/abs/hep-ph/9503291}{{\ttfamily hep-ph/9503291}}].

\bibitem{E665:1995xur}
{\scshape E665} collaboration, M.~R. Adams et~al., \textit{{Shadowing in
  inelastic scattering of muons on carbon, calcium and lead at low x(Bj)}},
  \href{https://doi.org/10.1007/BF01624583}{\textit{Z. Phys. C} {\bfseries 67}
  (1995) 403--410}, [\href{https://arxiv.org/abs/hep-ex/9505006}{{\ttfamily
  hep-ex/9505006}}].

\bibitem{NewMuon:1995tgs}
{\scshape New Muon} collaboration, M.~Arneodo et~al., \textit{{The Structure
  Function ratios F2(li) / F2(D) and F2(C) / F2(D) at small x}},
  \href{https://doi.org/10.1016/0550-3213(95)00023-2}{\textit{Nucl. Phys. B}
  {\bfseries 441} (1995) 12--30},
  [\href{https://arxiv.org/abs/hep-ex/9504002}{{\ttfamily hep-ex/9504002}}].

\bibitem{NewMuon:1996gam}
{\scshape New Muon} collaboration, M.~Arneodo et~al., \textit{{The Q**2
  dependence of the structure function ratio F2 Sn / F2 C and the difference R
  Sn - R C in deep inelastic muon scattering}},
  \href{https://doi.org/10.1016/S0550-3213(96)90119-4}{\textit{Nucl. Phys. B}
  {\bfseries 481} (1996) 23--39}.

\bibitem{NewMuon:1996yuf}
{\scshape New Muon} collaboration, M.~Arneodo et~al., \textit{{The A dependence
  of the nuclear structure function ratios}},
  \href{https://doi.org/10.1016/S0550-3213(96)90117-0}{\textit{Nucl. Phys. B}
  {\bfseries 481} (1996) 3--22}.

\bibitem{ATLAS:2012qdj}
{\scshape ATLAS} collaboration, G.~Aad et~al., \textit{{Measurement of $Z$
  boson Production in Pb+Pb Collisions at $\sqrt{s_{NN}}=2.76$ TeV with the
  ATLAS Detector}},
  \href{https://doi.org/10.1103/PhysRevLett.110.022301}{\textit{Phys. Rev.
  Lett.} {\bfseries 110} (2013) 022301},
  [\href{https://arxiv.org/abs/1210.6486}{{\ttfamily 1210.6486}}].

\bibitem{CMS:2011zfr}
{\scshape CMS} collaboration, S.~Chatrchyan et~al., \textit{{Study of Z boson
  production in PbPb collisions at $\sqrt{s_{NN}}$ = 2.76 TeV}},
  \href{https://doi.org/10.1103/PhysRevLett.106.212301}{\textit{Phys. Rev.
  Lett.} {\bfseries 106} (2011) 212301},
  [\href{https://arxiv.org/abs/1102.5435}{{\ttfamily 1102.5435}}].

\bibitem{CMS:2012fgk}
{\scshape CMS} collaboration, S.~Chatrchyan et~al., \textit{{Study of $W$ Boson
  Production in PbPb and $pp$ Collisions at $\sqrt{s_{NN}}=2.76$ TeV}},
  \href{https://doi.org/10.1016/j.physletb.2012.07.025}{\textit{Phys. Lett. B}
  {\bfseries 715} (2012) 66--87},
  [\href{https://arxiv.org/abs/1205.6334}{{\ttfamily 1205.6334}}].

\bibitem{ATLAS:2014sic}
{\scshape ATLAS} collaboration, G.~Aad et~al., \textit{{Measurement of the
  production and lepton charge asymmetry of $W$ bosons in Pb+Pb collisions at
  $\mathbf {\sqrt{\mathbf {s}_{\mathrm {\mathbf {NN}}}}=2.76\;TeV}$ with the
  ATLAS detector}},
  \href{https://doi.org/10.1140/epjc/s10052-014-3231-6}{\textit{Eur. Phys. J.
  C} {\bfseries 75} (2015) 23},
  [\href{https://arxiv.org/abs/1408.4674}{{\ttfamily 1408.4674}}].

\bibitem{CMS:2017hnw}
{\scshape CMS} collaboration, A.~M. Sirunyan et~al., \textit{{Observation of
  top quark production in proton-nucleus collisions}},
  \href{https://doi.org/10.1103/PhysRevLett.119.242001}{\textit{Phys. Rev.
  Lett.} {\bfseries 119} (2017) 242001},
  [\href{https://arxiv.org/abs/1709.07411}{{\ttfamily 1709.07411}}].

\bibitem{CMS:2020aem}
{\scshape CMS} collaboration, A.~M. Sirunyan et~al., \textit{{Evidence for Top
  Quark Production in Nucleus-Nucleus Collisions}},
  \href{https://doi.org/10.1103/PhysRevLett.125.222001}{\textit{Phys. Rev.
  Lett.} {\bfseries 125} (2020) 222001},
  [\href{https://arxiv.org/abs/2006.11110}{{\ttfamily 2006.11110}}].

\bibitem{ATLAS:2024qdu}
{\scshape ATLAS} collaboration, G.~Aad et~al., \textit{{Observation of $
  t\overline{t} $ production in the lepton+jets and dilepton channels in p+Pb
  collisions at $ \sqrt{s_{\textrm{NN}}} $ = 8.16 TeV with the ATLAS
  detector}}, \href{https://doi.org/10.1007/JHEP11(2024)101}{\textit{JHEP}
  {\bfseries 11} (2024) 101},
  [\href{https://arxiv.org/abs/2405.05078}{{\ttfamily 2405.05078}}].

\bibitem{Stelzer:1994ta}
T.~Stelzer and W.~F. Long, \textit{{Automatic generation of tree level helicity
  amplitudes}},
  \href{https://doi.org/10.1016/0010-4655(94)90084-1}{\textit{Comput. Phys.
  Commun.} {\bfseries 81} (1994) 357--371},
  [\href{https://arxiv.org/abs/hep-ph/9401258}{{\ttfamily hep-ph/9401258}}].

\bibitem{Alwall:2014hca}
J.~Alwall, R.~Frederix, S.~Frixione, V.~Hirschi, F.~Maltoni, O.~Mattelaer
  et~al., \textit{{The automated computation of tree-level and next-to-leading
  order differential cross sections, and their matching to parton shower
  simulations}}, \href{https://doi.org/10.1007/JHEP07(2014)079}{\textit{JHEP}
  {\bfseries 07} (2014) 079},
  [\href{https://arxiv.org/abs/1405.0301}{{\ttfamily 1405.0301}}].

\bibitem{Hou:2019qau}
T.-J. Hou et~al., \textit{{Progress in the CTEQ-TEA NNLO global QCD analysis}},
   \href{https://arxiv.org/abs/1908.11394}{{\ttfamily 1908.11394}}.

\bibitem{Buckley:2014ana}
A.~Buckley, J.~Ferrando, S.~Lloyd, K.~Nordstr\"om, B.~Page, M.~R\"ufenacht
  et~al., \textit{{LHAPDF6: parton density access in the LHC precision era}},
  \href{https://doi.org/10.1140/epjc/s10052-015-3318-8}{\textit{Eur. Phys. J.
  C} {\bfseries 75} (2015) 132},
  [\href{https://arxiv.org/abs/1412.7420}{{\ttfamily 1412.7420}}].

\bibitem{Pumplin:2001ct}
J.~Pumplin, D.~Stump, R.~Brock, D.~Casey, J.~Huston, J.~Kalk et~al.,
  \textit{{Uncertainties of predictions from parton distribution functions. 2.
  The Hessian method}},
  \href{https://doi.org/10.1103/PhysRevD.65.014013}{\textit{Phys. Rev. D}
  {\bfseries 65} (2001) 014013},
  [\href{https://arxiv.org/abs/hep-ph/0101032}{{\ttfamily hep-ph/0101032}}].

\bibitem{Pettersson:1995yyq}
{\scshape LHC Study Group} collaboration, \textit{{The Large Hadron Collider:
  Conceptual design}}, .

\bibitem{Bruning:2004ej}
\textit{{LHC Design Report Vol.1: The LHC Main Ring}}, .

\bibitem{Schaumann:2018qat}
M.~Schaumann et~al., \textit{{First Xenon-Xenon Collisions in the LHC}},  in
  \textit{{9th International Particle Accelerator Conference}}, 6, 2018,
  \href{https://doi.org/10.18429/JACoW-IPAC2018-MOPMF039}{DOI}.

\bibitem{Ruiz:2023ozv}
R.~Ruiz et~al., \textit{{Target mass corrections in lepton\textendash{}nucleus
  DIS: Theory and applications to nuclear PDFs}},
  \href{https://doi.org/10.1016/j.ppnp.2023.104096}{\textit{Prog. Part. Nucl.
  Phys.} {\bfseries 136} (2024) 104096},
  [\href{https://arxiv.org/abs/2301.07715}{{\ttfamily 2301.07715}}].

\bibitem{Kutak:2003bd}
K.~Kutak and J.~Kwiecinski, \textit{{Screening effects in the ultrahigh-energy
  neutrino interactions}},
  \href{https://doi.org/10.1140/epjc/s2003-01236-y}{\textit{Eur. Phys. J. C}
  {\bfseries 29} (2003) 521},
  [\href{https://arxiv.org/abs/hep-ph/0303209}{{\ttfamily hep-ph/0303209}}].

\bibitem{Jalilian-Marian:2005ccm}
J.~Jalilian-Marian and Y.~V. Kovchegov, \textit{{Saturation physics and
  deuteron-Gold collisions at RHIC}},
  \href{https://doi.org/10.1016/j.ppnp.2005.07.002}{\textit{Prog. Part. Nucl.
  Phys.} {\bfseries 56} (2006) 104--231},
  [\href{https://arxiv.org/abs/hep-ph/0505052}{{\ttfamily hep-ph/0505052}}].

\bibitem{Albacete:2014fwa}
J.~L. Albacete and C.~Marquet, \textit{{Gluon saturation and initial conditions
  for relativistic heavy ion collisions}},
  \href{https://doi.org/10.1016/j.ppnp.2014.01.004}{\textit{Prog. Part. Nucl.
  Phys.} {\bfseries 76} (2014) 1--42},
  [\href{https://arxiv.org/abs/1401.4866}{{\ttfamily 1401.4866}}].

\bibitem{STAR:2021fgw}
{\scshape STAR} collaboration, M.~S. Abdallah et~al., \textit{{Evidence for
  Nonlinear Gluon Effects in QCD and Their Mass Number Dependence at STAR}},
  \href{https://doi.org/10.1103/PhysRevLett.129.092501}{\textit{Phys. Rev.
  Lett.} {\bfseries 129} (2022) 092501},
  [\href{https://arxiv.org/abs/2111.10396}{{\ttfamily 2111.10396}}].

\bibitem{Eskola:2009uj}
K.~J. Eskola, H.~Paukkunen and C.~A. Salgado, \textit{{EPS09: A New Generation
  of NLO and LO Nuclear Parton Distribution Functions}},
  \href{https://doi.org/10.1088/1126-6708/2009/04/065}{\textit{JHEP} {\bfseries
  04} (2009) 065}, [\href{https://arxiv.org/abs/0902.4154}{{\ttfamily
  0902.4154}}].

\bibitem{Ethier:2020way}
J.~J. Ethier and E.~R. Nocera, \textit{{Parton Distributions in Nucleons and
  Nuclei}},
  \href{https://doi.org/10.1146/annurev-nucl-011720-042725}{\textit{Ann. Rev.
  Nucl. Part. Sci.} {\bfseries 70} (2020) 43--76},
  [\href{https://arxiv.org/abs/2001.07722}{{\ttfamily 2001.07722}}].

\bibitem{BCDMS:1994ala}
{\scshape BCDMS} collaboration, A.~C. Benvenuti et~al., \textit{{Nuclear
  structure function in carbon near x = 1}},
  \href{https://doi.org/10.1007/BF01577541}{\textit{Z. Phys. C} {\bfseries 63}
  (1994) 29--36}.

\bibitem{CLAS:2003eih}
{\scshape CLAS} collaboration, K.~S. Egiyan et~al., \textit{{Observation of
  nuclear scaling in the A(e, e-prime) reaction at x(B) greater than 1}},
  \href{https://doi.org/10.1103/PhysRevC.68.014313}{\textit{Phys. Rev. C}
  {\bfseries 68} (2003) 014313},
  [\href{https://arxiv.org/abs/nucl-ex/0301008}{{\ttfamily nucl-ex/0301008}}].

\bibitem{Fomin:2010ei}
N.~Fomin et~al., \textit{{Scaling of the $F_2$ structure function in nuclei and
  quark distributions at $x>1$}},
  \href{https://doi.org/10.1103/PhysRevLett.105.212502}{\textit{Phys. Rev.
  Lett.} {\bfseries 105} (2010) 212502},
  [\href{https://arxiv.org/abs/1008.2713}{{\ttfamily 1008.2713}}].

\bibitem{Freese:2015ebu}
A.~J. Freese, W.~Cosyn and M.~M. Sargsian, \textit{{QCD evolution of superfast
  quarks}}, \href{https://doi.org/10.1103/PhysRevD.99.114019}{\textit{Phys.
  Rev. D} {\bfseries 99} (2019) 114019},
  [\href{https://arxiv.org/abs/1511.06044}{{\ttfamily 1511.06044}}].

\bibitem{Segarra:2020gtj}
E.~P. Segarra et~al., \textit{{Extending nuclear PDF analyses into the high-$x$
  , low-$Q^2$ region}},
  \href{https://doi.org/10.1103/PhysRevD.103.114015}{\textit{Phys. Rev. D}
  {\bfseries 103} (2021) 114015},
  [\href{https://arxiv.org/abs/2012.11566}{{\ttfamily 2012.11566}}].

\bibitem{Schienbein:2007gr}
I.~Schienbein et~al., \textit{{A Review of Target Mass Corrections}},
  \href{https://doi.org/10.1088/0954-3899/35/5/053101}{\textit{J. Phys. G}
  {\bfseries 35} (2008) 053101},
  [\href{https://arxiv.org/abs/0709.1775}{{\ttfamily 0709.1775}}].

\bibitem{Collins:1984kg}
J.~C. Collins, D.~E. Soper and G.~F. Sterman, \textit{{Transverse Momentum
  Distribution in Drell-Yan Pair and W and Z Boson Production}},
  \href{https://doi.org/10.1016/0550-3213(85)90479-1}{\textit{Nucl. Phys. B}
  {\bfseries 250} (1985) 199--224}.

\bibitem{Collins:1985ue}
J.~C. Collins, D.~E. Soper and G.~F. Sterman, \textit{{Factorization for Short
  Distance Hadron - Hadron Scattering}},
  \href{https://doi.org/10.1016/0550-3213(85)90565-6}{\textit{Nucl. Phys. B}
  {\bfseries 261} (1985) 104--142}.

\bibitem{Collins:2011zzd}
J.~Collins, \textit{{Foundations of Perturbative QCD}}, vol.~32.
\newblock Cambridge University Press, 2011,
  \href{https://doi.org/10.1017/9781009401845}{10.1017/9781009401845}.

\bibitem{ATLASCollaboration:2012ilu}
\textit{{Letter of Intent for the Phase-II Upgrade of the ATLAS Experiment}}, .

\bibitem{Contardo:2015bmq}
\textit{{Technical Proposal for the Phase-II Upgrade of the CMS Detector}}, .

\bibitem{Bhattacharya:2015ada}
L.~Bhattacharya, R.~Ryblewski and M.~Strickland, \textit{{Photon production
  from a nonequilibrium quark-gluon plasma}},
  \href{https://doi.org/10.1103/PhysRevD.93.065005}{\textit{Phys. Rev. D}
  {\bfseries 93} (2016) 065005},
  [\href{https://arxiv.org/abs/1507.06605}{{\ttfamily 1507.06605}}].

\bibitem{Berges:2017eom}
J.~Berges, K.~Reygers, N.~Tanji and R.~Venugopalan, \textit{{Parametric
  estimate of the relative photon yields from the glasma and the quark-gluon
  plasma in heavy-ion collisions}},
  \href{https://doi.org/10.1103/PhysRevC.95.054904}{\textit{Phys. Rev. C}
  {\bfseries 95} (2017) 054904},
  [\href{https://arxiv.org/abs/1701.05064}{{\ttfamily 1701.05064}}].

\bibitem{Frixione:1998jh}
S.~Frixione, \textit{{Isolated photons in perturbative QCD}},
  \href{https://doi.org/10.1016/S0370-2693(98)00454-7}{\textit{Phys. Lett. B}
  {\bfseries 429} (1998) 369--374},
  [\href{https://arxiv.org/abs/hep-ph/9801442}{{\ttfamily hep-ph/9801442}}].

\bibitem{Altarelli:1978id}
G.~Altarelli, R.~K. Ellis and G.~Martinelli, \textit{{Leptoproduction and
  Drell-Yan Processes Beyond the Leading Approximation in Chromodynamics}},
  \href{https://doi.org/10.1016/0550-3213(78)90067-6}{\textit{Nucl. Phys. B}
  {\bfseries 143} (1978) 521}. [Erratum: Nucl.Phys.B 146, 544 (1978)].

\bibitem{Altarelli:1979ub}
G.~Altarelli, R.~K. Ellis and G.~Martinelli, \textit{{Large Perturbative
  Corrections to the Drell-Yan Process in QCD}},
  \href{https://doi.org/10.1016/0550-3213(79)90116-0}{\textit{Nucl. Phys. B}
  {\bfseries 157} (1979) 461--497}.

\bibitem{Hamberg:1990np}
R.~Hamberg, W.~L. van Neerven and T.~Matsuura, \textit{{A complete calculation
  of the order $\alpha-s^{2}$ correction to the Drell-Yan $K$ factor}},
  \href{https://doi.org/10.1016/0550-3213(91)90064-5}{\textit{Nucl. Phys. B}
  {\bfseries 359} (1991) 343--405}. [Erratum: Nucl.Phys.B 644, 403--404
  (2002)].

\bibitem{Harlander:2002wh}
R.~V. Harlander and W.~B. Kilgore, \textit{{Next-to-next-to-leading order Higgs
  production at hadron colliders}},
  \href{https://doi.org/10.1103/PhysRevLett.88.201801}{\textit{Phys. Rev.
  Lett.} {\bfseries 88} (2002) 201801},
  [\href{https://arxiv.org/abs/hep-ph/0201206}{{\ttfamily hep-ph/0201206}}].

\bibitem{Duhr:2020seh}
C.~Duhr, F.~Dulat and B.~Mistlberger, \textit{{Drell-Yan Cross Section to Third
  Order in the Strong Coupling Constant}},
  \href{https://doi.org/10.1103/PhysRevLett.125.172001}{\textit{Phys. Rev.
  Lett.} {\bfseries 125} (2020) 172001},
  [\href{https://arxiv.org/abs/2001.07717}{{\ttfamily 2001.07717}}].

\bibitem{Duhr:2020sdp}
C.~Duhr, F.~Dulat and B.~Mistlberger, \textit{{Charged current Drell-Yan
  production at N$^{3}$LO}},
  \href{https://doi.org/10.1007/JHEP11(2020)143}{\textit{JHEP} {\bfseries 11}
  (2020) 143}, [\href{https://arxiv.org/abs/2007.13313}{{\ttfamily
  2007.13313}}].

\bibitem{CMS:2014dyj}
{\scshape CMS} collaboration, S.~Chatrchyan et~al., \textit{{Study of Z
  production in PbPb and pp collisions at $ \sqrt{s_{\mathrm{NN}}}=2.76 $ TeV
  in the dimuon and dielectron decay channels}},
  \href{https://doi.org/10.1007/JHEP03(2015)022}{\textit{JHEP} {\bfseries 03}
  (2015) 022}, [\href{https://arxiv.org/abs/1410.4825}{{\ttfamily 1410.4825}}].

\bibitem{CMS:2021otx}
{\scshape CMS} collaboration, A.~M. Sirunyan et~al., \textit{{Using Z Boson
  Events to Study Parton-Medium Interactions in Pb-Pb Collisions}},
  \href{https://doi.org/10.1103/PhysRevLett.128.122301}{\textit{Phys. Rev.
  Lett.} {\bfseries 128} (2022) 122301},
  [\href{https://arxiv.org/abs/2103.04377}{{\ttfamily 2103.04377}}].

\bibitem{CMS:2021kvd}
{\scshape CMS} collaboration, A.~M. Sirunyan et~al., \textit{{Constraints on
  the Initial State of Pb-Pb Collisions via Measurements of $Z$-Boson Yields
  and Azimuthal Anisotropy at $\sqrt {s_{NN}}$=5.02\,\,TeV}},
  \href{https://doi.org/10.1103/PhysRevLett.127.102002}{\textit{Phys. Rev.
  Lett.} {\bfseries 127} (2021) 102002},
  [\href{https://arxiv.org/abs/2103.14089}{{\ttfamily 2103.14089}}].

\bibitem{ATLAS:2015mwq}
{\scshape ATLAS} collaboration, G.~Aad et~al., \textit{{$Z$ boson production in
  $p+$Pb collisions at $\sqrt{s_{NN}}=5.02$ TeV measured with the ATLAS
  detector}}, \href{https://doi.org/10.1103/PhysRevC.92.044915}{\textit{Phys.
  Rev. C} {\bfseries 92} (2015) 044915},
  [\href{https://arxiv.org/abs/1507.06232}{{\ttfamily 1507.06232}}].

\bibitem{CMS:2015ehw}
{\scshape CMS} collaboration, V.~Khachatryan et~al., \textit{{Study of W boson
  production in pPb collisions at $\sqrt{s_{\mathrm{NN}}} =$ 5.02 TeV}},
  \href{https://doi.org/10.1016/j.physletb.2015.09.057}{\textit{Phys. Lett. B}
  {\bfseries 750} (2015) 565--586},
  [\href{https://arxiv.org/abs/1503.05825}{{\ttfamily 1503.05825}}].

\bibitem{CMS:2015zlj}
{\scshape CMS} collaboration, V.~Khachatryan et~al., \textit{{Study of Z boson
  production in pPb collisions at $\sqrt {s_{NN}} = 5.02$ TeV}},
  \href{https://doi.org/10.1016/j.physletb.2016.05.044}{\textit{Phys. Lett. B}
  {\bfseries 759} (2016) 36--57},
  [\href{https://arxiv.org/abs/1512.06461}{{\ttfamily 1512.06461}}].

\bibitem{ALICE:2016rzo}
{\scshape ALICE} collaboration, J.~Adam et~al., \textit{{W and Z boson
  production in p-Pb collisions at $\sqrt{s_{\rm NN}}$ = 5.02 TeV}},
  \href{https://doi.org/10.1007/JHEP02(2017)077}{\textit{JHEP} {\bfseries 02}
  (2017) 077}, [\href{https://arxiv.org/abs/1611.03002}{{\ttfamily
  1611.03002}}].

\bibitem{CMS:2019leu}
{\scshape CMS} collaboration, A.~M. Sirunyan et~al., \textit{{Observation of
  nuclear modifications in W$^\pm$ boson production in pPb collisions at
  $\sqrt{s_\mathrm{NN}} =$ 8.16 TeV}},
  \href{https://doi.org/10.1016/j.physletb.2019.135048}{\textit{Phys. Lett. B}
  {\bfseries 800} (2020) 135048},
  [\href{https://arxiv.org/abs/1905.01486}{{\ttfamily 1905.01486}}].

\bibitem{Denniston:2023dwd}
{\scshape nCTEQ} collaboration, A.~W. Denniston et~al., \textit{{Modification
  of Quark-Gluon Distributions in Nuclei by Correlated Nucleon Pairs}},
  \href{https://doi.org/10.1103/PhysRevLett.133.152502}{\textit{Phys. Rev.
  Lett.} {\bfseries 133} (2024) 152502},
  [\href{https://arxiv.org/abs/2312.16293}{{\ttfamily 2312.16293}}].

\bibitem{Ohnemus:1990za}
J.~Ohnemus and J.~F. Owens, \textit{{An Order $\alpha^- s$ calculation of
  hadronic $Z Z$ production}},
  \href{https://doi.org/10.1103/PhysRevD.43.3626}{\textit{Phys. Rev. D}
  {\bfseries 43} (1991) 3626--3639}.

\bibitem{Mele:1990bq}
B.~Mele, P.~Nason and G.~Ridolfi, \textit{{QCD radiative corrections to Z boson
  pair production in hadronic collisions}},
  \href{https://doi.org/10.1016/0550-3213(91)90475-D}{\textit{Nucl. Phys. B}
  {\bfseries 357} (1991) 409--438}.

\bibitem{Ohnemus:1991kk}
J.~Ohnemus, \textit{{An Order $\alpha^- s$ calculation of hadronic $W^{-}
  W^{+}$ production}},
  \href{https://doi.org/10.1103/PhysRevD.44.1403}{\textit{Phys. Rev. D}
  {\bfseries 44} (1991) 1403--1414}.

\bibitem{Ohnemus:1991gb}
J.~Ohnemus, \textit{{An Order $\alpha^- s$ calculation of hadronic $W^\pm Z$
  production}}, \href{https://doi.org/10.1103/PhysRevD.44.3477}{\textit{Phys.
  Rev. D} {\bfseries 44} (1991) 3477--3489}.

\bibitem{Frixione:1992pj}
S.~Frixione, P.~Nason and G.~Ridolfi, \textit{{Strong corrections to W Z
  production at hadron colliders}},
  \href{https://doi.org/10.1016/0550-3213(92)90668-2}{\textit{Nucl. Phys. B}
  {\bfseries 383} (1992) 3--44}.

\bibitem{Frixione:1993yp}
S.~Frixione, \textit{{A Next-to-leading order calculation of the cross-section
  for the production of W+ W- pairs in hadronic collisions}},
  \href{https://doi.org/10.1016/0550-3213(93)90435-R}{\textit{Nucl. Phys. B}
  {\bfseries 410} (1993) 280--324}.

\bibitem{Gehrmann:2014fva}
T.~Gehrmann, M.~Grazzini, S.~Kallweit, P.~Maierh\"ofer, A.~von Manteuffel,
  S.~Pozzorini et~al., \textit{{$W^+W^-$ Production at Hadron Colliders in Next
  to Next to Leading Order QCD}},
  \href{https://doi.org/10.1103/PhysRevLett.113.212001}{\textit{Phys. Rev.
  Lett.} {\bfseries 113} (2014) 212001},
  [\href{https://arxiv.org/abs/1408.5243}{{\ttfamily 1408.5243}}].

\bibitem{Cascioli:2014yka}
F.~Cascioli, T.~Gehrmann, M.~Grazzini, S.~Kallweit, P.~Maierh\"ofer, A.~von
  Manteuffel et~al., \textit{{ZZ production at hadron colliders in NNLO QCD}},
  \href{https://doi.org/10.1016/j.physletb.2014.06.056}{\textit{Phys. Lett. B}
  {\bfseries 735} (2014) 311--313},
  [\href{https://arxiv.org/abs/1405.2219}{{\ttfamily 1405.2219}}].

\bibitem{Grazzini:2016swo}
M.~Grazzini, S.~Kallweit, D.~Rathlev and M.~Wiesemann, \textit{{$W^{\pm}Z$
  production at hadron colliders in NNLO QCD}},
  \href{https://doi.org/10.1016/j.physletb.2016.08.017}{\textit{Phys. Lett. B}
  {\bfseries 761} (2016) 179--183},
  [\href{https://arxiv.org/abs/1604.08576}{{\ttfamily 1604.08576}}].

\bibitem{Mikaelian:1977ux}
K.~O. Mikaelian, \textit{{Photoproduction of Charged Intermediate Vector
  Bosons}}, \href{https://doi.org/10.1103/PhysRevD.17.750}{\textit{Phys. Rev.
  D} {\bfseries 17} (1978) 750}.

\bibitem{Brown:1979ux}
R.~W. Brown, D.~Sahdev and K.~O. Mikaelian, \textit{{W+- Z0 and W+- gamma Pair
  Production in Neutrino e, p p, and anti-p p Collisions}},
  \href{https://doi.org/10.1103/PhysRevD.20.1164}{\textit{Phys. Rev. D}
  {\bfseries 20} (1979) 1164}.

\bibitem{Mikaelian:1979nr}
K.~O. Mikaelian, M.~A. Samuel and D.~Sahdev, \textit{{The Magnetic Moment of
  Weak Bosons Produced in p p and p anti-p Collisions}},
  \href{https://doi.org/10.1103/PhysRevLett.43.746}{\textit{Phys. Rev. Lett.}
  {\bfseries 43} (1979) 746}.

\bibitem{Zhu:1980sz}
D.-p. Zhu, \textit{{Zeros in Scattering Amplitudes and the Structure of
  Nonabelian Gauge Theories}},
  \href{https://doi.org/10.1103/PhysRevD.22.2266}{\textit{Phys. Rev. D}
  {\bfseries 22} (1980) 2266}.

\bibitem{Brodsky:1982sh}
S.~J. Brodsky and R.~W. Brown, \textit{{Zeros in Amplitudes: Gauge Theory and
  Radiation Interference}},
  \href{https://doi.org/10.1103/PhysRevLett.49.966}{\textit{Phys. Rev. Lett.}
  {\bfseries 49} (1982) 966}.

\bibitem{Brown:1982xx}
R.~W. Brown, K.~L. Kowalski and S.~J. Brodsky, \textit{{Classical Radiation
  Zeros in Gauge Theory Amplitudes}},
  \href{https://doi.org/10.1103/PhysRevD.29.2100}{\textit{Phys. Rev. D}
  {\bfseries 28} (1983) 624}. [Addendum: Phys.Rev.D 29, 2100--2104 (1984)].

\bibitem{Ohnemus:1992jn}
J.~Ohnemus, \textit{{Order $\alpha^- s$ calculations of hadronic $W^\pm \gamma$
  and $Z \gamma$ production}},
  \href{https://doi.org/10.1103/PhysRevD.47.940}{\textit{Phys. Rev. D}
  {\bfseries 47} (1993) 940--955}.

\bibitem{Baur:1997kz}
U.~Baur, T.~Han and J.~Ohnemus, \textit{{QCD corrections and anomalous
  couplings in $Z \gamma$ production at hadron colliders}},
  \href{https://doi.org/10.1103/PhysRevD.57.2823}{\textit{Phys. Rev. D}
  {\bfseries 57} (1998) 2823--2836},
  [\href{https://arxiv.org/abs/hep-ph/9710416}{{\ttfamily hep-ph/9710416}}].

\bibitem{Grazzini:2015nwa}
M.~Grazzini, S.~Kallweit and D.~Rathlev, \textit{{$W\gamma$ and $Z\gamma$
  production at the LHC in NNLO QCD}},
  \href{https://doi.org/10.1007/JHEP07(2015)085}{\textit{JHEP} {\bfseries 07}
  (2015) 085}, [\href{https://arxiv.org/abs/1504.01330}{{\ttfamily
  1504.01330}}].

\bibitem{Han:1991ia}
T.~Han and S.~Willenbrock, \textit{{QCD correction to the p p ---\ensuremath{>}
  W H and Z H total cross-sections}},
  \href{https://doi.org/10.1016/0370-2693(91)90572-8}{\textit{Phys. Lett. B}
  {\bfseries 273} (1991) 167--172}.

\bibitem{Brein:2003wg}
O.~Brein, A.~Djouadi and R.~Harlander, \textit{{NNLO QCD corrections to the
  Higgs-strahlung processes at hadron colliders}},
  \href{https://doi.org/10.1016/j.physletb.2003.10.112}{\textit{Phys. Lett. B}
  {\bfseries 579} (2004) 149--156},
  [\href{https://arxiv.org/abs/hep-ph/0307206}{{\ttfamily hep-ph/0307206}}].

\bibitem{Baglio:2022wzu}
J.~Baglio, C.~Duhr, B.~Mistlberger and R.~Szafron, \textit{{Inclusive
  production cross sections at N$^{3}$LO}},
  \href{https://doi.org/10.1007/JHEP12(2022)066}{\textit{JHEP} {\bfseries 12}
  (2022) 066}, [\href{https://arxiv.org/abs/2209.06138}{{\ttfamily
  2209.06138}}].

\bibitem{Berger:2018mtg}
E.~L. Berger, J.~Gao, A.~Jueid and H.~Zhang, \textit{{Production and hadronic
  decays of Higgs bosons in heavy ion collisions}},
  \href{https://doi.org/10.1103/PhysRevLett.122.041803}{\textit{Phys. Rev.
  Lett.} {\bfseries 122} (2019) 041803},
  [\href{https://arxiv.org/abs/1804.06858}{{\ttfamily 1804.06858}}].

\bibitem{dEnterria:2019cps}
D.~d'Enterria, \textit{{Higgs boson production in partonic and electromagnetic
  interactions with heavy ions}},  in \textit{{54th Rencontres de Moriond on
  QCD and High Energy Interactions}}, pp.~25--29, ARISF, 6, 2019,
  \href{https://arxiv.org/abs/1906.07536}{{\ttfamily 1906.07536}}.

\bibitem{Lazopoulos:2007ix}
A.~Lazopoulos, K.~Melnikov and F.~Petriello, \textit{{QCD corrections to
  tri-boson production}},
  \href{https://doi.org/10.1103/PhysRevD.76.014001}{\textit{Phys. Rev. D}
  {\bfseries 76} (2007) 014001},
  [\href{https://arxiv.org/abs/hep-ph/0703273}{{\ttfamily hep-ph/0703273}}].

\bibitem{Hankele:2007sb}
V.~Hankele and D.~Zeppenfeld, \textit{{QCD corrections to hadronic WWZ
  production with leptonic decays}},
  \href{https://doi.org/10.1016/j.physletb.2008.02.014}{\textit{Phys. Lett. B}
  {\bfseries 661} (2008) 103--108},
  [\href{https://arxiv.org/abs/0712.3544}{{\ttfamily 0712.3544}}].

\bibitem{Binoth:2008kt}
T.~Binoth, G.~Ossola, C.~G. Papadopoulos and R.~Pittau, \textit{{NLO QCD
  corrections to tri-boson production}},
  \href{https://doi.org/10.1088/1126-6708/2008/06/082}{\textit{JHEP} {\bfseries
  06} (2008) 082}, [\href{https://arxiv.org/abs/0804.0350}{{\ttfamily
  0804.0350}}].

\bibitem{Bozzi:2009ig}
G.~Bozzi, F.~Campanario, V.~Hankele and D.~Zeppenfeld, \textit{{NLO QCD
  corrections to W+W- gamma and Z Z gamma production with leptonic decays}},
  \href{https://doi.org/10.1103/PhysRevD.81.094030}{\textit{Phys. Rev. D}
  {\bfseries 81} (2010) 094030},
  [\href{https://arxiv.org/abs/0911.0438}{{\ttfamily 0911.0438}}].

\bibitem{Baglio:2011juf}
J.~Baglio et~al., \textit{{VBFNLO: A parton level Monte Carlo for processes
  with electroweak bosons -- Manual for Version 3.0}},
  \href{https://arxiv.org/abs/1107.4038}{{\ttfamily 1107.4038}}.

\bibitem{Baur:2010zf}
U.~Baur, D.~Wackeroth and M.~M. Weber, \textit{{Radiative corrections to W
  gamma gamma production at the LHC}},
  \href{https://doi.org/10.22323/1.092.0067}{\textit{PoS} {\bfseries
  RADCOR2009} (2010) 067}, [\href{https://arxiv.org/abs/1001.2688}{{\ttfamily
  1001.2688}}].

\bibitem{Bozzi:2010sj}
G.~Bozzi, F.~Campanario, M.~Rauch, H.~Rzehak and D.~Zeppenfeld, \textit{{NLO
  QCD corrections to $W^\pm Z\gamma$ production with leptonic decays}},
  \href{https://doi.org/10.1016/j.physletb.2010.12.051}{\textit{Phys. Lett. B}
  {\bfseries 696} (2011) 380--385},
  [\href{https://arxiv.org/abs/1011.2206}{{\ttfamily 1011.2206}}].

\bibitem{Bozzi:2011wwa}
G.~Bozzi, F.~Campanario, M.~Rauch and D.~Zeppenfeld, \textit{{$W^{+-}\gamma
  \gamma$ production with leptonic decays at NLO QCD}},
  \href{https://doi.org/10.1103/PhysRevD.83.114035}{\textit{Phys. Rev. D}
  {\bfseries 83} (2011) 114035},
  [\href{https://arxiv.org/abs/1103.4613}{{\ttfamily 1103.4613}}].

\bibitem{Bozzi:2011en}
G.~Bozzi, F.~Campanario, M.~Rauch and D.~Zeppenfeld, \textit{{$Z \gamma\gamma$
  production with leptonic decays and triple photon production at
  next-to-leading order QCD}},
  \href{https://doi.org/10.1103/PhysRevD.84.074028}{\textit{Phys. Rev. D}
  {\bfseries 84} (2011) 074028},
  [\href{https://arxiv.org/abs/1107.3149}{{\ttfamily 1107.3149}}].

\bibitem{Song:2013sex}
M.~Song, N.~Wan, G.~Li, W.-G. Ma, R.-Y. Zhang, L.~Guo et~al.,
  \textit{{Next-to-leading order QCD corrections to $HW^{\pm}\gamma$ production
  at the LHC}}, \href{https://doi.org/10.1103/PhysRevD.88.076002}{\textit{Phys.
  Rev. D} {\bfseries 88} (2013) 076002},
  [\href{https://arxiv.org/abs/1310.0946}{{\ttfamily 1310.0946}}].

\bibitem{Frederix:2014hta}
R.~Frederix, S.~Frixione, V.~Hirschi, F.~Maltoni, O.~Mattelaer, P.~Torrielli
  et~al., \textit{{Higgs pair production at the LHC with NLO and parton-shower
  effects}},
  \href{https://doi.org/10.1016/j.physletb.2014.03.026}{\textit{Phys. Lett. B}
  {\bfseries 732} (2014) 142--149},
  [\href{https://arxiv.org/abs/1401.7340}{{\ttfamily 1401.7340}}].

\bibitem{Ciafaloni:1998xg}
P.~Ciafaloni and D.~Comelli, \textit{{Sudakov enhancement of electroweak
  corrections}},
  \href{https://doi.org/10.1016/S0370-2693(98)01541-X}{\textit{Phys. Lett. B}
  {\bfseries 446} (1999) 278--284},
  [\href{https://arxiv.org/abs/hep-ph/9809321}{{\ttfamily hep-ph/9809321}}].

\bibitem{Beccaria:1998qe}
M.~Beccaria, G.~Montagna, F.~Piccinini, F.~M. Renard and C.~Verzegnassi,
  \textit{{Rising bosonic electroweak virtual effects at high-energy e+ e-
  colliders}}, \href{https://doi.org/10.1103/PhysRevD.58.093014}{\textit{Phys.
  Rev. D} {\bfseries 58} (1998) 093014},
  [\href{https://arxiv.org/abs/hep-ph/9805250}{{\ttfamily hep-ph/9805250}}].

\bibitem{Ciafaloni:2000df}
M.~Ciafaloni, P.~Ciafaloni and D.~Comelli, \textit{{Bloch-Nordsieck violating
  electroweak corrections to inclusive TeV scale hard processes}},
  \href{https://doi.org/10.1103/PhysRevLett.84.4810}{\textit{Phys. Rev. Lett.}
  {\bfseries 84} (2000) 4810--4813},
  [\href{https://arxiv.org/abs/hep-ph/0001142}{{\ttfamily hep-ph/0001142}}].

\bibitem{Melles:2001ye}
M.~Melles, \textit{{Electroweak radiative corrections in high-energy
  processes}},
  \href{https://doi.org/10.1016/S0370-1573(02)00550-1}{\textit{Phys. Rept.}
  {\bfseries 375} (2003) 219--326},
  [\href{https://arxiv.org/abs/hep-ph/0104232}{{\ttfamily hep-ph/0104232}}].

\bibitem{Bauer:2016kkv}
C.~W. Bauer and N.~Ferland, \textit{{Resummation of electroweak Sudakov
  logarithms for real radiation}},
  \href{https://doi.org/10.1007/JHEP09(2016)025}{\textit{JHEP} {\bfseries 09}
  (2016) 025}, [\href{https://arxiv.org/abs/1601.07190}{{\ttfamily
  1601.07190}}].

\bibitem{Chen:2016wkt}
J.~Chen, T.~Han and B.~Tweedie, \textit{{Electroweak Splitting Functions and
  High Energy Showering}},
  \href{https://doi.org/10.1007/JHEP11(2017)093}{\textit{JHEP} {\bfseries 11}
  (2017) 093}, [\href{https://arxiv.org/abs/1611.00788}{{\ttfamily
  1611.00788}}].

\bibitem{Baur:1997bn}
U.~Baur, T.~Han, N.~Kauer, R.~Sobey and D.~Zeppenfeld, \textit{{$W \gamma
  \gamma$ production at the Fermilab Tevatron collider: Gauge invariance and
  radiation amplitude zero}},
  \href{https://doi.org/10.1103/PhysRevD.56.140}{\textit{Phys. Rev. D}
  {\bfseries 56} (1997) 140--150},
  [\href{https://arxiv.org/abs/hep-ph/9702364}{{\ttfamily hep-ph/9702364}}].

\bibitem{Bell:2009vh}
P.~J. Bell, \textit{{Quartic Gauge Couplings and the Radiation Zero in pp
  ---\ensuremath{>} l+- nu gamma gamma events at the LHC}},
  \href{https://doi.org/10.1140/epjc/s10052-009-1129-5}{\textit{Eur. Phys. J.
  C} {\bfseries 64} (2009) 25--33},
  [\href{https://arxiv.org/abs/0907.5299}{{\ttfamily 0907.5299}}].

\bibitem{Berger:1983yi}
E.~L. Berger, E.~Braaten and R.~D. Field, \textit{{Large p(T) Production of
  Single and Double Photons in Proton Proton and Pion-Proton Collisions}},
  \href{https://doi.org/10.1016/0550-3213(84)90084-1}{\textit{Nucl. Phys. B}
  {\bfseries 239} (1984) 52--92}.

\bibitem{Field:1989uq}
R.~D. Field, \textit{{Applications of Perturbative QCD}}, vol.~77.
\newblock 1989.

\bibitem{Baer:1990ra}
H.~Baer, J.~Ohnemus and J.~F. Owens, \textit{{A Next-to-leading Logarithm
  Calculation of Direct Photon Production}},
  \href{https://doi.org/10.1103/PhysRevD.42.61}{\textit{Phys. Rev. D}
  {\bfseries 42} (1990) 61--71}.

\bibitem{Gordon:1994ut}
L.~E. Gordon and W.~Vogelsang, \textit{{Polarized and unpolarized isolated
  prompt photon production beyond the leading order}},
  \href{https://doi.org/10.1103/PhysRevD.50.1901}{\textit{Phys. Rev. D}
  {\bfseries 50} (1994) 1901--1916}.

\bibitem{Binoth:1999qq}
T.~Binoth, J.~P. Guillet, E.~Pilon and M.~Werlen, \textit{{A Full
  next-to-leading order study of direct photon pair production in hadronic
  collisions}}, \href{https://doi.org/10.1007/s100520050024}{\textit{Eur. Phys.
  J. C} {\bfseries 16} (2000) 311--330},
  [\href{https://arxiv.org/abs/hep-ph/9911340}{{\ttfamily hep-ph/9911340}}].

\bibitem{Catani:2002ny}
S.~Catani, M.~Fontannaz, J.~P. Guillet and E.~Pilon, \textit{{Cross-section of
  isolated prompt photons in hadron hadron collisions}},
  \href{https://doi.org/10.1088/1126-6708/2002/05/028}{\textit{JHEP} {\bfseries
  05} (2002) 028}, [\href{https://arxiv.org/abs/hep-ph/0204023}{{\ttfamily
  hep-ph/0204023}}].

\bibitem{Mandal:2014vpa}
M.~K. Mandal, P.~Mathews, V.~Ravindran and S.~Seth, \textit{{Three photon
  production to NLO+PS accuracy at the LHC}},
  \href{https://doi.org/10.1140/epjc/s10052-014-3044-7}{\textit{Eur. Phys. J.
  C} {\bfseries 74} (2014) 3044},
  [\href{https://arxiv.org/abs/1403.2917}{{\ttfamily 1403.2917}}].

\bibitem{CMS:2012oiv}
{\scshape CMS} collaboration, S.~Chatrchyan et~al., \textit{{Measurement of
  Isolated Photon Production in $pp$ and PbPb Collisions at
  $\sqrt{s_{NN}}=2.76$ TeV}},
  \href{https://doi.org/10.1016/j.physletb.2012.02.077}{\textit{Phys. Lett. B}
  {\bfseries 710} (2012) 256--277},
  [\href{https://arxiv.org/abs/1201.3093}{{\ttfamily 1201.3093}}].

\bibitem{ATLAS:2015rlt}
{\scshape ATLAS} collaboration, G.~Aad et~al., \textit{{Centrality, rapidity
  and transverse momentum dependence of isolated prompt photon production in
  lead-lead collisions at $\sqrt{s_{\mathrm{NN}}} = 2.76$ TeV measured with the
  ATLAS detector}},
  \href{https://doi.org/10.1103/PhysRevC.93.034914}{\textit{Phys. Rev. C}
  {\bfseries 93} (2016) 034914},
  [\href{https://arxiv.org/abs/1506.08552}{{\ttfamily 1506.08552}}].

\bibitem{CMS:2020oen}
{\scshape CMS} collaboration, A.~M. Sirunyan et~al., \textit{{The production of
  isolated photons in PbPb and pp collisions at $\sqrt{s_\mathrm{NN}} =$ 5.02
  TeV}}, \href{https://doi.org/10.1007/JHEP07(2020)116}{\textit{JHEP}
  {\bfseries 07} (2020) 116},
  [\href{https://arxiv.org/abs/2003.12797}{{\ttfamily 2003.12797}}].

\bibitem{ALICE:2023ode}
{\scshape ALICE} collaboration, S.~Acharya et~al., \textit{{Inclusive photon
  production at forward rapidities in pp and p\textendash{}Pb collisions at
  $\sqrt{{{s}}_{\textrm{NN}}}={5.02}$~TeV}},
  \href{https://doi.org/10.1140/epjc/s10052-023-11729-y}{\textit{Eur. Phys. J.
  C} {\bfseries 83} (2023) 661},
  [\href{https://arxiv.org/abs/2303.00590}{{\ttfamily 2303.00590}}].

\bibitem{ATLAS:2019ery}
{\scshape ATLAS} collaboration, M.~Aaboud et~al., \textit{{Measurement of
  prompt photon production in $\sqrt{s_\mathrm{NN}} = 8.16$ TeV $p$+Pb
  collisions with ATLAS}},
  \href{https://doi.org/10.1016/j.physletb.2019.07.031}{\textit{Phys. Lett. B}
  {\bfseries 796} (2019) 230--252},
  [\href{https://arxiv.org/abs/1903.02209}{{\ttfamily 1903.02209}}].

\bibitem{Catani:2011qz}
S.~Catani, L.~Cieri, D.~de~Florian, G.~Ferrera and M.~Grazzini,
  \textit{{Diphoton production at hadron colliders: a fully-differential QCD
  calculation at NNLO}},
  \href{https://doi.org/10.1103/PhysRevLett.108.072001}{\textit{Phys. Rev.
  Lett.} {\bfseries 108} (2012) 072001},
  [\href{https://arxiv.org/abs/1110.2375}{{\ttfamily 1110.2375}}]. [Erratum:
  Phys.Rev.Lett. 117, 089901 (2016)].

\bibitem{Campbell:2016yrh}
J.~M. Campbell, R.~K. Ellis, Y.~Li and C.~Williams, \textit{{Predictions for
  diphoton production at the LHC through NNLO in QCD}},
  \href{https://doi.org/10.1007/JHEP07(2016)148}{\textit{JHEP} {\bfseries 07}
  (2016) 148}, [\href{https://arxiv.org/abs/1603.02663}{{\ttfamily
  1603.02663}}].

\bibitem{Campbell:2016lzl}
J.~M. Campbell, R.~K. Ellis and C.~Williams, \textit{{Direct Photon Production
  at Next-to\textendash{}Next-to-Leading Order}},
  \href{https://doi.org/10.1103/PhysRevLett.118.222001}{\textit{Phys. Rev.
  Lett.} {\bfseries 118} (2017) 222001},
  [\href{https://arxiv.org/abs/1612.04333}{{\ttfamily 1612.04333}}]. [Erratum:
  Phys.Rev.Lett. 124, 259901 (2020)].

\bibitem{Chen:2019zmr}
X.~Chen, T.~Gehrmann, N.~Glover, M.~H\"ofer and A.~Huss, \textit{{Isolated
  photon and photon+jet production at NNLO QCD accuracy}},
  \href{https://doi.org/10.1007/JHEP04(2020)166}{\textit{JHEP} {\bfseries 04}
  (2020) 166}, [\href{https://arxiv.org/abs/1904.01044}{{\ttfamily
  1904.01044}}].

\bibitem{ATLAS:2017cvh}
{\scshape ATLAS} collaboration, M.~Aaboud et~al., \textit{{Measurements of
  integrated and differential cross sections for isolated photon pair
  production in $pp$ collisions at $\sqrt{s}=8$ TeV with the ATLAS detector}},
  \href{https://doi.org/10.1103/PhysRevD.95.112005}{\textit{Phys. Rev. D}
  {\bfseries 95} (2017) 112005},
  [\href{https://arxiv.org/abs/1704.03839}{{\ttfamily 1704.03839}}].

\bibitem{CMS:2018dqv}
{\scshape CMS} collaboration, A.~M. Sirunyan et~al., \textit{{Search for
  physics beyond the standard model in high-mass diphoton events from
  proton-proton collisions at $\sqrt{s} =$ 13 TeV}},
  \href{https://doi.org/10.1103/PhysRevD.98.092001}{\textit{Phys. Rev. D}
  {\bfseries 98} (2018) 092001},
  [\href{https://arxiv.org/abs/1809.00327}{{\ttfamily 1809.00327}}].

\bibitem{CMS:2019jlq}
{\scshape CMS} collaboration, A.~M. Sirunyan et~al., \textit{{Measurements of
  triple-differential cross sections for inclusive isolated-photon+jet events
  in pp collisions at $\sqrt{s} = 8\,\text {TeV} $}},
  \href{https://doi.org/10.1140/epjc/s10052-019-7451-7}{\textit{Eur. Phys. J.
  C} {\bfseries 79} (2019) 969},
  [\href{https://arxiv.org/abs/1907.08155}{{\ttfamily 1907.08155}}].

\bibitem{ATLAS:2019buk}
{\scshape ATLAS} collaboration, G.~Aad et~al., \textit{{Measurement of the
  inclusive isolated-photon cross section in $pp$ collisions at $\sqrt{s}=13$
  TeV using 36 fb$^{-1}$ of ATLAS data}},
  \href{https://doi.org/10.1007/JHEP10(2019)203}{\textit{JHEP} {\bfseries 10}
  (2019) 203}, [\href{https://arxiv.org/abs/1908.02746}{{\ttfamily
  1908.02746}}].

\bibitem{Nason:1987xz}
P.~Nason, S.~Dawson and R.~K. Ellis, \textit{{The Total Cross-Section for the
  Production of Heavy Quarks in Hadronic Collisions}},
  \href{https://doi.org/10.1016/0550-3213(88)90422-1}{\textit{Nucl. Phys. B}
  {\bfseries 303} (1988) 607--633}.

\bibitem{Beenakker:1988bq}
W.~Beenakker, H.~Kuijf, W.~L. van Neerven and J.~Smith, \textit{{QCD
  Corrections to Heavy Quark Production in p anti-p Collisions}},
  \href{https://doi.org/10.1103/PhysRevD.40.54}{\textit{Phys. Rev. D}
  {\bfseries 40} (1989) 54--82}.

\bibitem{Beenakker:2002nc}
W.~Beenakker, S.~Dittmaier, M.~Kramer, B.~Plumper, M.~Spira and P.~M. Zerwas,
  \textit{{NLO QCD corrections to t anti-t H production in hadron collisions}},
  \href{https://doi.org/10.1016/S0550-3213(03)00044-0}{\textit{Nucl. Phys. B}
  {\bfseries 653} (2003) 151--203},
  [\href{https://arxiv.org/abs/hep-ph/0211352}{{\ttfamily hep-ph/0211352}}].

\bibitem{Dawson:2002tg}
S.~Dawson, L.~H. Orr, L.~Reina and D.~Wackeroth, \textit{{Associated top quark
  Higgs boson production at the LHC}},
  \href{https://doi.org/10.1103/PhysRevD.67.071503}{\textit{Phys. Rev. D}
  {\bfseries 67} (2003) 071503},
  [\href{https://arxiv.org/abs/hep-ph/0211438}{{\ttfamily hep-ph/0211438}}].

\bibitem{Lazopoulos:2008de}
A.~Lazopoulos, T.~McElmurry, K.~Melnikov and F.~Petriello,
  \textit{{Next-to-leading order QCD corrections to $t \bar{t} Z$ production at
  the LHC}},
  \href{https://doi.org/10.1016/j.physletb.2008.06.073}{\textit{Phys. Lett. B}
  {\bfseries 666} (2008) 62--65},
  [\href{https://arxiv.org/abs/0804.2220}{{\ttfamily 0804.2220}}].

\bibitem{Badger:2010mg}
S.~Badger, J.~M. Campbell and R.~K. Ellis, \textit{{QCD Corrections to the
  Hadronic Production of a Heavy Quark Pair and a W-Boson Including Decay
  Correlations}}, \href{https://doi.org/10.1007/JHEP03(2011)027}{\textit{JHEP}
  {\bfseries 03} (2011) 027},
  [\href{https://arxiv.org/abs/1011.6647}{{\ttfamily 1011.6647}}].

\bibitem{Melnikov:2011ta}
K.~Melnikov, M.~Schulze and A.~Scharf, \textit{{QCD corrections to top quark
  pair production in association with a photon at hadron colliders}},
  \href{https://doi.org/10.1103/PhysRevD.83.074013}{\textit{Phys. Rev. D}
  {\bfseries 83} (2011) 074013},
  [\href{https://arxiv.org/abs/1102.1967}{{\ttfamily 1102.1967}}].

\bibitem{Campbell:2012dh}
J.~M. Campbell and R.~K. Ellis, \textit{{$t \bar{t} W^{+-}$ production and
  decay at NLO}}, \href{https://doi.org/10.1007/JHEP07(2012)052}{\textit{JHEP}
  {\bfseries 07} (2012) 052},
  [\href{https://arxiv.org/abs/1204.5678}{{\ttfamily 1204.5678}}].

\bibitem{Czakon:2013goa}
M.~Czakon, P.~Fiedler and A.~Mitov, \textit{{Total Top-Quark Pair-Production
  Cross Section at Hadron Colliders Through $O(\alpha^4_S)$}},
  \href{https://doi.org/10.1103/PhysRevLett.110.252004}{\textit{Phys. Rev.
  Lett.} {\bfseries 110} (2013) 252004},
  [\href{https://arxiv.org/abs/1303.6254}{{\ttfamily 1303.6254}}].

\bibitem{Buonocore:2023ljm}
L.~Buonocore, S.~Devoto, M.~Grazzini, S.~Kallweit, J.~Mazzitelli, L.~Rottoli
  et~al., \textit{{Precise Predictions for the Associated Production of a W
  Boson with a Top-Antitop Quark Pair at the LHC}},
  \href{https://doi.org/10.1103/PhysRevLett.131.231901}{\textit{Phys. Rev.
  Lett.} {\bfseries 131} (2023) 231901},
  [\href{https://arxiv.org/abs/2306.16311}{{\ttfamily 2306.16311}}].

\bibitem{dEnterria:2015mgr}
D.~d'Enterria, K.~Krajcz\'ar and H.~Paukkunen, \textit{{Top-quark production in
  proton\textendash{}nucleus and nucleus\textendash{}nucleus collisions at LHC
  energies and beyond}},
  \href{https://doi.org/10.1016/j.physletb.2015.04.044}{\textit{Phys. Lett. B}
  {\bfseries 746} (2015) 64--72},
  [\href{https://arxiv.org/abs/1501.05879}{{\ttfamily 1501.05879}}].

\bibitem{dEnterria:2017jyt}
D.~d'Enterria, \textit{{Top-quark and Higgs boson perspectives at heavy-ion
  colliders}},
  \href{https://doi.org/10.1016/j.nuclphysbps.2017.05.053}{\textit{Nucl. Part.
  Phys. Proc.} {\bfseries 289-290} (2017) 237--240},
  [\href{https://arxiv.org/abs/1701.08047}{{\ttfamily 1701.08047}}].

\bibitem{Lee:1977yc}
B.~W. Lee, C.~Quigg and H.~B. Thacker, \textit{{The Strength of Weak
  Interactions at Very High-Energies and the Higgs Boson Mass}},
  \href{https://doi.org/10.1103/PhysRevLett.38.883}{\textit{Phys. Rev. Lett.}
  {\bfseries 38} (1977) 883--885}.

\bibitem{Chanowitz:1985hj}
M.~S. Chanowitz and M.~K. Gaillard, \textit{{The TeV Physics of Strongly
  Interacting W's and Z's}},
  \href{https://doi.org/10.1016/0550-3213(85)90580-2}{\textit{Nucl. Phys. B}
  {\bfseries 261} (1985) 379--431}.

\bibitem{vonBuddenbrock:2020ter}
S.~von Buddenbrock, R.~Ruiz and B.~Mellado, \textit{{Anatomy of inclusive
  $t\bar t W$ production at hadron colliders}},
  \href{https://doi.org/10.1016/j.physletb.2020.135964}{\textit{Phys. Lett. B}
  {\bfseries 811} (2020) 135964},
  [\href{https://arxiv.org/abs/2009.00032}{{\ttfamily 2009.00032}}].

\bibitem{Frederix:2021agh}
R.~Frederix and I.~Tsinikos, \textit{{On improving NLO merging for $
  \mathrm{t}\overline{\mathrm{t}}\mathrm{W} $ production}},
  \href{https://doi.org/10.1007/JHEP11(2021)029}{\textit{JHEP} {\bfseries 11}
  (2021) 029}, [\href{https://arxiv.org/abs/2108.07826}{{\ttfamily
  2108.07826}}].

\end{thebibliography}\endgroup

\end{document}